\documentclass[acmsmall]{acmart}

\setcopyright{acmlicensed}
\copyrightyear{2024}
\acmYear{2023}
\acmDOI{XXXXXXX.XXXXXXX}

\acmJournal{TORS}
\acmVolume{1}
\acmNumber{1}
\acmArticle{1}
\acmMonth{1}
\usepackage{adjustbox}

\usepackage{listings}
\usepackage{multirow} 
\usepackage{caption}
\usepackage{subcaption}
\usepackage{amsmath}
\usepackage{algorithm}
\usepackage{algorithmic}
\usepackage{graphicx}

\usepackage{float}

\begin{document}
%\nolinenumbers
\title{Efficient and Responsible Adaptation of Large Language Models for Robust Top-k Recommendations}\titlenote{This is a substantially expanded version of \citet{kaur2024efficient} that appeared in the 1st Workshop on Risks, Opportunities, and Evaluation of Generative Models in Recommender Systems (ROEGEN@RECSYS'24).This version includes expanded evaluation across three datasets (including BookCrossing and MovieLens), integration of fairness-aware baselines such as BPR+FairRec and GRU4Rec+FairRec, statistical significance testing using paired t-tests, and a refined methodology for weak user identification. These additions enhance both the empirical scope and theoretical depth of the work.}

\author{Kirandeep Kaur}
\affiliation{%
  \institution{University of Washington}
  \city{Seattle}
  \country{United States}}
\email{kaur13@cs.washington.edu}

\author{Vinayak Gupta}
\affiliation{%
  \institution{University of Washington}
  \city{Seattle}
  \country{United States}}
\email{guptavinayak51@gmail.com}

\author{Manya Chadha}
\affiliation{%
  \institution{University of Washington}
  \city{Seattle}
  \country{United States}}
\email{manyac@uw.edu}

\author{Chirag Shah}
\affiliation{%
  \institution{University of Washington}
  \city{Seattle}
  \country{United States}}
\email{chirags@uw.edu}
\renewcommand{\shortauthors}{Kaur et al.}

\begin{abstract}
Conventional recommendation systems (RSs) often optimize for aggregate accuracy, inadvertently underserving users with sparse interaction histories. While large language models (LLMs) offer strong zero-shot capabilities for ranking, their widespread adoption in RS pipelines is hindered by high computational cost, inference latency, and limited scalability. As a result, most existing works evaluate LLMs on a small, randomly sampled subset of users, raising questions about generalizability and real-world applicability in recommendation scenarios. To address these challenges, we propose a hybrid task allocation framework that responsibly allocates ranking tasks between traditional RSs and LLMs to enhance robustness and efficiency. Our strategy works by first identifying weak and inactive users that receive suboptimal ranking performance from RSs. Next, we use an in-context learning approach for such users, wherein each user's interaction history is contextualized as a distinct ranking task. We evaluate our hybrid framework by incorporating eight different recommendation algorithms and three different LLMs -- both open- and closed-sourced. Our results show that the proposed framework significantly reduces the number of weak users (\textasciitilde12\%) while maintaining cost-effectiveness through targeted LLM usage.
\end{abstract}

\maketitle
\section{Introduction} 
\label{sec:introduction}

Recommendation systems (RSs) have become an integral part of numerous online platforms, assisting users in navigating vast amounts of content to relieve information overload~\cite{jacoby1984perspectives}. While Collaborative Filtering based RSs~\cite{papadakis2022collaborative} primarily rely on user-item interactions to predict users' preferences for certain candidate items, the utilization of language in recommendations has been prevalent for decades in hybrid and content-based recommenders, mainly through item descriptions and text-based reviews~\cite{lops2011content}. Furthermore, conversational recommenders~\cite{sun2018conversational} have highlighted language as a primary mechanism for allowing users to naturally and intuitively express their preferences~\cite{google_LLM+RS}. Deep recommendation models, trained under the Empirical Risk Minimization (ERM) framework, uniformly minimize the loss function across all training samples. However, these models fail to cater to a diverse set of sub-populations, affecting robustness and fairness across user groups~\cite{geifman2017deep,wang2023exploring,chaney2018algorithmic,beutel2017beyond,yao2021measuring,zhang2021model,beutel2019fairness}.~\citet{li2021user}  show that active users who have rated many items receive better recommendations on average than inactive users. This performance gap motivates the need for robust RS frameworks that can generalize more robustly across different levels of user activity~\cite{gunawardana2012evaluating}.

%%How LLMs play a role here? Coherent to para1, what's missing in LLMs+RS
Large Language Models (LLMs) like GPT~\cite{achiam2023gpt}, LLaMA~\cite{touvron2023llama}, LaMDA~\cite{collins2021lamda}, Mixtral~\cite{jiang2024mixtral} can effectively analyze and interpret textual data, thus enabling a better understanding of user preferences. These foundation models demonstrate remarkable versatility, adeptly tackling various tasks across multiple domains~\cite{chen2023exploring,bansal2024llm,yang2023harnessing}. However, the field of recommendations is highly domain-specific and requires in-domain knowledge. Consequently, many researchers have sought to adapt LLMs for recommendation tasks~\cite{mohanty2023recommendation,huang2024foundation,fan2023recommender,lin2023can}.~\citet{lin2023can} outline four key stages in integrating LLMs into the recommendation pipeline: user interaction, feature encoding, feature engineering, and scoring/ranking. The purpose of using LLMs as a ranking function aligns closely with general-purpose recommendation models. The transition from traditional library-based book searches to evaluating various products, job applicants, opinions, and potential romantic partners signifies an important societal transformation, emphasizing the considerable responsibility incumbent upon ranking systems~\cite{singh2018}. Existing works that deploy LLMs for ranking~\cite{google_LLM+RS,hou2024large,xu2024prompting,wang2023multiple,ghosh2023jobrecogpt,zhang2023agentcf,tamber2023scaling,wang2023drdt,yue2023llamarec} have proven excellence of LLMs as zero-shot or few-shot re-rankers demonstrating their capabilities in re-ranking with frozen parameters. These works use traditional RSs as candidate item retrieval models to limit the candidate items that need to be ranked by LLM due to a limited context window. Furthermore,~\citet{xu2024prompting,hou2024large} interpret user interaction histories as prompts for LLMs and show that LLMs perform well only when the interaction length is up to a few items, demonstrating the ability of LLMs for (near) cold-start users. Since adapting LLMs can raise concerns around economic and efficiency factors, most of these works train RS on entire datasets but randomly sample interaction histories of some users to evaluate the performance of LLMs, questioning the generalizability of results for all users. This leads us to two important research questions. 
\begin{figure*}[t!]
\centering
\includegraphics[width=0.98\textwidth]{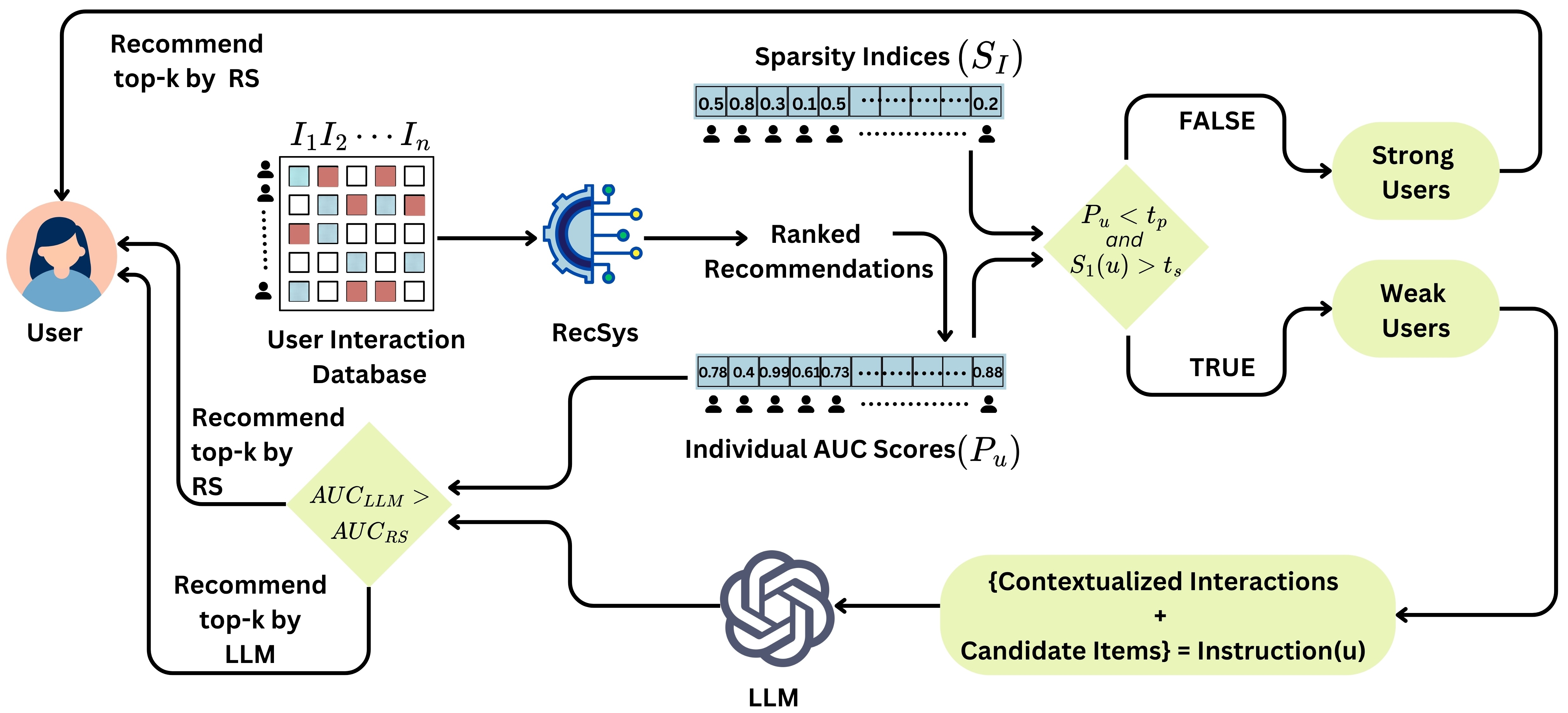}
    \caption{An overview of our framework that uses task allocation to adapt LLMs responsibly. We compute each user's sparsity index ($S_I$), evaluate recommendations retrieved from RS using performance metric ($P(u_m)$), and plot $P(u_m)$ against $S_I$. Interaction histories of highly sparse users with low $P(u_m)$ are contextualized and given to LLM for ranking. Strong users receive RS recommendations, while weak users get LLM recommendations if LLM outperforms RS.}
    \vspace{-10pt}
\label{fig:mainlabel}
\end{figure*}
\begin{itemize}
    \item \textbf{RQ1}: How can we leverage LLMs to enhance inclusivity and robustness to sub-populations in RSs?
    \item \textbf{RQ2}: How can we ensure the scalability of LLMs for ranking in a cost-effective manner? 
\end{itemize}

To address these RQs, we propose a task allocation strategy that leverages LLM and RS's capabilities in a hybrid framework (Fig.~\ref{fig:mainlabel}). Our strategy operates in two phases based on the responsible and strategic selection of tasks for the cost-effective usage of LLMs. First, we identify the users with highly sparse interaction histories on whom the ranking performance of RS is below a certain threshold $t_p$. All such users are termed as \emph{weak users}. In the second phase, interaction histories of weak users are contextualized using in-context learning to demonstrate user preferences as instruction inputs for LLM. While the strong users receive the final recommendations retrieved by RS, weak users receive the recommendations ranked by the LLM if the quality of the ranked list is better than the one given by the RS. 

Our approach aims to improve the robustness and adaptability of LLM-based recommendation systems by targeting users with sparse interaction histories, for whom traditional RSs often perform poorly. These users are prevalent across real-world applications where engagement may be infrequent, episodic, or context-dependent. For instance, educational platforms rely on recommendation engines to personalize learning resources such as videos and practice exercises~\cite{mahendra2024}, but many learners engage sporadically, making accurate personalization challenging. Similarly, job recommendation systems like LinkedIn's talent engine~\cite{geyik2018talent} must surface relevant opportunities for users with limited prior activity, especially those entering the platform for the first time. In healthcare applications, users may interact infrequently with health tracking systems, yet require timely and accurate suggestions related to lifestyle or treatment options~\cite{americanprogress2022}. By selectively applying LLMs to these cases, our framework delivers higher-quality recommendations where RSs struggle, while also minimizing unnecessary model overhead. This leads to a more efficient and scalable hybrid system for diverse recommendation tasks. Unlike prior works that uniformly evaluate LLMs on randomly sampled user sets, our contribution lies in a \emph{strategic task allocation framework} that identifies exactly which users are likely to benefit from LLM augmentation. This improves both computational efficiency and ranking robustness, while avoiding unnecessary inference costs. We focus on how existing RS and LLM systems can be orchestrated more effectively under real-world constraints.

We test our framework based on eight different recommendation models and our results show the efficacy of our strategy, both with open-source as well as closed-source LLMs, in boosting the model robustness to sub-population and data sparsity and improving the quality of recommendations. %For reproducibility and to support research community, our code is available on \hyperlink{https://anonymous.4open.science/r/resp-llmsRS/}{https://anonymous.4open.science/r/resp-llmsRS/} and the link to the video is \hyperlink{https://youtu.be/uqriMZHL-Ng}{https://youtu.be/uqriMZHL-Ng}
In short, the following are our contributions in this paper.

\begin{itemize}
    \item We introduce a {\bf novel hybrid task allocation strategy} that combines the strengths of LLMs and traditional RSs to improve robustness to subpopulations and data sparsity.
    \item Our {\bf unique method} for pinpointing weak users based upon two criteria (user activity and the received recommendation quality below a set threshold) facilitates interventions using LLMs for equitable recommendations.
    \item Our {\bf proposed framework improves the robustness of traditional recommendation models} by reducing weak user count, enhancing recommendation quality, and addressing high costs associated with adapting LLMs.
    \item Our {\bf experiments, both on closed-source and open-source LLMs}, show the efficacy of our framework in improving the model robustness to sub-populations by $(\approx 12\%)$ for varying levels of sparsity and reducing the count of weak users significantly.
\end{itemize}

\section{Related Work}
\label{sec:related_work}
In this section, we review the existing literature on robustness in machine learning, with a particular focus on recommendation systems. We explore various challenges such as data sparsity, distribution shifts, and adversarial attacks, and discuss strategies to enhance model robustness. Additionally, we examine the role of large language models (LLMs) in improving recommendation quality, highlighting their advantages and the associated trade-offs. This review provides a comprehensive overview of the current state of research and identifies gaps that our work aims to address.

\subsection{Robustness in Recommendation Systems}
Robustness in machine learning (ML) targets developing models capable of withstanding the challenges posed by imperfect data in diverse forms~\cite{zhang2019building}. Within the paradigm of recommendations, some existing work developed models resilient to shifts in popularity distribution~\cite{zhang2023invariant, zhang2024robust,10.1145/3583780.3615492} which refers to how changes in item popularity over time can adversely affect the performance of recommendation systems that rely on historical data. Additionally, distribution disparity in train and test datasets~\cite{yang2023generic,wang2024distributionally} which can lead to models that perform well during training but poorly in deployment. These works aim to ensure that models generalize better to the real world by incorporating temporal dynamics and "worst-case" scenarios during training. Some works have even focused on creating frameworks to protect recommendation models against adversarial and data poisoning~\cite{jia2023pore,burke2015robust,wu2023influence,tang2019adversarial,wu2021fight}, which are commonly used with the intention of deceiving models. 
Our work aims to tackle the robustness of recommendation models to data sparsity~\cite{song2022data} and subpopulations~\cite{ovaisi2022rgrecsys}. 

\subsubsection{Addressing Robustness to Data Sparsity and Subpopulations: }
Data sparsity is defined as a condition in which available training data are insufficient for deep learning models to efficiently estimate parameters and achieve high performance. Particularly in the context of sequential recommendation systems, research suggests using data augmentation strategies like noise injection and item masking, to pad the size of the training dataset in order to get better performance~\cite{song2022data, axolotl}. Data sparsity manifested within user interactions can result in poorly performing recommendation models for groups with low interactions. Consequently, it is crucial to evaluate the performance of recommendation models across various sub-populations. A high overall performance metrics may mask poor performance on particular groups, exacerbating the issue of unequal recommendations. Evaluating model performance on specific subsets of users or items within a dataset that share common characteristics or features is one way to measure inequity within such systems~\cite{ovaisi2022rgrecsys}.

Transitioning from data sparsity, it is also crucial to consider the performance of recommendation models across various subpopulations. Evaluating model performance on specific subsets of users or items within a dataset that share common characteristics or features is important, as a high overall performance metric may mask poor performance on particular groups, potentially leading to issues of inequality~\cite{ovaisi2022rgrecsys}.

In line with these issues of data sparsity and robustness to subpopulations,~\citet{li2021user} illustrated that while RSs excel in catering to \emph{active users}, i.e., users who have rated a lot of times, they generally fall short in meeting the overall needs of inactive users. To address this inequality, they proposed a reranking technique that reduced the disparity among active and inactive users. Their results show that such post-processing techniques~\cite{DBLP:journals/corr/YangS16a,DBLP:journals/corr/ZehlikeB0HMB17,DBLP:journals/corr/CelisSV17, revamp, reformd} can harm the average performance in advantaged users to reduce the disparity or reduce the overall utility of the models. Though the in-processing techniques~\cite{pmlr-v81-kamishima18a, DBLP:journals/corr/abs-1805-08716,DBLP:journals/corr/YaoH17} for improving equitable recommendations across various sub-populations can tackle fairness-utility trade-offs, simply adding regularizer term results in sub-optimal performance~\cite{wang2023survey}. In their work, ~\cite{DBLP:journals/corr/abs-1809-09030} extend the HyPER recommendation model to address the issue of data imbalance between what they describe as "protected" and "unprotected" groups. To avoid bias due to imbalanced datasets, they add certain rules to the HyPER framework that offer a structured way to integrate robustness across subpopulations into RSs. ~\cite{DBLP:journals/corr/abs-2104-10671} also contribute to the robustness of RS across subpopulations by introducing a post-processing technique called User-Oriented Group Fairness or UGF that re-ranks recommendation outputs by any kind of RS under certain fairness constraints to balance performance across advantaged and disadvantaged user groups. Most of these works have shown disparity and evaluated existing models by grouping users according to their activity, demographics, and preferences. Similarly,~\citet{wen2022distributionally} developed a Streaming-Distributionally Robust Optimization (S-DRO) framework to improve performance between user subgroups, particularly by accommodating their preferences for popular items. Unlike these, our work first builds upon the existing literature that elicits the issue of performance disparities among active and inactive users and then indicates that although inactive users receive lower-quality recommendations on average, this degradation only affects a subset of inactive users, rather than all inactive users. Unlike these works, our framework identifies weak users - inactive individuals whose preferences traditional recommendation systems struggle to capture effectively.

\subsection{Large Language Models in Recommendation Systems
}
Many researchers have turned to LLMs to address some of these problems because, in recent years, LLMs have proven to be excellent re-rankers and have often outperformed existing SOTA recommendation models in zero-shot and few-shot settings without requiring fine-tuning. %For example,~\citet{Gao2023ChatRECTI} proposed an enhanced recommender system that integrates ChatGPT with traditional RS by synthesizing user-item history, profiles, queries, and dialogue to provide personalized explanations to the recommendations through iterative refinement based on user feedback. 
AgentCF~\cite{zhang2023agentcf}, designed to rank items for users, involves treating users and items as agents and optimizing their interactions collaboratively. While user agents capture user preferences, item agents reflect item characteristics and potential adopters’ preferences. They used collaborative memory-based optimization to ensure agents align better with real-world behaviours. While the retrieval-ranker framework in~\cite{wang2023multiple} remains similar to previous works, authors generate instructions with key values obtained from both users (e.g., gender, age, occupation) and items (e.g., title, rating, category). 

\subsubsection{Challenges and Adaptation Techniques: }Despite the excellence of LLMs as ranking agents, adapting LLMs can involve processing lengthy queries containing numerous interactions from millions of users. Additionally, since LLMs are not specifically trained for recommendation tasks, many works leverage concepts like instruction tuning~\cite{Bao_2023, zhang2023recommendationinstructionfollowinglarge}, pre-training~\cite{geng2023recommendationlanguageprocessingrlp} and prompt tuning~\cite{wang2023zeroshotnextitemrecommendationusing, lyu2024llmrecpersonalizedrecommendationprompting} for LLMs to elicit reasonably good performance from them. Furthermore, each query can raise various economic and latency concerns. Thus, all these works randomly select a few users from the original datasets to evaluate the performance of LLMs. In practice, this user base can involve many more users, which questions the practical applicability of large models for recommendations. However, some recent studies have shown the efficacy of large language models (LLMs) as re-ranking agents to cater to queries with shorter interaction histories compared to lengthy instructions that constitute hundreds of interactions. 

For example, %~\citet{hou2024large} trained recommendation systems to generate candidate item sets and then used user-item interactions to develop instructions. The authors sorted users' rating histories based on timestamps and used in-context learning to design recency-focused prompts. They prompted LLMs to re-rank the candidate items retrieved by the recommendation systems. Their analysis showed decreased performance of LLMs if the candidate item set had more than 20 items. 
ProLLM4Rec \cite{xu2024prompting} adopted a unified framework for prompting LLMs for recommendation. The authors integrated existing recommendation systems and works that use LLMs for recommendations within a single framework. They provided a detailed comparison of the capabilities of LLMs and recommendation systems. Their empirical analysis showed that while state-of-the-art sequential recommendation models like SASRec \cite{kang2018self} improve with a growing number of interactions, LLMs start to perform worse when the number of interactions grows. Furthermore, both of these works sampled some users to evaluate the performance of LLMs due to the high adaptation costs. To investigate the effectiveness of various prompting strategies %\citet{google_LLM+RS} focused on a (near) cold-start scenario where minimal interaction data is available. They used various prompting techniques to provide a natural language summary of preferences to enhance user satisfaction by offering a personalized experience. By exploiting rich positive and negative descriptive content and item preferences within a unified framework, they compared the efficacy of prompting paradigms with large language models against collaborative filtering baselines that rely solely on item ratings. Similarly, other works like 
~\cite{wang2023zeroshotnextitemrecommendationusing, zhang2023recommendationinstructionfollowinglarge} have also analyzed the efficacy of different prompting strategies on recommendation quality. For instance, ~\cite{wang2023zeroshotnextitemrecommendationusing} use a 3-step prompting strategy to generate a ranked list of 10 recommendations from a pre-constructed candidate set. The candidate set was constructed using user filtering and item filtering to narrow down the recommendation space and was found to improve performance. It was also observed that the ideal size for a candidate set was around 19 items for optimal performance. ~\cite{zhang2023recommendationinstructionfollowinglarge} leveraged pre-trained, instruction tuned LLMS like FLAN-T5 to generate recommendations through prompting using task-specific templates alone, and avoided the computational expenses of fine-tuning altogether. 

Furthermore, ~\cite{deldjoo2024understandingbiaseschatgptbasedrecommender} demonstrates how tailored prompt designs significantly influence the accuracy, robustness, and diversity of recommendations. This work categorizes prompts into accuracy-oriented, beyond-accuracy, and reasoning-oriented groups, showcasing how design choices can target diverse objectives such as improving novelty, temporal recency, or logical reasoning behind recommendations. Notably, diversification prompts improved genre diversity and temporal freshness but came at the cost of reduced accuracy, illustrating the trade-offs inherent in prompt-based strategies.

In summary, past works suggest that despite the high costs associated with adapting LLMs for recommendations, these models can outperform existing recommendation models significantly. Moreover, we acknowledge that the literature shows the contrasting capabilities of both RSs and LLMs -- RSs fail to perform well on inactive users due to sparse interaction vectors, and in contrast, LLMs can be prompted to cater to inactive users in near cold-start settings without requiring any fine-tuning.

Building upon these crucial insights, our framework first aims to identify the weak users for whom RS finds it hard to capture their preferences accurately. We then use in-context learning to prompt LLMs to generate recommendations for such users. While past works like ProLLM4Rec by~\cite{xu2024prompting}, dynamic reflection with
divergent thinking within a retriever-reranked by~\cite{wang2023drdt}, recency-focused prompting by~\cite{hou2024large} and aligning ChatGPT with conventional ranking techniques such as point-wise, pair-wise, and list-wise ranking by~\cite{dai2023uncovering} are all different techniques to design prompts with different variations, our main contribution lies in the responsible task allocation within recommendation systems and all such techniques can be used within our framework for designing prompts. In the next section, we discuss our methodology in detail.

\section{Methodology}
\label{sec:methodology}
We begin here by providing a formal definition of the existing problem. We then discuss our framework, which adopts a hybrid structure by leveraging the capabilities of both traditional RSs and LLMs. For this, we first identify users for whom RSs do not perform well and then leverage LLMs for these users to demonstrate user preferences using in-context learning.

\subsection{Problem Formulation}
Consider a recommendation dataset $\mathcal{D}$ with $k$ data points. Let $U = \{u_1,u_2,\dots,u_M\}$ be the set of users and $|U| = M$ represents the number of users in $\mathcal{D}$. Let $I = \{i_1,i_2,\dots,i_N\}$ be the set of all the items and $|I| = N$ represents the number of items in $\mathcal{D}$.

\begin{align}
\mathcal{D} = \{(u_m,i_n,r_{mn}): m=1,2,\dots,M; n = 1,2,\dots,N\}
\end{align}

Here, the triplet $d_{mn} = (u_m,i_n,r_{mn})$  represents one data point where a user $u_m$ provided a rating of $r_{mn}$ to an item $i_n$. Now, if a user $u_m$ has rated a set of items, then let $[r_{mn}]_{n=1}^{N}$ denote the rating vector consisting of explicit rating values ranging from $1$ to $5$ if a user provided a rating and $0$ indicating no rating. 

Our goal is to enhance the robustness of recommendation systems by addressing the challenges posed by weak users — those who possess extremely sparse rating vectors and for whom traditional recommendation models, denoted as $\theta^r$, perform poorly. The first step involves determining criteria to rank users based on the performance of the recommendation system on each user. Subsequently, we aim to understand user characteristics to identify extremely weak users.

For each identified weak user, we contextualize their interaction history as a distinct recommendation task. These tasks are then allocated to a Large Language Model (LLM) to generate personalized recommendations. The objective is to optimize task allocation to the LLM, thereby improving recommendation performance for sub-populations and addressing data sparsity while managing computational costs effectively.

\subsection{Identifying Weak Users}
We consider two criteria for identifying weak users for recommendation model $\theta^r$. First, given $K$ users and their associated rating vectors, we evaluate how well the model could rank the relevant user items, often termed as \emph{positive items} above the irrelevant or \emph{negative items}. Let $r$ denote the rank of the relevant item, and $r'$ be the rank of irrelevant items. The indicator function $\delta(r < r')$ outputs one if the rank of the relevant item $r$ is higher than that of the irrelevant item $r'$:

\begin{align}
    \delta(r < r')
\end{align}

\noindent
Let $N$ denote the total number of items and $|R|$ be the set of all relevant items.  Then, similar to~\citet{rendle2012bpr}, we use AUC measure to evaluate how hard it was for $\theta^r$ to rank items preferred by a certain user, given by

\begin{align}
    \mathrm{\mathcal{P}(u)} &= \frac{1}{|R|(N-|R|)} \sum_{r \in R} \sum_{r' \in \{1, \dots, N\} \setminus R} \delta(r < r')
    \label{eq:p(u)}
\end{align}

Here, $|R|(n-|R|)$ denotes all possible pairs of relevant and irrelevant items.

We acknowledge that various metrics like NDCG, F1, precision and recall have been used to measure the quality of ranking ability of recommendation models. However, these metrics place significant importance on the outcomes of the top-k items in the list and completely ignore the tail. For identifying weak users, we require a metric consistent under sampling i.e. if a recommendation model tends to give better recommendations than another on average across all the data, it should still tend to do so even if we only look at a smaller part of the data. The aim of our framework is a clear task distribution. The performance of top-k metrics varies with $k$, and this might raise uncertainty as $k$ varies with varying users and platforms. Nevertheless, AUC is the only metric which remains consistent under-sampling, and as $k$ reduces, all top-k metrics collapse to AUC. For more details, we refer the readers to~\cite{krichene2020sampled}. %For continuity, we use the definition and proof~\cite{krichene2020sampled} to show that our metric is consistent under-sampling.

Past works~\cite{li2021user} have shown that \emph{active} users that provide more ratings receive better recommendations than the \emph{inactive} users on average. However, only a few inactive users might receive irrelevant recommendations individually (Fig.~\ref{fig:scatterplot}). Thus, we evaluate each user's activity. Let a user $u$ rated $|R|$ items out of a total of $N$ items. Then sparsity index $\mathcal{S}_I$ associated with a given user $u$ can be calculated as:

\begin{align}
    \mathcal{S}_I(u) = 1 - \frac{|R|}{N} 
    \label{eq:s(u)}
\end{align}

If this value falls above a certain threshold $t_s$, the user is considered as inactive. Combining with the weak user identification, we obtain,

\noindent\textbf{Definition 1.} Given dataset \( \mathcal{D} \) and a recommendation model \( \theta^r \), 
we say that a user \( u_m \) is extremely weak if the likelihood of \( \theta^r \) being able to 
rank the relevant items above the irrelevant items is below \( t_p \), and the user’s 
interaction vector \( \mathbf{r}_{m*} \) exhibits extremely high sparsity i.e., exceeds \( t_s \).
\begin{equation}
    \mathcal{P}(u_m) \leq t_p \quad \text{and} \quad S_I(u_m) > t_s
\end{equation}

It is important to note that a higher AUC value implies better performance, and the value always lies between 0 to 1. Studies have shown that an AUC value below 0.5 indicates that the model performs worse than random guessing, which is undesirable in ranking models. For instance, research on reducing popularity bias in recommender systems highlights that an AUC value below 0.5 would indicate poor ranking performance \cite{ReducingPopularityBias2023}. Additionally, a critical analysis of AUC variants emphasizes that such values suggest the model is not effectively distinguishing between positive and negative instances \cite{CriticalAnalysisAUC2008}. Furthermore, discussions on solving AUC less than 0.5 problems indicate that these values often point to significant issues like underfitting or incorrect model structure \cite{SolveAUCProblem2019}. Thus we use, $t_s = 0.5$ as threshold.  In addition, we use $t_s = avg(\mathcal{S}_I(D))$, the average sparsity of all users in $\mathcal{D}$ i.e.,\\ $avg(\mathcal{S}_I(D)) = 1/m * \sum_{j=1}^m \mathcal{S}_I(u_j)$ for determining this threshold. This approach is supported by literature that addresses sparsity issues in recommender systems. For example, a systematic literature review highlights the importance of managing user sparsity to improve recommendation accuracy, suggesting that average sparsity can serve as a useful benchmark \cite{A_Systematic_Literature_Review_of_Sparsity_Issues_in_Recommender_Systems}. Additionally, research on cold start and data sparsity problems emphasizes the significance of using average sparsity as a threshold to identify sparse users and enhance system performance \cite{aggarwal2016introduction}.

\begin{figure}
    \centering
   \includegraphics[width=0.8\textwidth]{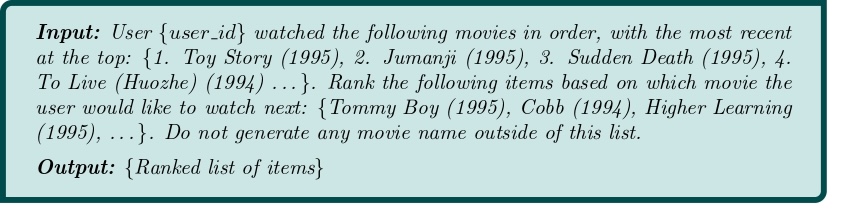}
    \caption{Instruction template for contextualizing interaction histories of weak users.}
    \label{fig:template1}
\end{figure}
\subsection{Designing Natural Language Instructions for Ranking} 
Closest to our work,~\citet{hou2024large} formalized the recommendation problem as a \emph{conditional} ranking task considering sequential interaction histories as conditions and uses the items retrieved by traditional RS as $candidate$ items. While we aim to design the conditional ranking tasks, our approach differs significantly from theirs as instead of using LLMs as a re-ranking agent for all users; we instruct LLM with the preferences of weak users preferences (\emph{sorted in descending order of decreased preference}). This technique is detailed below. 

For each user, we use in-context learning to instruct LLM about user preferences \emph{conditions} and assign the task of ranking the \emph{candidate} items. For a user $u$, let $\mathcal{H}_{u} = \{i_1,i_2,\dots,i_n\}$ depict the user interaction histories sorted in decreasing order of preference and $\mathcal{C}_{u} = \{i_1,i_2,\dots,i_j\}$ be the candidate items to be ranked. Then, each instruction can be generated as a sum of conditions and candidate items, i.e., 

\begin{align}
    \mathcal{I}_{u} = \mathcal{H}_{u} + \mathcal{C}_{u} 
    \label{eq:I(u)}
\end{align}

\textbf{In-context learning:} We use in-context learning to provide a demonstration of the user preferences to LLM using certain examples. As suggested by~\citet{hou2024large}, providing examples of other users may introduce extra noise if a user has different preferences. Therefore, we sort every weak user's preferences based on the timestamp. or example, \emph{"User ${user_id}$ watched the following movies in order, with the most recent at the top: 1. Harry Potter, 2. Jurassic Park, \dots"}. This forms the condition part of the instruction. We then select items top items recommended by traditional RS as test items as candidate items and instruct LLM to recommend which movie user would like to watch next and rank all movies accordingly: \emph{"Now, rank the following items based on which movie the user would like to watch next: Multiplicity, Dune, \dots"}.
It is important to note that while the presentation order in conditions plays a significant role in demonstrating user preferences to LLM, we deliberately shuffle the candidate items to test the ability of LLM to rank correctly. Since LLMs can generate items out of the set, we specially instruct to restrict recommendations to the candidate set. Fig.~\ref{fig:template1} shows the final template of the instruction given to LLM for a particular user. We use the same template for all identified weak users to contextualize their past interactions into a ranking task.

\begin{algorithm}[t!]
\caption{Hybrid LLM-RecSys Algorithm for Ranking}
\label{alg:framework}
\begin{algorithmic}[1]
    \STATE \textbf{Input}: $\mathcal{D}_{train}$: training dataset; $\mathcal{D}_{test}$: test dataset; $\mathcal{U}$: set of users; $\mathcal{S}_I$: Sparsity index for all users; $\theta^r$: recommendation algorithm; $\theta^l$: large language model, $t_{s}$: sparsity threshold, $t_{p}$: performance threshold.
    \STATE \textbf{Output}: $ranked\_pred_{strong}$: ranked lists of items for strong users,  $ranked\_pred_{weak}$: ranked lists of items for weak users.
    \STATE $ranked\_pred \leftarrow \theta^r(\mathcal{D}_{train}$)
    \FOR{ each user $u_{m} \in \mathcal{U}$}
        \STATE Calculate $\mathcal{P}(u_m)$ using Eq.~\ref{eq:p(u)}
        \STATE Calculate $\mathcal{S}(u_m)$ using Eq.~\ref{eq:s(u)}
        \IF{$\mathcal{P}(u_m) < t_{p}$ $\&\&$ $\mathcal{S}_I(u_m)$ }
            \STATE $\mathcal{U}_{weak} \leftarrow u_m$
        \ELSE
            \STATE $\mathcal{U}_{strong} \leftarrow u_m$
            \STATE $ranked\_pred_{strong} \leftarrow ranked\_pred[u_m]$
        \ENDIF
        
    \ENDFOR
    \FOR{each $u_{i} \in \mathcal{U}_{weak} $}
        \STATE Generate  instruction $\mathcal{I}_{u_i}$ using Eq. \ref{eq:I(u)}
        \STATE $ranked\_list_{u_{i}}$ = $\theta^{L}(\mathcal{I}_u)$
        \STATE $ranked\_list_{weak} \leftarrow ranked\_list_{u_{i}}$ 
    \ENDFOR
\end{algorithmic}
\end{algorithm}
\subsection{Our Framework}
This section discusses the workflow adopted by our framework as depicted in Fig.~\ref{fig:mainlabel} and corresponding Algorithm~\ref{alg:framework}. Initially, the model takes as input the training dataset $\mathcal{D}_{train}$ and test dataset $\mathcal{D}_{test}$, a set of users $\mathcal{U}$, a recommendation model $\theta^r$, a large language model $\theta^l$, and two thresholds: sparsity threshold $t_s$ and performance threshold $t_p$. These thresholds depict the minimum sparsity and performance values for a user to be classified as a strong user. It is important to note that splitting data will not yield a mutually exclusive set of users in both sets, but item ratings for each user in $\mathcal{D}_{train}$ will differ from those in $\mathcal{D}_{test}$.

The algorithm begins by training the recommendation model $\theta^{r}$ on the training set $\mathcal{D}_{train}$ and provides ranked items for all users. Using $\mathcal{D}_{test}$, we test the ranking ability of the model for each user by evaluating $\mathcal{P}(u_m)$ using Eq.~\ref{eq:p(u)}. Further, each user is also assigned a sparsity score $\mathcal{S}_I(u)$ evaluated using Eq.~\ref{eq:s(u)}. If $\mathcal{P}(u)$ has a value less than $t_p$ and the sparsity index $\mathcal{S}_I(u)$ for a particular user falls below $t_s$, the user is termed as a weak user. While previous works have shown that, on average, inactive users receive poor performance, we pinpoint weak users by evaluating both the sparsity and performance.

For all such weak users, we convert rating histories from $\mathcal{D}_{train}$ as conditions $\mathcal{H}_{u}$ using in-context learning and use test items as candidate items $\mathcal{C}_u$ for testing purposes. However, in practice, these candidate items can be replaced by unobserved items. The final instructions are generated by combining conditions and candidate items as depicted by (Eq.\ref{eq:I(u)}). These instructions are given to the LLM, which provides a ranked list of items for each user. For all the strong users, the recommendations presented are the ones ranked by the conventional recommendation model. However, the weak users receive final ranked lists generated by the LLM.

\section{Experiments}
This section discusses our experimental setup with details of the datasets and models used, followed by the implementation details of all these models and various metrics used. We finally present empirical results and a comparative analysis of various recommendation models and LLMs.
\subsection{Experimental Setup}
\subsubsection{Datasets.}  
To evaluate the effectiveness of our framework, we conducted experiments on three real-world datasets, each presenting unique challenges and characteristics.  

\begin{itemize}  
    \item \textbf{ML-1M (MovieLens 1M):} Widely used in recommendation system research, this dataset contains over 1 million ratings from users on movies, with each user rating at least 20 movies. The dataset includes metadata such as movie titles, genres, and timestamps\footnote{\url{https://grouplens.org/datasets/movielens/1m/}}.  

    \item \textbf{Amazon Software:} A subset of the Amazon Review Dataset focused on software products, featuring user reviews, ratings, and metadata like timestamps and product information. This dataset is particularly challenging due to its inherent sparsity, with many products receiving limited reviews\footnote{\url{https://datarepo.eng.ucsd.edu/mcauley_group/data/amazon_2023/raw/review_categories/Software}}.  

    \item \textbf{Amazon Video Games:} Another subset of the Amazon Review Dataset, this dataset concentrates on video games and provides user reviews, ratings, and product metadata. Its diversity and long-tail nature make it ideal for experiments on niche product recommendations\footnote{\url{https://datarepo.eng.ucsd.edu/mcauley_group/data/amazon_2023/raw/review_categories/Video_Games}}.  
\end{itemize}  
\begin{table}[t!]  
\centering  
\begin{tabular}{lccc}  
\toprule  
\textbf{Dataset} & \textbf{\# Users} & \textbf{\# Items} & \textbf{Sparsity} \\  
\midrule  
ML-1M            & 6,036             & 3,416             & 0.9518           \\  
Amazon Software  & 6,102             & 6,049             & 0.9951           \\  
Amazon Video Games & 2,289           & 4,754             & 0.9932           \\  
\bottomrule  
\end{tabular}  
\caption{Dataset statistics after preprocessing.}  
\label{tab:dataset_summary}  
\vspace{-9pt}
\end{table} 
To preprocess the data, we utilized the same methodology as outlined in most of our existing baseline models. \cite{rendle2010factorizing,he2017translation,he2016fusing,kang2018self}. \noindent Table~\ref{tab:dataset_summary} provides a summary of the dataset statistics after preprocessing. The ML-1M dataset contains 6,036 users and 3,416 items, resulting in a sparsity of 0.9518, indicating that 4.8\% of potential interactions are observed. The Amazon Software dataset includes 6,102 users and 6,049 items, with a sparsity of 0.9951, meaning only 0.49\% of possible interactions are recorded. Similarly, the Amazon Video Games dataset has 2,289 users and 4,754 items, with a sparsity of 0.9932 (0.68\% observed interactions). The sparsity of the Amazon datasets highlights the challenge of dealing with limited interactions, while the ML-1M dataset serves as a relatively denser benchmark. Together, these datasets cover diverse domains and challenges, ensuring a comprehensive evaluation of our framework.

\subsubsection{Baselines and Models.} To leverage the strengths of both traditional recommendation systems and advanced large language models, our hybrid framework integrates a diverse set of baseline models. These models are carefully selected to cover various recommendation strategies, ensuring a comprehensive evaluation of our framework's capabilities. We mainly evaluate three different types of recommendation models as discussed below.
:
\begin{itemize}
    \item \textbf{Collaborative-filtering based models}: Collaborative filtering (CF) methods predict a user's interests by collecting preferences from many users and operate on the principle that users who agreed in the past will agree in the future and that they will like similar kinds of items. We mainly use the following two CF-based RS:
    \begin{itemize} 
        \item \textbf{ItemKNN}~\cite{sarwar2001item} uses a k-nearest neighbors approach to find items that are most similar to the items a user has rated or interacted with. The similarity between items is typically calculated using cosine similarity or Pearson correlation. ItemKNN is efficient and scalable, making it suitable for large datasets.
        \item \textbf{NeuMF (Neural Matrix Factorization)}~\cite{he2017neural} combines Generalized Matrix Factorization (GMF) and Multi-Layer Perceptron (MLP) to capture both linear and non-linear user-item interactions. GMF models the linear interactions between users and items by performing element-wise multiplication of user and item latent vectors. MLP, on the other hand, captures the non-linear interactions by concatenating user and item embeddings and passing them through multiple hidden layers. The outputs of GMF and MLP are then concatenated and fed into a final prediction layer.
        \item \textbf{NNCF (Neighborhood-based Neural Collaborative Filtering)}~\cite{10.1145/3132847.3133083} integrates \\neighborhood-based methods with neural collaborative filtering to enhance recommendation accuracy. It captures localized information by considering the interactions within a user's neighborhood, complementing the user-item interaction data. This model is particularly effective in scenarios where local context plays a significant role in user preferences. 
        \item \textbf{DMF (Deep Matrix Factorization)}~\cite{xue2017deep} constructs a user-item matrix with explicit ratings and non-preference implicit feedback, and uses a deep learning structure to learn a common low-dimensional space for user and item representations. This model effectively captures both explicit and implicit feedback, optimizing a new loss function based on binary cross-entropy.
    \end{itemize} 

    \item \textbf{Sequential recommendation models}: Such models focus on capturing the order and timing of user interactions to predict future behavior. These models are particularly useful in scenarios where the sequence of actions provides significant information about user preferences. By understanding the patterns in user behavior over time, sequential models can make more accurate and contextually relevant recommendations. We evaluate our model on three different types of sequential RSs. 
    \begin{itemize}
        \item \textbf{BERT4Rec}~\cite{sun2019bert4rec} uses Bidirectional Encoder Representations from Transformers (BERT) to capture sequential patterns in user interactions. It predicts the next item by considering both past and future contexts, improving recommendation accuracy.
        \item \textbf{GRU4Rec}~\cite{10.1145/2988450.2988452} employs Gated Recurrent Units (GRUs) to model session-based recommendations. It is effective in capturing the sequential dependencies in user behavior within a session.
        \item \textbf{SASRec}~\cite{kang2018self} utilizes self-attention mechanisms to identify relevant items from a user's action history. It balances the ability to capture long-term dependencies with computational efficiency.
    \end{itemize}
    \item \textbf{Learning-to-rank model} uses techniques for training a model to solve ranking tasks. These models are typically used to rank items in a way that maximizes relevance to the user. These models are typically used to rank items in a way that maximizes relevance to the user. Such models are particularly effective in scenarios where the goal is to order items based on their predicted relevance or preference. We use the most popular learning-to-rank RS for our framework: 
    \begin{itemize}
        \item \textbf{Bayesian Personalized Ranking (BPR)}~\cite{rendle2012bpr} is a pairwise ranking model optimized for implicit feedback. t assumes that users prefer observed items over unobserved ones and is effective in generating personalized ranked lists. BPR is particularly useful for collaborative filtering tasks where explicit ratings are not available, and it focuses on maximizing the difference in ranking between observed and unobserved items.
    \end{itemize}

    \item \textbf{Fairness-Aware Baseline:} Since the adaptation of LLMs for ranking tasks can be interpreted as a post-hoc re-ranking strategy, we include a well-established fairness-oriented post-hoc baseline for comparison:
    \begin{itemize}
        \item \textbf{FairRec}~\cite{li2021user}: FairRec is a user-oriented fairness framework that aims to balance recommendation quality across user groups, particularly between advantaged (active) and disadvantaged (inactive) users. It formulates the fairness-constrained re-ranking problem as a 0-1 integer programming task, ensuring that the final ranking optimizes utility while satisfying fairness constraints. FairRec has been shown to significantly improve fairness as performance disparity in active and inactive users.
    \end{itemize}

\end{itemize}

While these models identify weak users and generate candidate items, LLMs are further deployed to enhance performance for such users. We use both open (LLaMA 3-70B-Instruct) and closed-source (Claude 3.5, GPT-4) LLMs to test the capability of the proposed framework.

Developed by OpenAI, \textbf{GPT-4}~\cite{openai2023gpt4} is a multimodal large language model known for its advanced reasoning and instruction-following capabilities. It has been trained on a diverse dataset and is designed to produce safer and more useful responses. Compared to its predecessors, GPT-4 features a larger model size, improved context retention, and enhanced safety measures. \textbf{LLaMA 3-70B-Instruct}~\cite{touvron2024llama3}, created by Meta AI, is optimized for efficiency and performance. It is designed to handle a wide range of tasks with a focus on providing accurate and contextually relevant responses. The LLaMA 3-70B-Instruct model is particularly noted for its instruction-following capabilities and customizable nature. \textbf{Claude 3.5}~\cite{anthropic2024claude}, developed by Anthropic, is a family of large language models that includes versions optimized for speed, performance, and complex reasoning tasks. 

\subsubsection{Implementation details.} 

For ease of reproducibility, we use the open-source recommendation library \textbf{RecBole}~\cite{recbole[2.0]} for implementing all recommendation models. RecBole is a comprehensive and flexible library that supports over 100 widely used recommendation algorithms, covering general, sequential, context-aware, and knowledge-based recommendations. This ensures that our implementation is robust and aligns with current research standards.
To access the LLMs, we utilize API calls. GPT-4 was accessed using the API provided by OpenAI~\footnote{https://platform.openai.com/docs/api-reference}. This API allowed us to integrate GPT4 -4's advanced capabilities into our framework seamlessly. For the other two LLMs, we used \textbf{AWS Bedrock}. AWS Bedrock is a fully managed service that makes high-performing foundation models from leading AI companies and Amazon available through a unified API~\cite{awsbedrock}. This service allows us to experiment with and evaluate top foundation models, customize them with our data, and integrate them into our hybrid framework efficiently. \\

\noindent\emph{Evaluation Protocol.} For evaluation, we adopt the leave-one-out (LOO) setting as utilized by all the baseline models~\cite{ he2017neural, 10.1145/3132847.3133083, sun2019bert4rec, kang2018self, rendle2012bpr}. For each user, we split their interaction history as follows: the most recent interaction is designated as the test item, the second most recent interaction is used as the validation item, and all remaining interactions are utilized for training. This approach maximizes the use of available data for training while providing a realistic and stringent test of the model's predictive capabilities. By focusing on the most recent interactions for testing and validation, we closely mimic real-world scenarios where the goal is to predict the next item a user will interact with. This method allows us to evaluate the model's performance in a detailed and user-specific manner, ensuring that our framework is both effective and reliable in practical applications.\\

\noindent\emph{Hyperparameter Search.} We conducted an extensive hyperparameter search for each model to ensure optimal performance. The hyperparameter configurations were chosen based on the model’s characteristics and the specific requirements of our recommendation task. 
For the \emph{BPR (Bayesian Personalized Ranking)} model, we focused on tuning the learning rate, as it directly influences the convergence speed and stability of the model. We explored a wide range of learning rates: \([5e\text{-}5, 1e\text{-}4, 5e\text{-}4, 7e\text{-}4, 1e\text{-}3, 5e\text{-}3, 7e\text{-}3]\) to balance the trade-off between fast convergence and maintaining model accuracy.
In the case of \emph{ItemKNN (Item-based K-Nearest Neighbors)}, we varied two important hyperparameters: \(k\), the number of neighbors considered for each item, and the shrinkage parameter, which controls the weight of the neighbors. The values for \(k\) ranged from 10 to 400 (\([10, 50, 100, 200, 250, 300, 400]\)), and the shrinkage parameter took values \([0.0, 0.1, 0.5, 1, 2]\), as adjusting these parameters helps in fine-tuning the balance between overfitting and generalization. For \emph{NeuMF (Neural Matrix Factorization)}, the hyperparameters include the learning rate, which influences the speed of model optimization, and the architecture of the neural network, specifically the hidden layer sizes. We tested learning rates from \([5e\text{-}7, 1e\text{-}6, 5e\text{-}6, 1e\text{-}5, 1e\text{-}4, 1e\text{-}3]\) and used a multi-layer perceptron (MLP) structure with hidden sizes of \([64, 32, 16]\). Dropout probabilities (\([0.0, 0.1, 0.3]\)) were also explored to prevent overfitting by randomly deactivating neurons during training. For \emph{NNCF (Neural Collaborative Filtering)}, we explored a wider range of hyperparameters, as this model is more complex. We varied the number of neighbors (\([20, 50, 100]\)) and the size of the neighbor embeddings (\([64, 32]\)), which affect how much information from neighboring items is used. Additionally, the number of convolution kernels (\([128, 64]\)) was adjusted to optimize the feature extraction process, while the learning rate was tuned with values \([5e\text{-}5, 1e\text{-}4, 5e\text{-}4]\). We also experimented with two methods for aggregating neighborhood information: \texttt{random} and \texttt{knn}, which can influence the model's ability to capture complex patterns in the data. For \emph{BERT4Rec}, we focused on the learning rate (\([0.0003, 0.0005, 0.001, 0.003]\)), which is crucial for fine-tuning the pre-trained model. We also experimented with different fine-tuning ratios (\([0, 0.1, 0.5]\)) to control the degree of adaptation to the specific task at hand, which can help in leveraging domain-specific knowledge while retaining the pre-trained features. For \emph{GRU4Rec}, we varied the learning rate (\([0.005, 0.001, 0.0005, 0.0001]\)) to determine the optimal step size for model training, and the number of layers (\([1, 2]\)) to assess the impact of model depth on recommendation quality. Finally, for \emph{SASRec}, we focused on the learning rate alone, testing values in \([0.0003, 0.0005, 0.001, 0.003, 0.005]\), as it plays a critical role in controlling the model's learning speed and preventing overfitting. This comprehensive hyperparameter search allowed us to explore a wide range of configurations for each model, ensuring that we could optimize each one for the specific characteristics of our datasets and tasks.\\

\noindent\emph{Evaluation Metrics.} We adopt the evaluation protocol presented by the recently released toolkit, RGRecSys~\cite{ovaisi2022rgrecsys}, to assess the robustness of recommendation models to sub-populations. Specifically, we utilize two widely recognized metrics—Normalized Discounted Cumulative Gain (NDCG) and Area Under the Receiver Operating Characteristic Curve (AUC)—for this evaluation. The choice of AUC is particularly important in measuring the difficulty associated with each user for a given recommendation model, as AUC exhibits a consistency property. This property ensures that the ranking quality is invariant to changes in the distribution of positive and negative samples, making it a reliable metric for comparing model performance across different user sub-populations.

To calculate AUC, we leverage the popular CatBoost\footnote{https://github.com/catboost/} library, which provides a robust and efficient AUC implementation specifically designed for ranking tasks. Additionally, we report final NDCG@10 scores, which focus on the quality of the top 10 ranked items, a critical aspect in recommender systems where relevance at the top of the list is prioritized.\\

\noindent\emph{Language Model Parameters.} For the language models (LLMs) used in the experiments, we set two key parameters: \textbf{temperature} and \textbf{top-p sampling}. The \textbf{temperature} parameter controls the randomness of the predictions generated by the model. A temperature of 0 results in deterministic outputs, where the model selects the highest probability item, effectively reducing randomness in the generated items. By setting the temperature to 0, we minimize the chances of generating irrelevant or hallucinated items, ensuring that only contextually appropriate recommendations are made.
The \textbf{top-p sampling} parameter, also known as nucleus sampling, is used to control the diversity of the generated outputs. Top-p sampling ensures that the model only considers a subset of the most likely next tokens, where the cumulative probability of the chosen tokens is less than or equal to a specified threshold \(p\). We set \(p\) to 0.1, meaning that the model considers only the top 10\% of the probability distribution, effectively filtering out lower probability, less relevant items. This reduces the likelihood of generating out-of-list items or hallucinations, which are not part of the original candidate list. Additionally, we removed any items that were not originally present in the candidate list to maintain consistency with the recommendation process. This step ensures that the evaluation is conducted strictly within the boundaries of the dataset’s available items.

In the following sections, we present and discuss the empirical inferences derived from these experiments, which were conducted following the aforementioned protocols and settings.

\subsection{Empirical Evaluation}
In this section, we conduct an in-depth empirical evaluation to analyze the performance of our proposed framework. Our analysis focuses on building criteria to identify weak users and categorize weak and strong users, to examine how large language models (LLMs) can enhance robustness, and to assess the impact on different types of users.

\subsubsection{Phase One: Identifying Weak and Strong Users} 
\begin{figure*}[t!]
    \centering
    \begin{subfigure}[b]{0.3\textwidth}
        \centering
        \includegraphics[width=\textwidth]{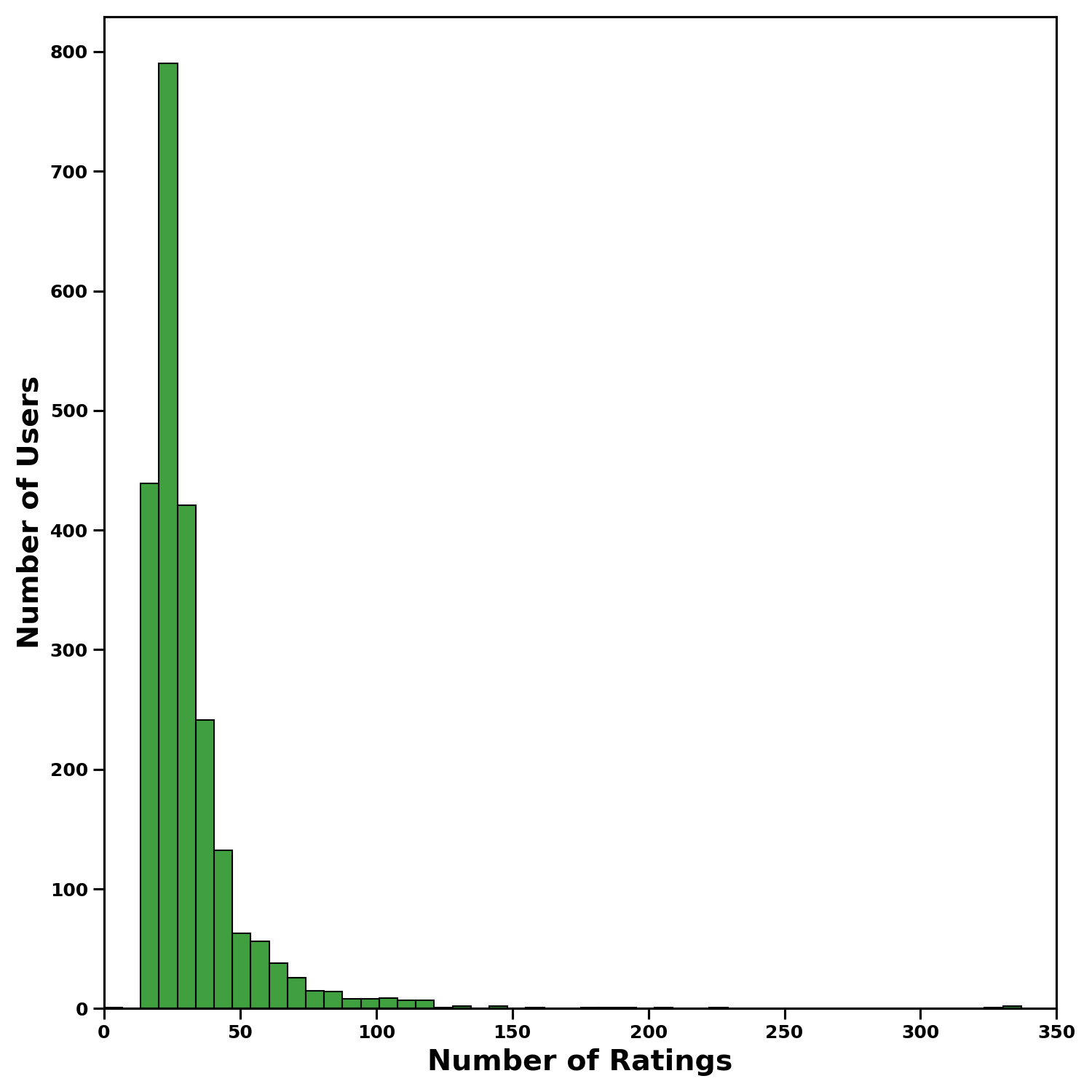}
        \caption{Amazon Software}
        \label{sfig:games_dist}
    \end{subfigure}
    %\hfill
    \begin{subfigure}[b]{0.3\textwidth}
        \centering
        \includegraphics[width=\textwidth]{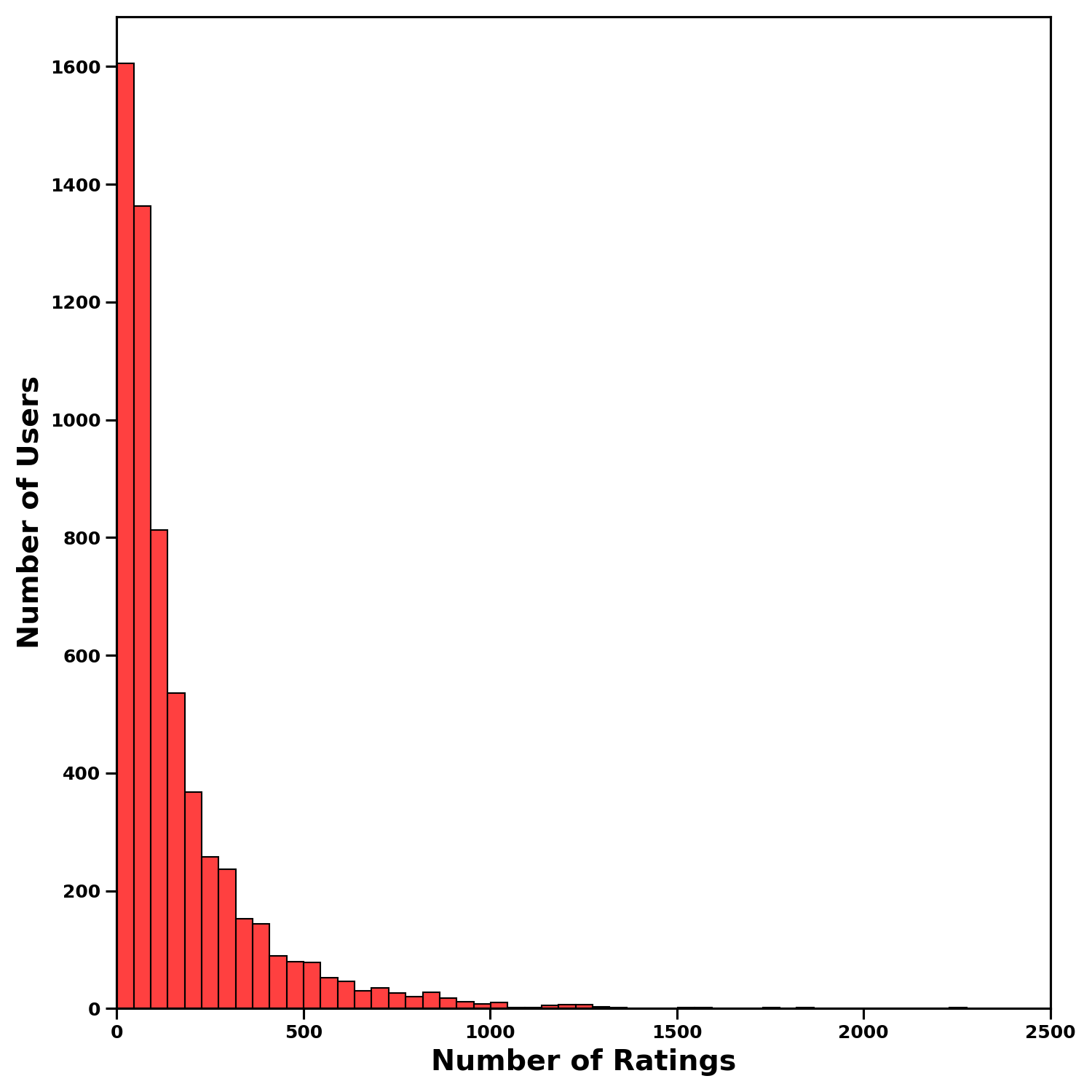}
        \caption{ML1M}
        \label{sfig:ml1m_dist}
    \end{subfigure}
    %\hfill
    \begin{subfigure}[b]{0.3\textwidth}
        \centering
        \includegraphics[width=\textwidth]{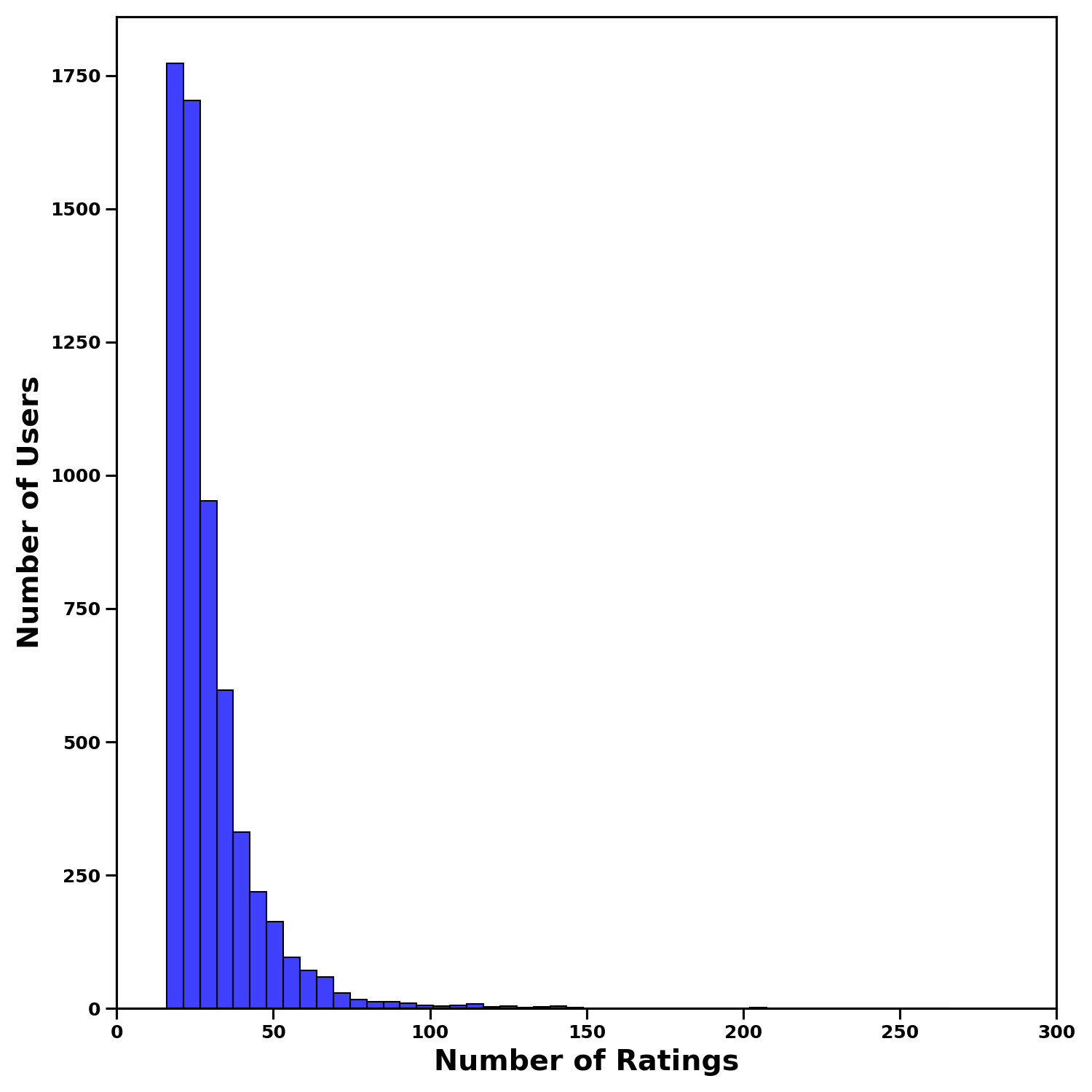}
        \caption{Amazon Video Games}
        \label{sfig:software_dist}
    \end{subfigure}
    \caption{Histograms illustrating the distribution of the number of ratings per user for three datasets.}
    \label{fig:dist_plot}
\end{figure*}

The \emph{first phase} begins by identifying the inactive users. For this, we first analyze the users and varying distributions across all three datasets. The histograms in Fig.~\ref{fig:dist_plot} detail a glimpse into the user engagement patterns across three distinct datasets: Amazon Software (Fig.~\ref{sfig:software_dist}), Amazon Video Games (Fig.~\ref{sfig:games_dist}), and ML1M (Fig.~\ref{sfig:ml1m_dist}). Each plot reveals a common theme of sparsity, where the majority of users contribute only a handful of ratings. In the Amazon Software dataset, we observe a steep decline in user numbers as the rating count increases, highlighting that most users are infrequent raters, with only a few prolific contributors. This trend is mirrored in the Amazon Video Games dataset, where the user base predominantly consists of sparse raters, creating a pronouncedlong-taill distribution. Interestingly, the ML1M dataset, while still exhibiting sparsity, shows a slightly more balanced distribution. The decline in user numbers is less abrupt, suggesting a higher proportion of active users who engage more frequently. By understanding these user behavior nuances, we can better tailor our recommendation strategies to enhance user experience and satisfaction across diverse datasets.

\begin{figure*}[t!]
    \centering
    \begin{subfigure}[b]{0.3\textwidth}
        \centering
        \includegraphics[width=\textwidth]{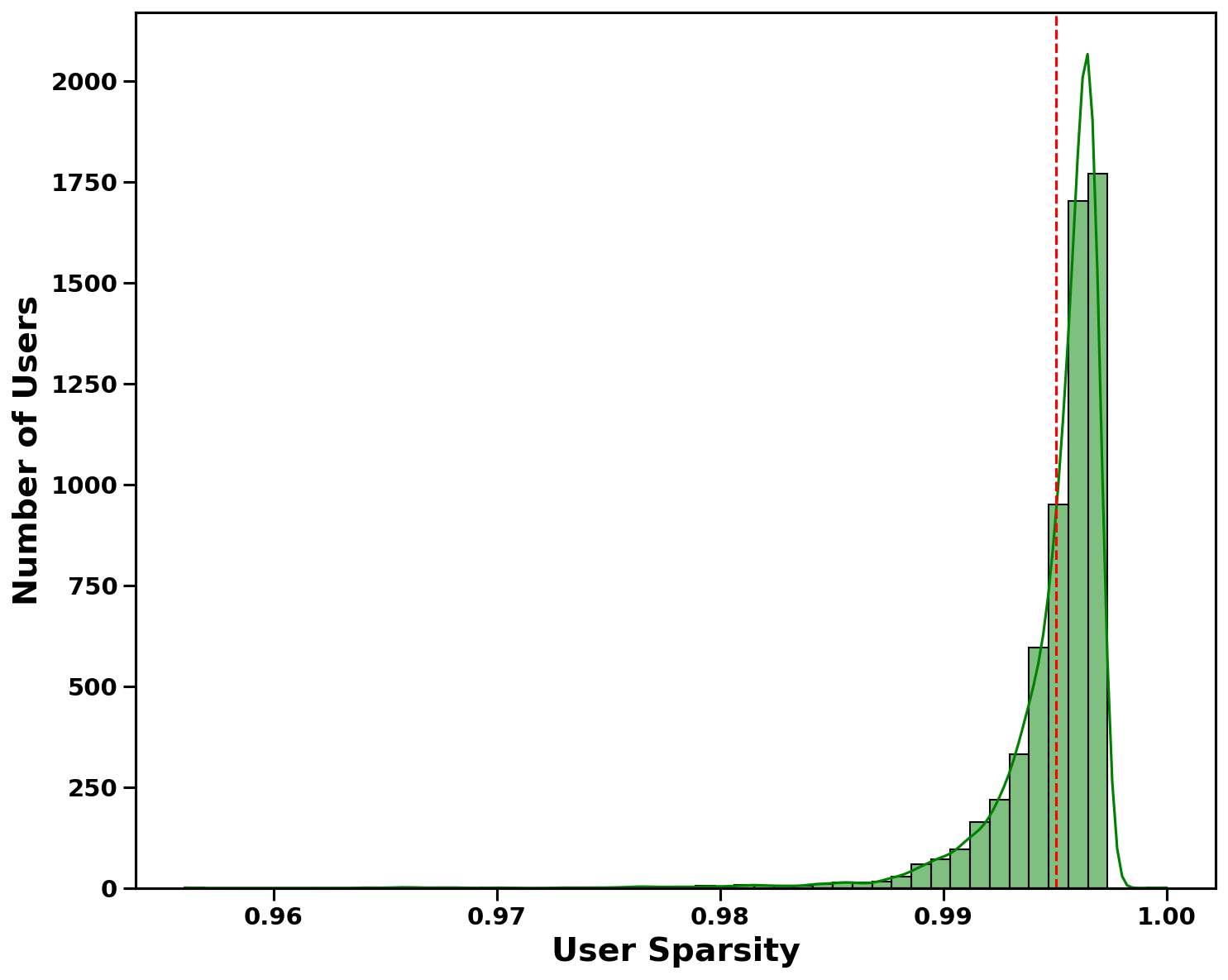}
        \caption{Amazon Software}
        \label{sfig:software_sparsity_dist}
    \end{subfigure}
    %\hfill
    \begin{subfigure}[b]{0.3\textwidth}
        \centering
        \includegraphics[width=\textwidth]{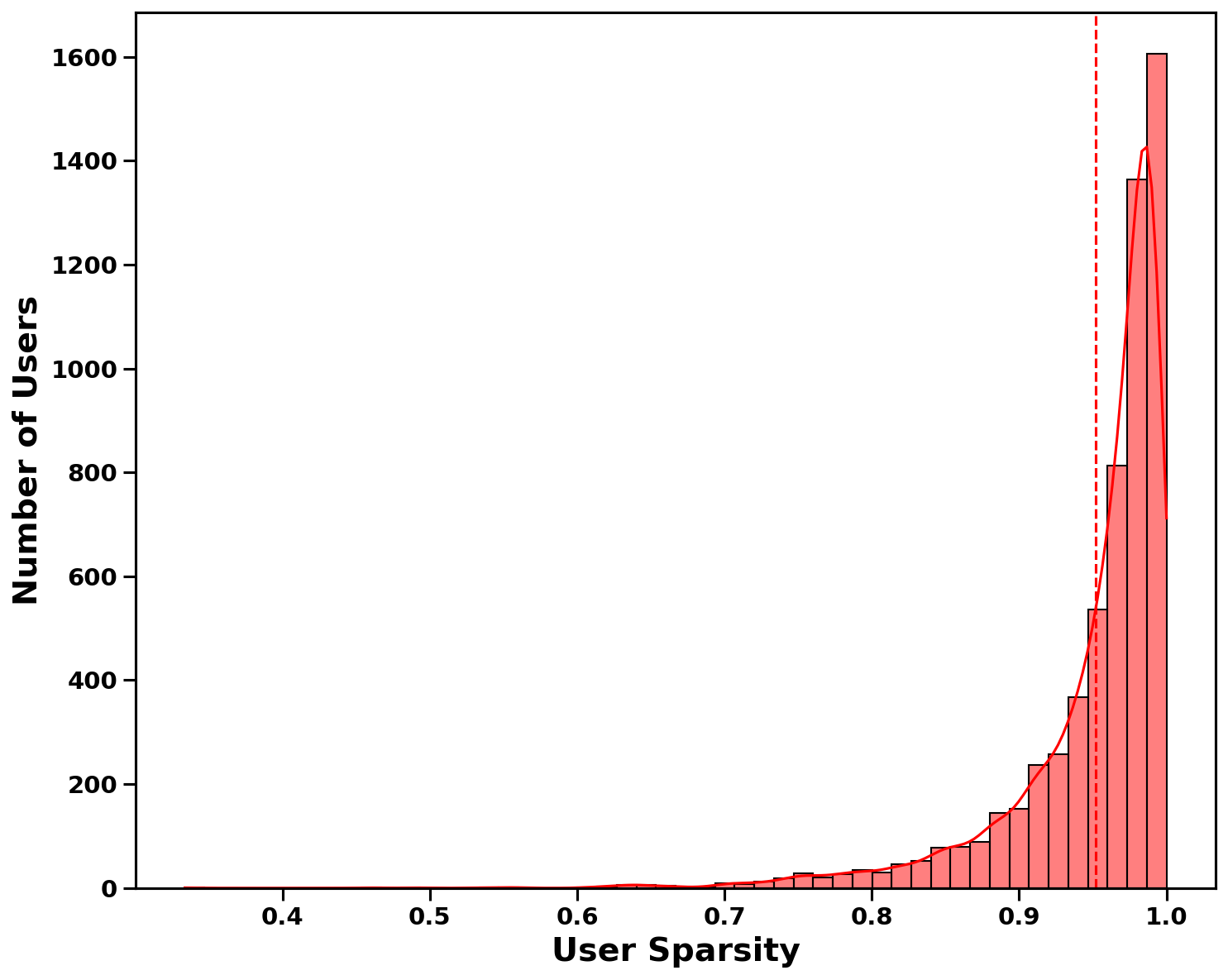}
        \caption{ML1M}
        \label{sfig:ml1m_sparsity_dist}
    \end{subfigure}
    %\hfill
    \begin{subfigure}[b]{0.3\textwidth}
        \centering
        \includegraphics[width=\textwidth]{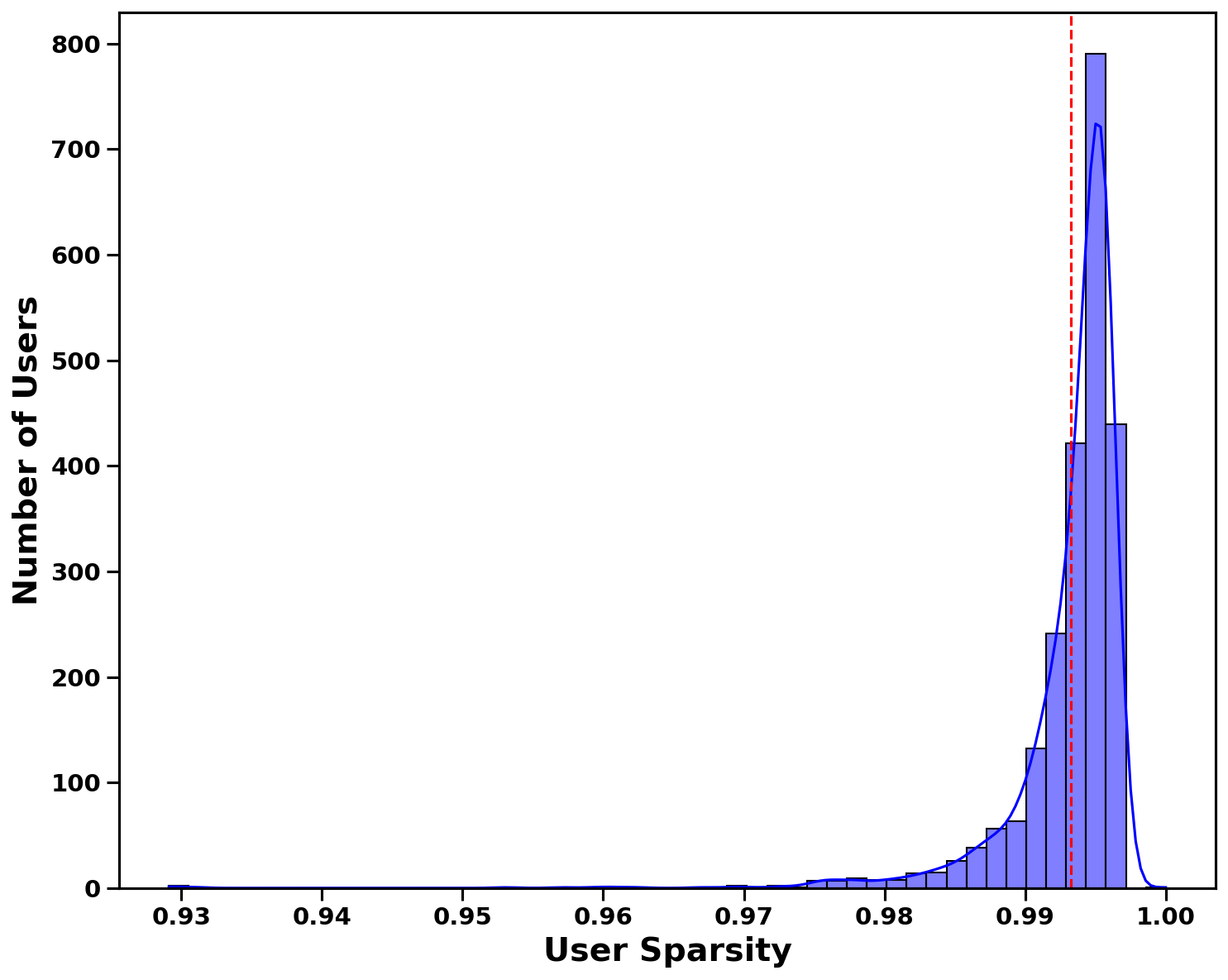}
        \caption{Amazon Video Games}
        \label{sfig:games_sparsity_dist}
    \end{subfigure}
    \caption{Distribution of user sparsity across three datasets. }
    \label{fig:dist_sparsity_plot}
\end{figure*}

Further, we analyze the distribution of user sparsity across all three datasets in  Fig.~\ref{fig:dist_sparsity_plot} to deepen our analysis of user engagement patterns. This figure determines the proportion of users who are infrequent raters versus more active participants.
The histogram in Fig.~\ref{sfig:software_sparsity_dist} shows that the majority of users have a sparsity value very close to $1.0$, indicating that most users have rated very few items. The distribution is heavily skewed towards the right, with a sharp peak near 1.0. This confirms the high average sparsity of 0.995, suggesting that almost all users are infrequent raters. Similar to the Software dataset, the histogram for the Games dataset (Fig.~\ref{sfig:games_sparsity_dist}) also shows a high concentration of users with sparsity values close to $1.0$. The distribution is slightly more spread out compared to the Software dataset but still exhibits a strong right skew. The average sparsity is $0.993$, indicating a very high level of sparsity among users, with most contributing minimal ratings. The histogram for the ML1M dataset (Fig.~\ref{sfig:ml1m_sparsity_dist}) presents a slightly different pattern. While there is still a significant number of users with high sparsity values, the distribution is more balanced compared to the Software and Games datasets. There is a noticeable spread of users across a wider range of sparsity values, with fewer users clustered near $1.0$. The average sparsity of $0.951$ reflects this more balanced distribution, suggesting a higher proportion of active users who engage more frequently. To determine a sparsity threshold, we consider the average sparsity values as suggested by several previous works~\cite{A_Systematic_Literature_Review_of_Sparsity_Issues_in_Recommender_Systems,aggarwal2016introduction} and the distribution patterns observed. A threshold set across the average sparsity value can capture the majority of infrequent raters while distinguishing more active users.

\begin{figure*}[t!]
    \centering
    \begin{subfigure}[b]{0.3\textwidth}
        \centering
        \includegraphics[width=\textwidth]{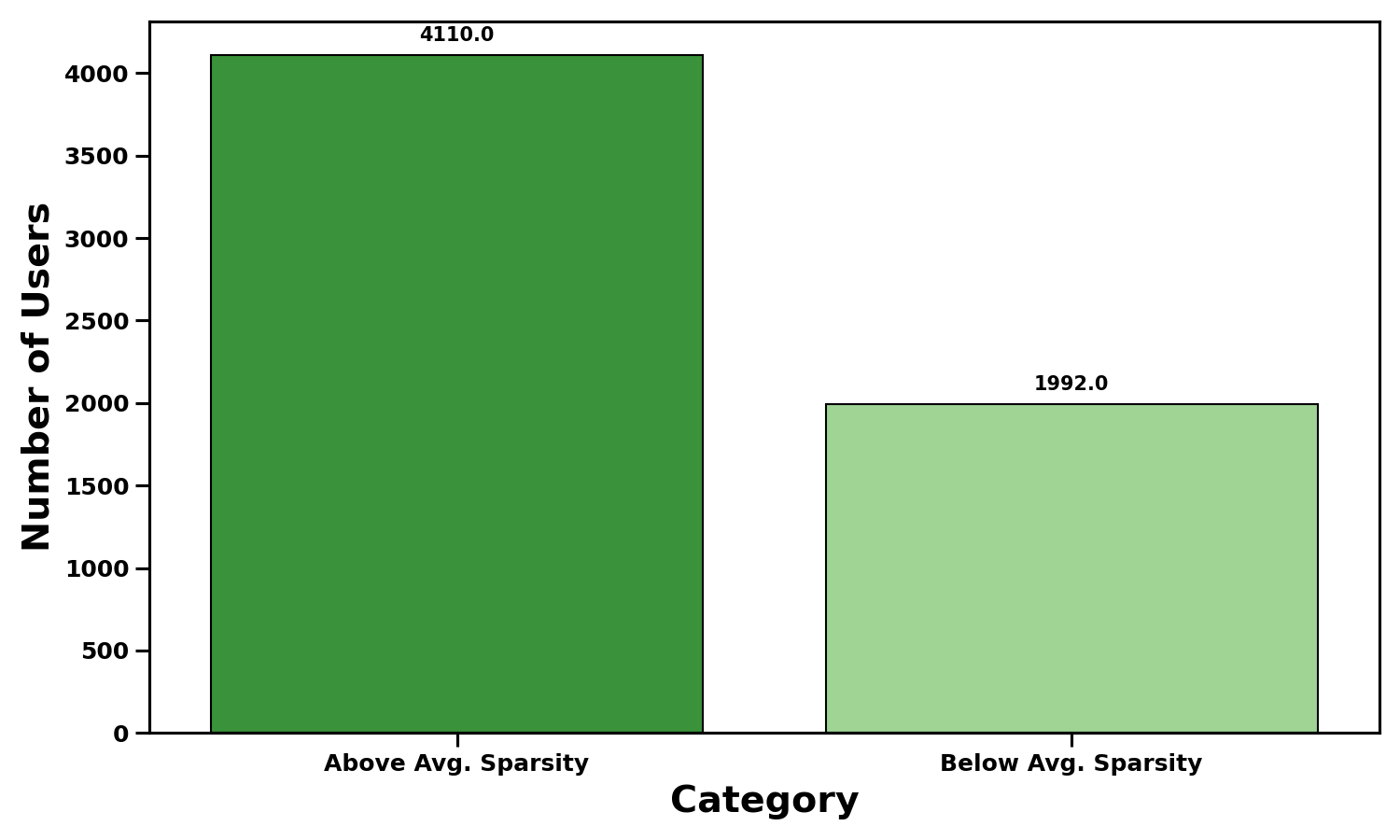}
        \caption{Amazon Software}
        \label{sfig:software_mean}
    \end{subfigure}
    %\hfill
    \begin{subfigure}[b]{0.3\textwidth}
        \centering
        \includegraphics[width=\textwidth]{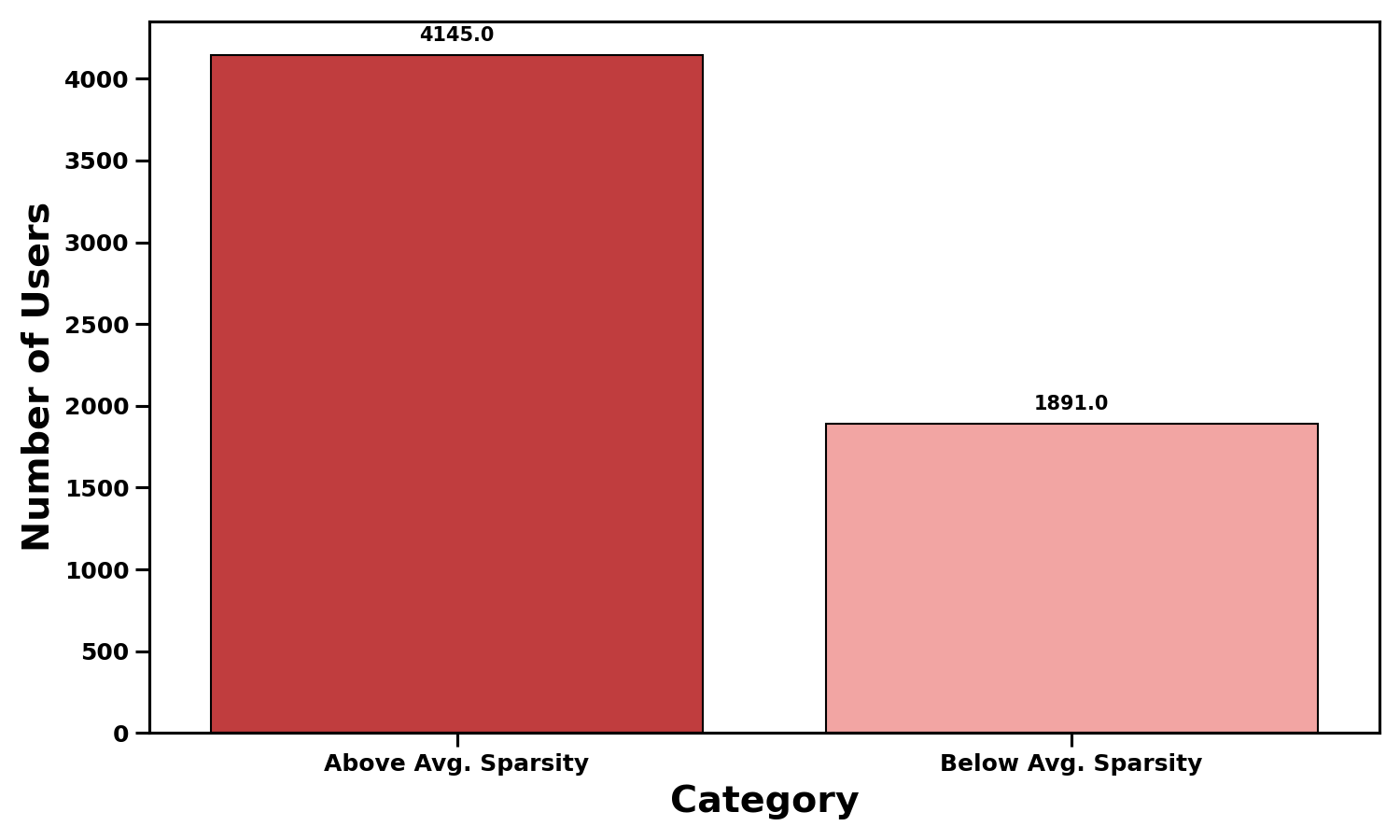}
        \caption{ML1M}
        \label{sfig:ml1m_mean}
    \end{subfigure}
    %\hfill
    \begin{subfigure}[b]{0.3\textwidth}
        \centering
        \includegraphics[width=\textwidth]{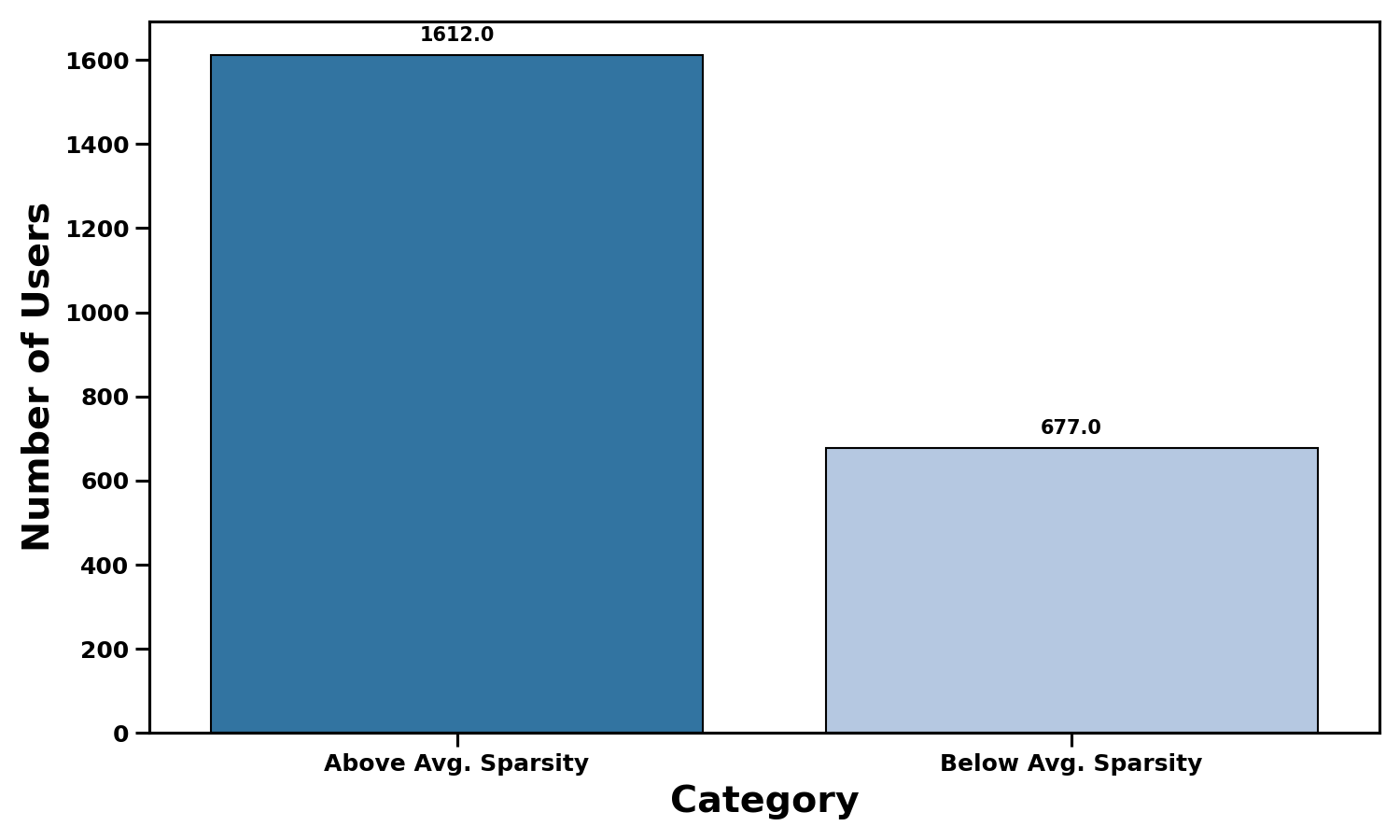}
        \caption{Amazon Video Games}
        \label{sfig:games_mean}
    \end{subfigure}
    \caption{ Comparative analysis of user distribution based on average sparsity levels. }
    \label{fig:mean}
\end{figure*}

We consider the average sparsity value for each dataset and plot the results in Fig.~\ref{fig:mean} to determine the count of active and inactive users. The three plots illustrate the distribution of users across the average sparsity threshold. The first plot (Fig.~\ref{sfig:software_mean}) shows a significant difference in user numbers between above-average and below-average sparsity, with $4110$ users in the above-average category compared to $1992$ in the below-average category. The second plot (Fig.~\ref{sfig:ml1m_mean}), with a comparatively smaller scale, also highlights a disparity, with $1612$ users above average and $677$ below average. The third plot (Fig.~\ref{sfig:games_mean}) somewhat mirrors the first, showing $4145$ users above average and $1891$ below average. Overall, the data consistently indicates a higher number of users in the above-average sparsity category across all plots. While different values can be used as a threshold, using the average sparsity as a benchmark retains the true distribution of data by ensuring that the comparison is grounded in a central tendency measure. This approach allows for a balanced assessment of how data points deviate from the mean, providing a clear and consistent reference point that reflects the overall pattern and variability within the dataset.

\begin{figure*}[t!]
    \centering
    \begin{subfigure}[b]{0.22\textwidth}
        \centering
        \includegraphics[width=\textwidth]{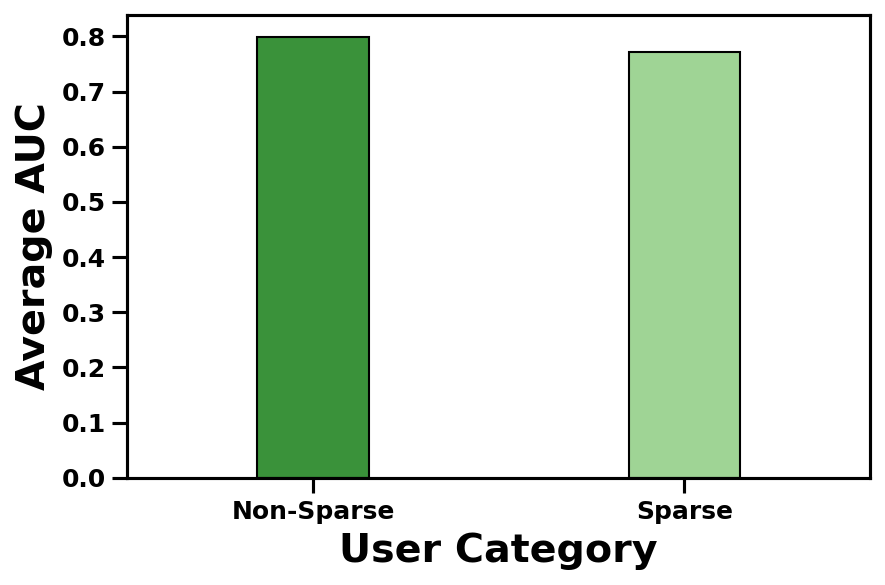}
        \caption{ItemKNN}
        \label{sfig:software_bar_ItemKNN}
    \end{subfigure}
    %\hfill
    \begin{subfigure}[b]{0.22\textwidth}
        \centering
        \includegraphics[width=\textwidth]{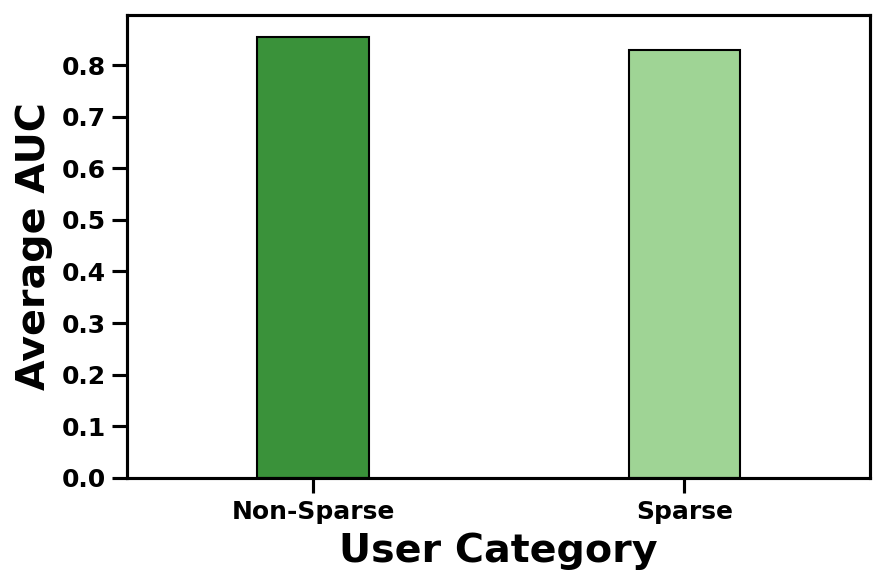}
        \caption{NeuMF}
        \label{sfig:software_bar_NeuMF}
    \end{subfigure}
    %\hfill
    \begin{subfigure}[b]{0.22\textwidth}
        \centering
        \includegraphics[width=\textwidth]{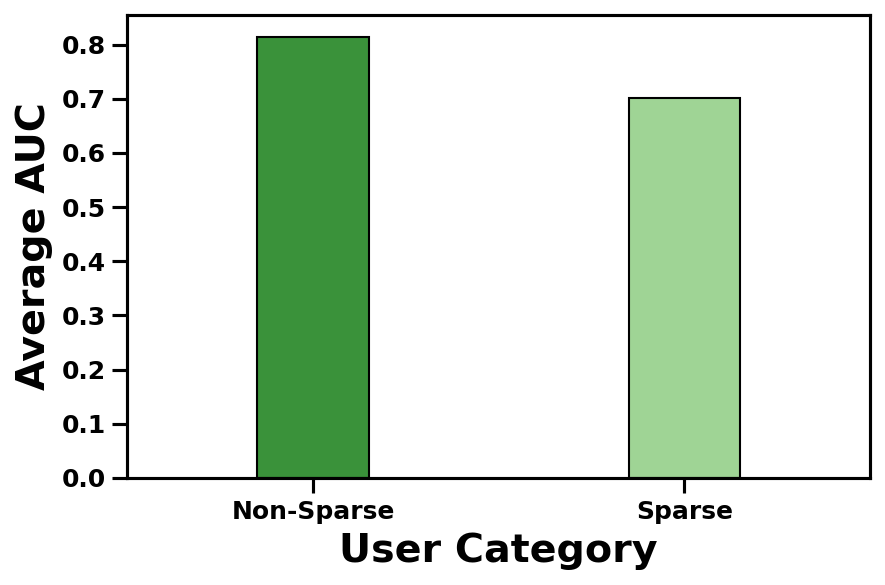}
        \caption{DMF}
        \label{sfig:software_bar_DMF}
    \end{subfigure}
    \begin{subfigure}[b]{0.22\textwidth}
        \centering
        \includegraphics[width=\textwidth]{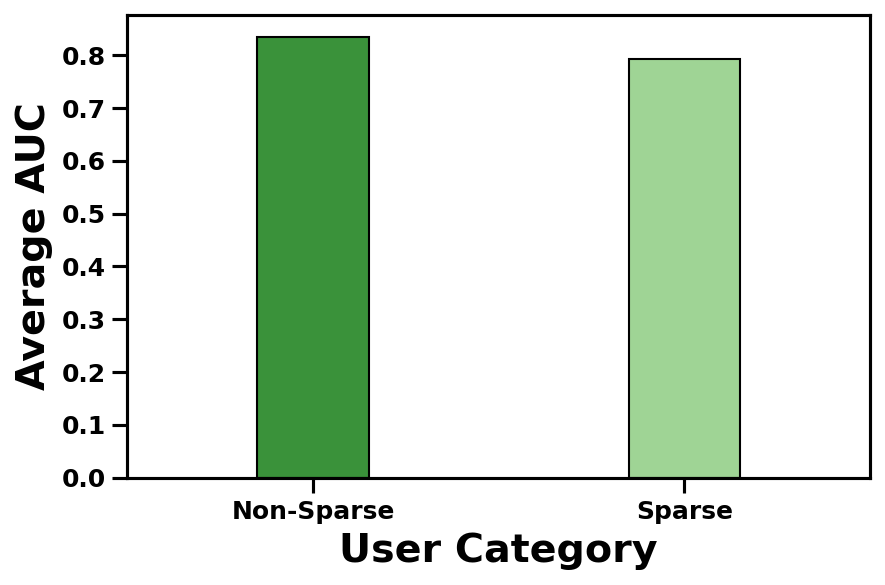}
        \caption{NNCF}
        \label{sfig:software_bar_NNCF}
    \end{subfigure}
    %\hfill
    \begin{subfigure}[b]{0.22\textwidth}
        \centering
        \includegraphics[width=\textwidth]{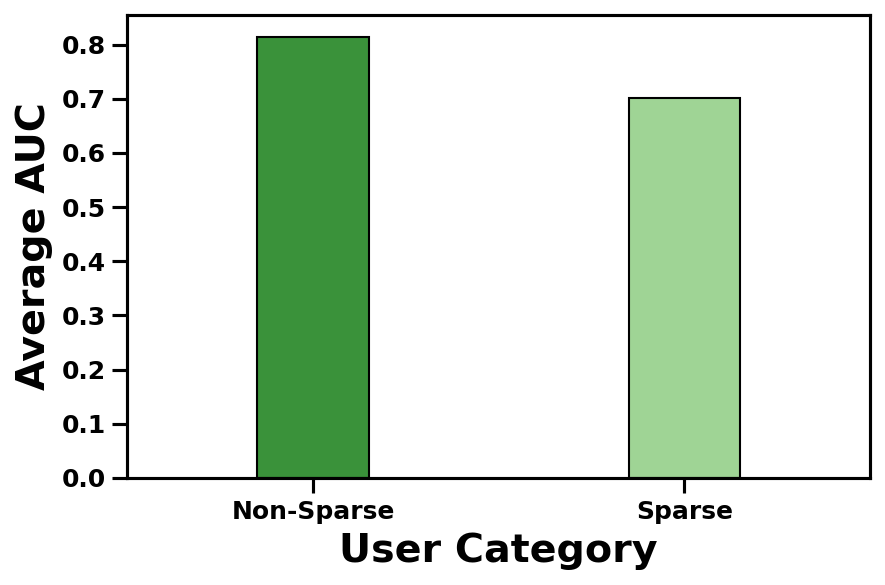}
        \caption{BPR}
        \label{sfig:software_bar_BPR}
    \end{subfigure}
    \begin{subfigure}[b]{0.22\textwidth}
        \centering
        \includegraphics[width=\textwidth]{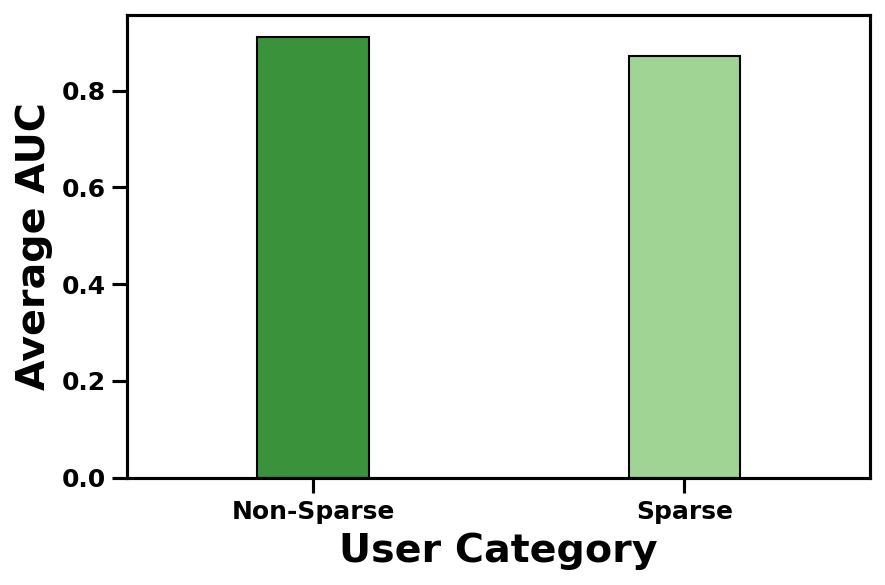}
        \caption{SASRec}
        \label{sfig:software_bar_SASRec}
    \end{subfigure}
    %\hfill
    \begin{subfigure}[b]{0.22\textwidth}
        \centering
        \includegraphics[width=\textwidth]{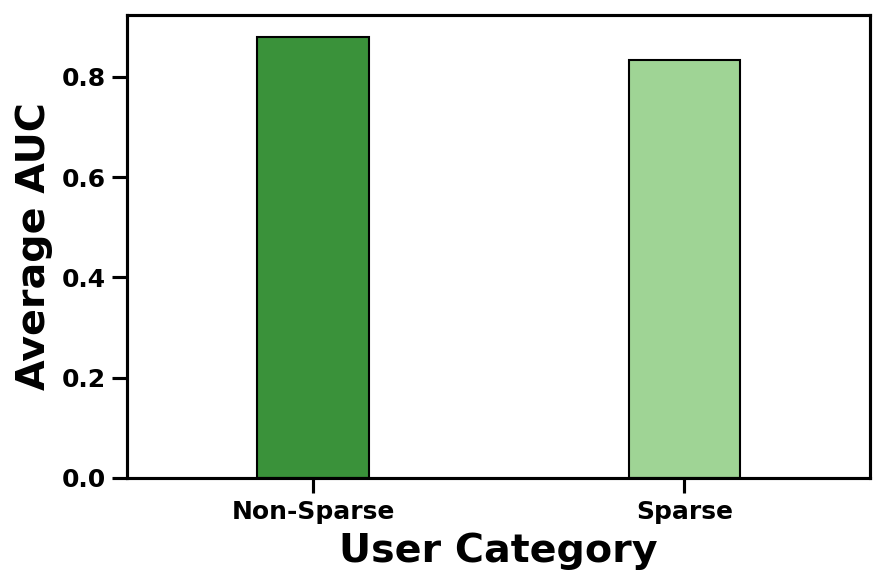}
        \caption{BERT4Rec}
        \label{sfig:software_bar_BERT4Rec}
    \end{subfigure}
    \begin{subfigure}[b]{0.22\textwidth}
        \centering
        \includegraphics[width=\textwidth]{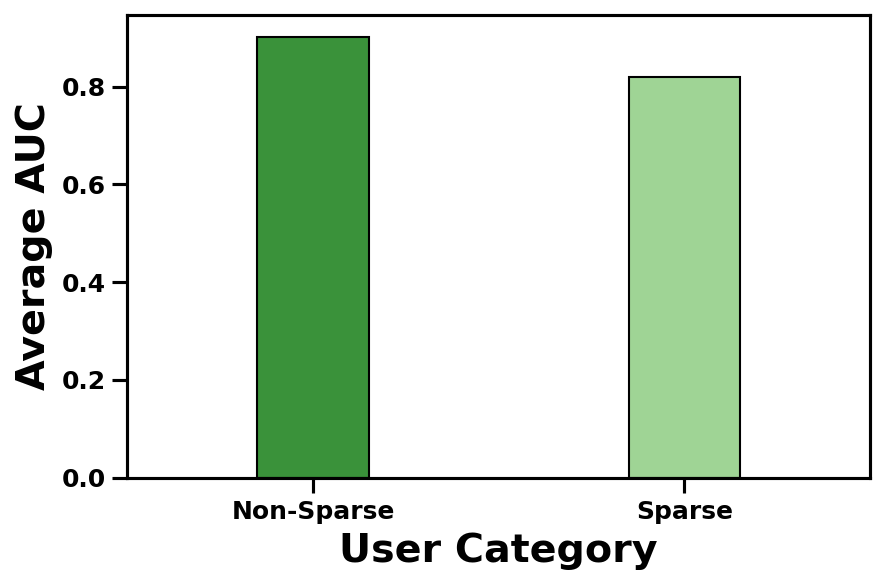}
        \caption{GRU4Rec}
        \label{sfig:software_bar_GRU4Rec}
    \end{subfigure}
    %\hfill
    \begin{subfigure}[b]{0.22\textwidth}
        \centering
        \includegraphics[width=\textwidth]{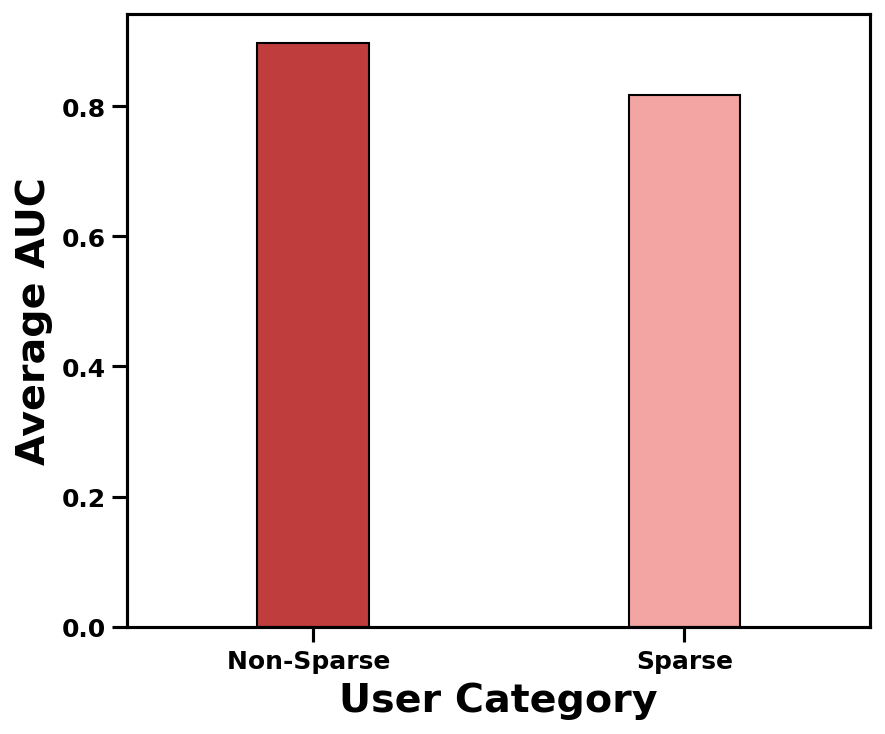}
        \caption{ItemKNN}
        \label{sfig:ml1m_bar_ItemKNN}
    \end{subfigure}
    %\hfill
    \begin{subfigure}[b]{0.22\textwidth}
        \centering
        \includegraphics[width=\textwidth]{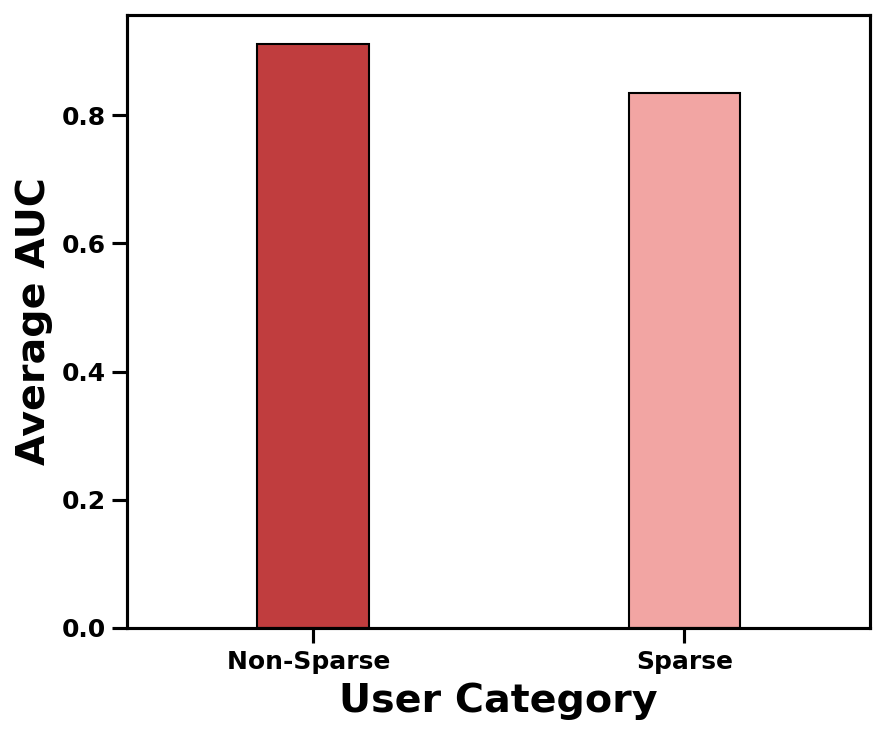}
        \caption{NeuMF}
        \label{sfig:ml1m_bar_NeuMF}
    \end{subfigure}
    %\hfill
    \begin{subfigure}[b]{0.22\textwidth}
        \centering
        \includegraphics[width=\textwidth]{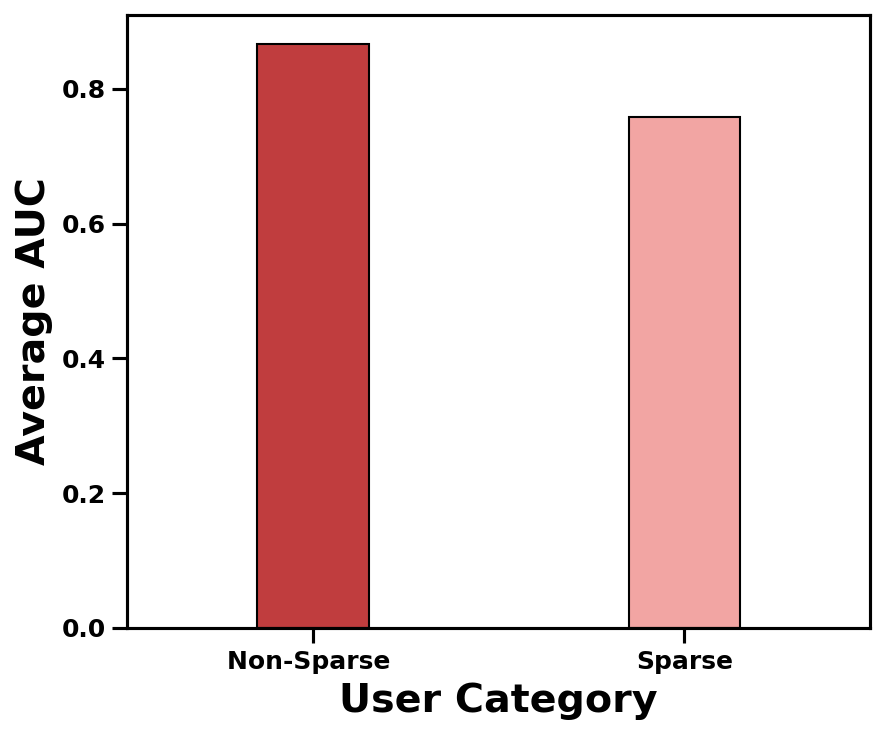}
        \caption{DMF}
        \label{sfig:ml1m_bar_DMF}
    \end{subfigure}
    \begin{subfigure}[b]{0.22\textwidth}
        \centering
        \includegraphics[width=\textwidth]{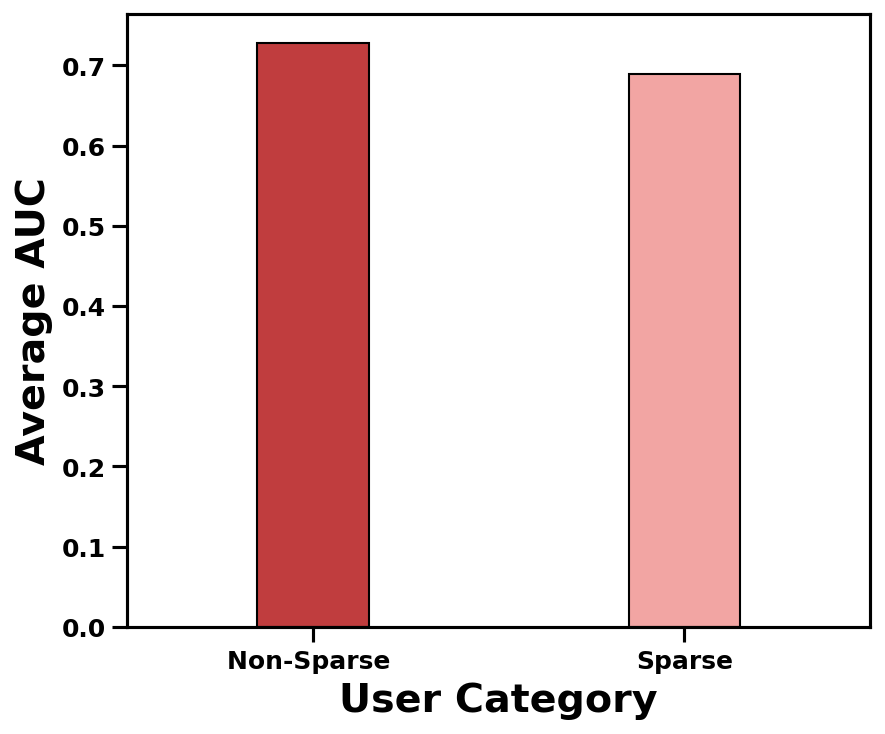}
        \caption{NNCF}
        \label{sfig:ml1m_bar_NNCF}
    \end{subfigure}
    %\hfill
    \begin{subfigure}[b]{0.22\textwidth}
        \centering
        \includegraphics[width=\textwidth]{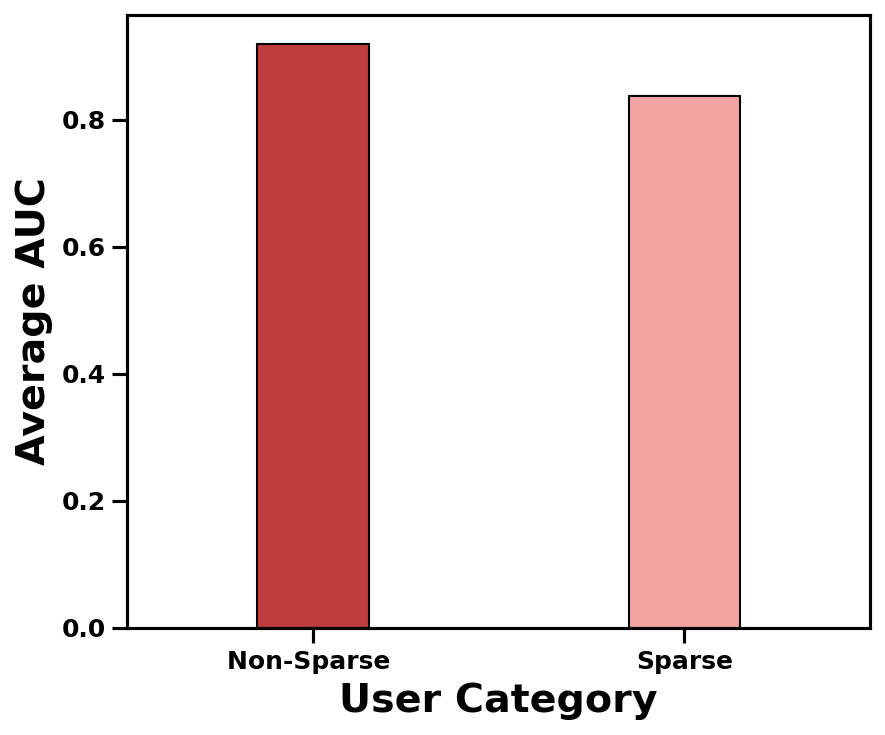}
        \caption{BPR}
        \label{sfig:ml1m_bar_BPR}
    \end{subfigure}
    \begin{subfigure}[b]{0.22\textwidth}
        \centering
        \includegraphics[width=\textwidth]{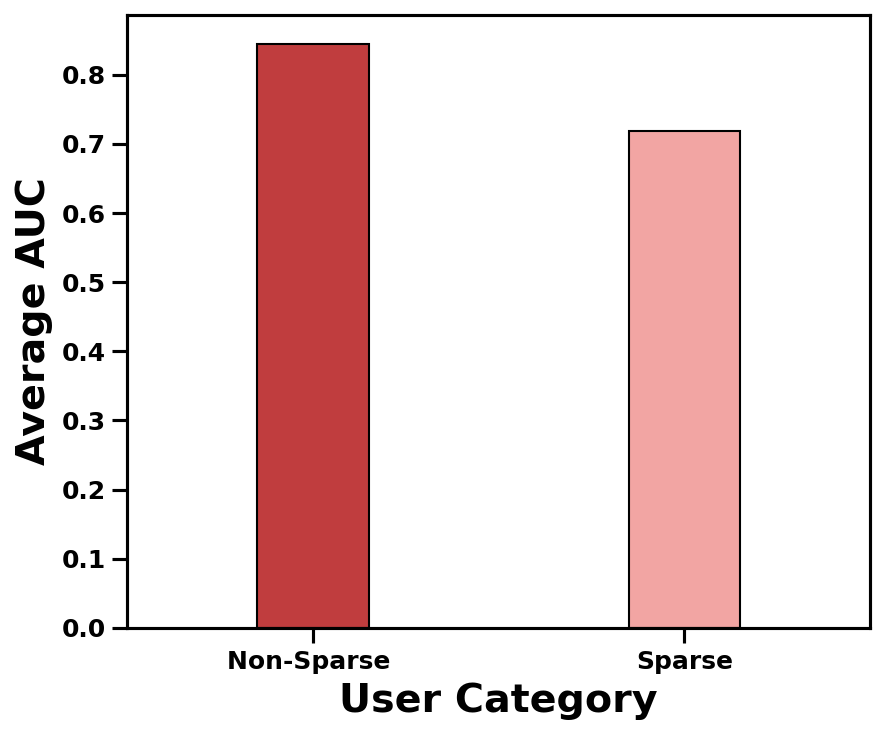}
        \caption{SASRec}
        \label{sfig:ml1m_bar_SASRec}
    \end{subfigure}
    %\hfill
    \begin{subfigure}[b]{0.22\textwidth}
        \centering
        \includegraphics[width=\textwidth]{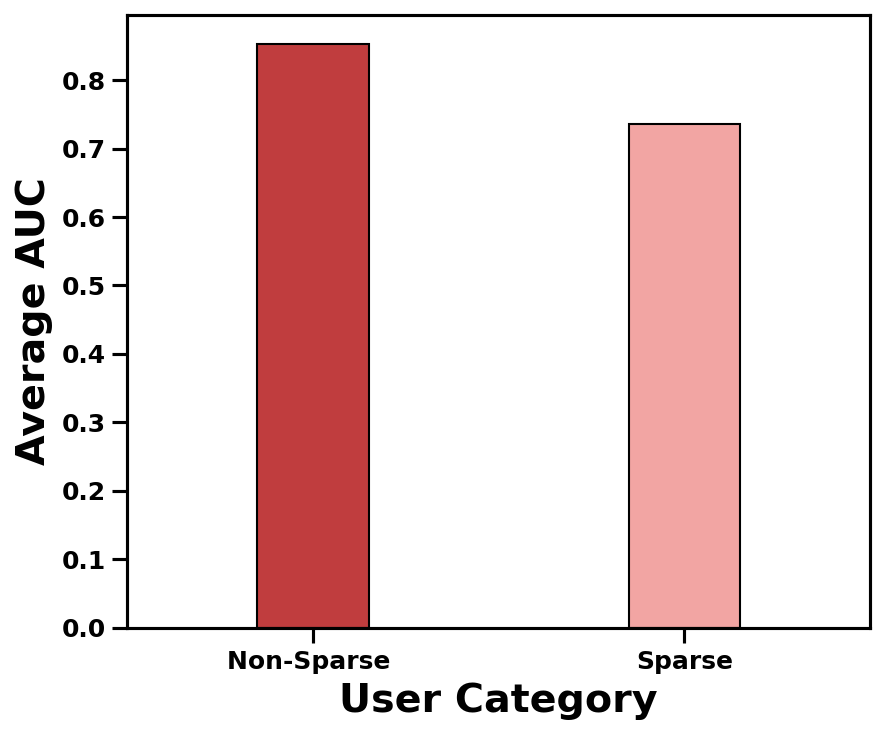}
        \caption{BERT4Rec}
        \label{sfig:ml1m_bar_BERT4Rec}
    \end{subfigure}
    \begin{subfigure}[b]{0.22\textwidth}
        \centering
        \includegraphics[width=\textwidth]{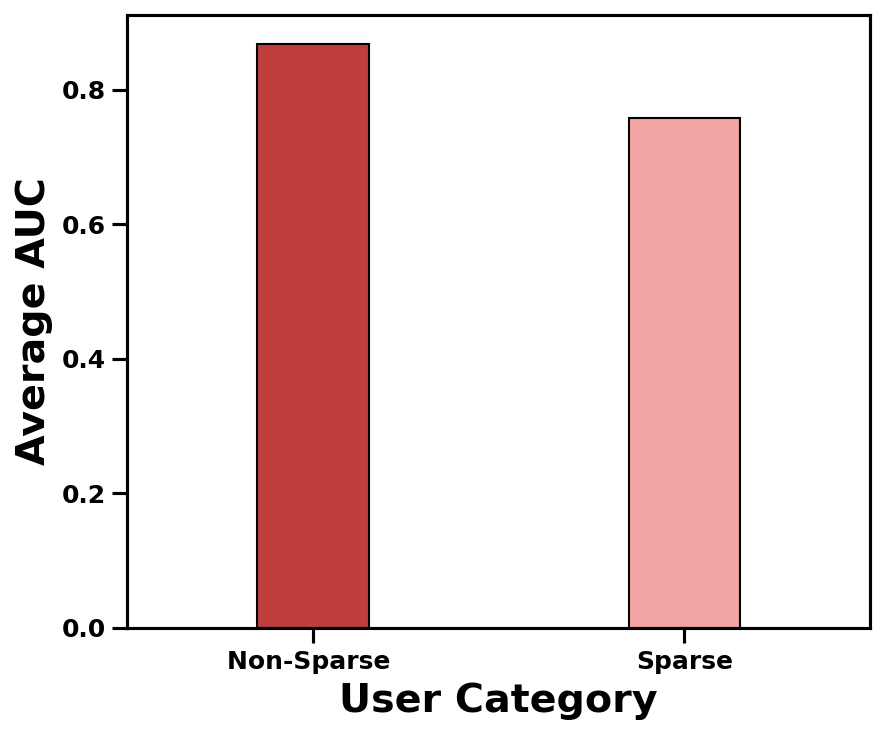}
        \caption{GRU4Rec}
        \label{sfig:ml1m_bar_GRU4Rec}
    \end{subfigure}
    %\hfill
    \begin{subfigure}[b]{0.22\textwidth}
        \centering
        \includegraphics[width=\textwidth]{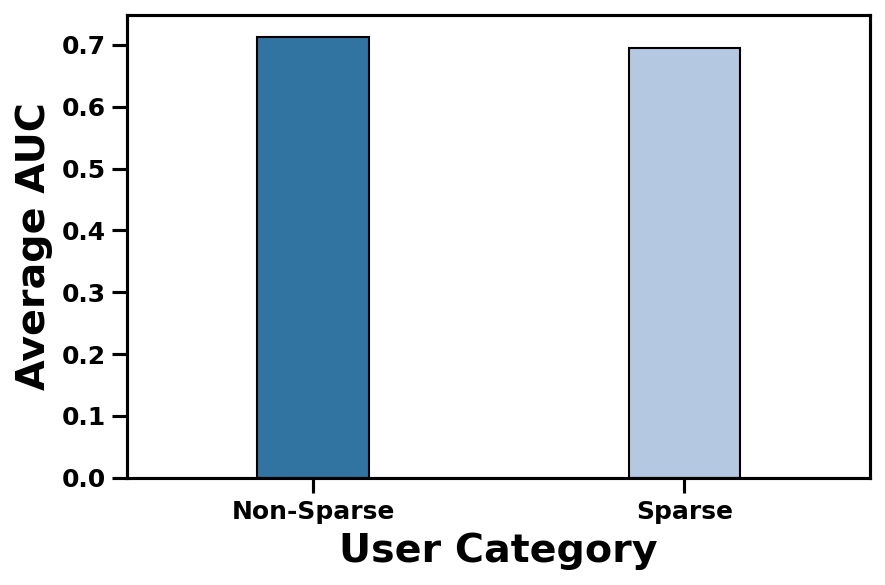}
        \caption{ItemKNN}
        \label{sfig:Amazon_Video_Games_bar_ItemKNN}
    \end{subfigure}
    %\hfill
    \begin{subfigure}[b]{0.22\textwidth}
        \centering
        \includegraphics[width=\textwidth]{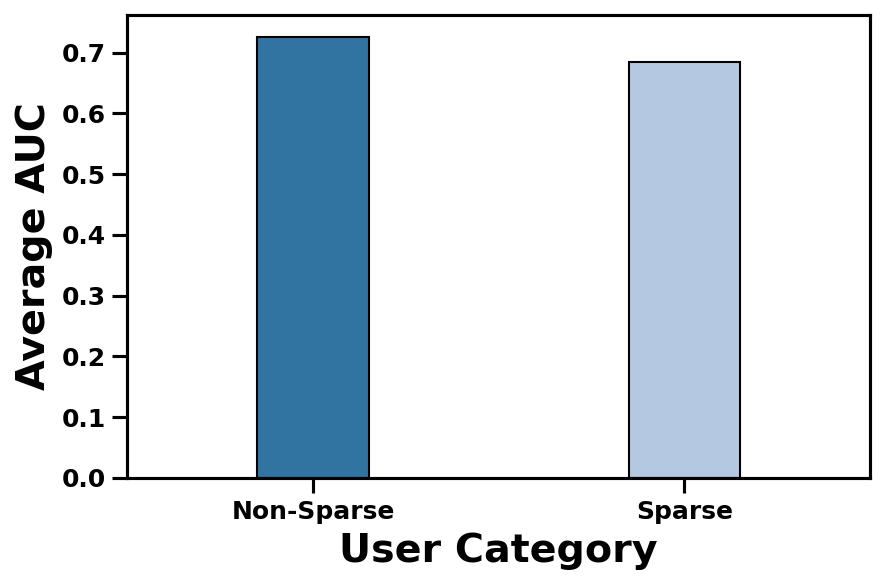}
        \caption{NeuMF}
        \label{sfig:Amazon_Video_Games_bar_NeuMF}
    \end{subfigure}
    %\hfill
    \begin{subfigure}[b]{0.22\textwidth}
        \centering
        \includegraphics[width=\textwidth]{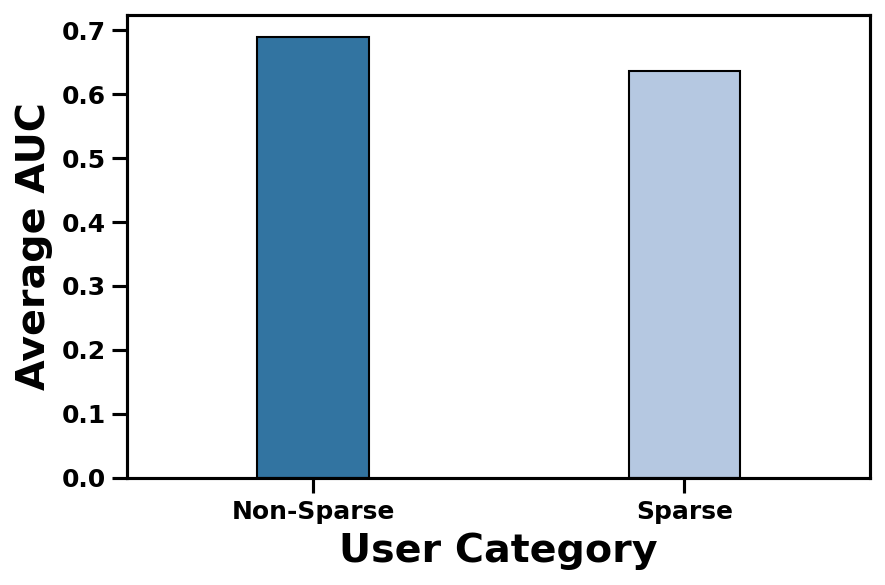}
        \caption{DMF}
        \label{sfig:Amazon_Video_Games_bar_DMF}
    \end{subfigure}
    \begin{subfigure}[b]{0.22\textwidth}
        \centering
        \includegraphics[width=\textwidth]{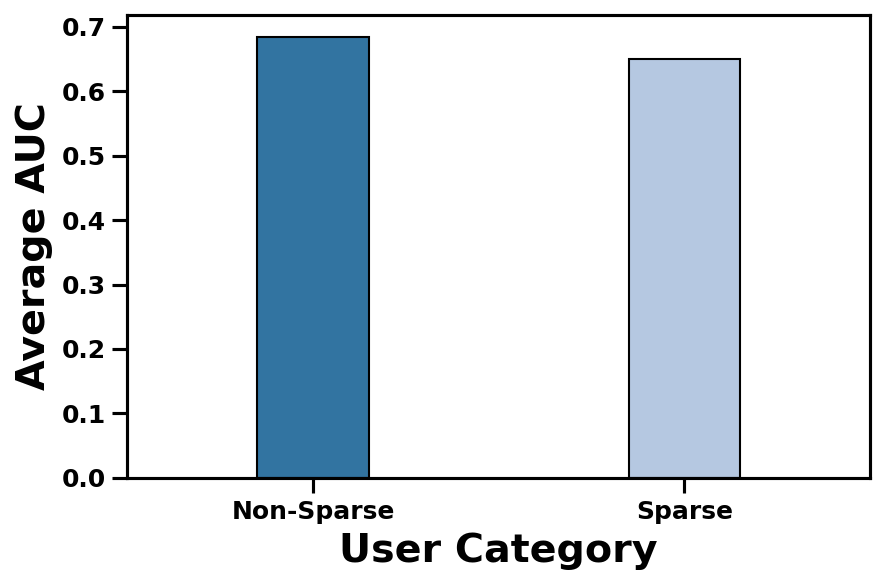}
        \caption{NNCF}
        \label{sfig:Amazon_Video_Games_bar_NNCF}
    \end{subfigure}
    %\hfill
    \begin{subfigure}[b]{0.22\textwidth}
        \centering
        \includegraphics[width=\textwidth]{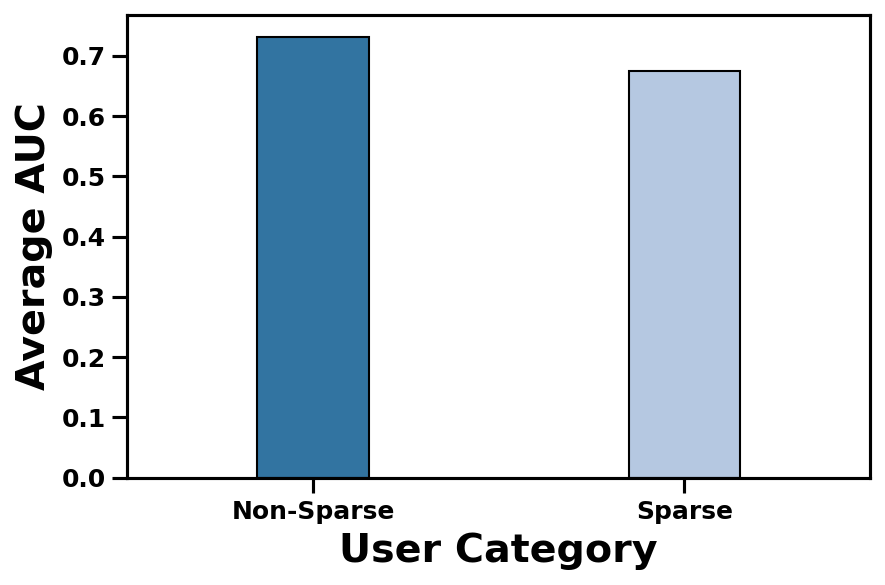}
        \caption{BPR}
        \label{sfig:Amazon_Video_Games_bar_BPR}
    \end{subfigure}
    \begin{subfigure}[b]{0.22\textwidth}
        \centering
        \includegraphics[width=\textwidth]{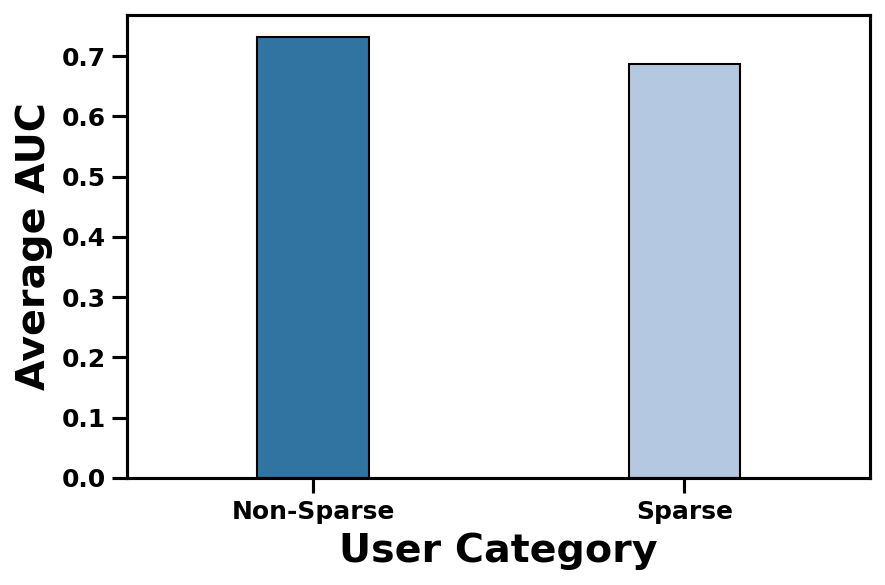}
        \caption{SASRec}
        \label{sfig:Amazon_Video_Games_bar_SASRec}
    \end{subfigure}
    %\hfill
    \begin{subfigure}[b]{0.22\textwidth}
        \centering
        \includegraphics[width=\textwidth]{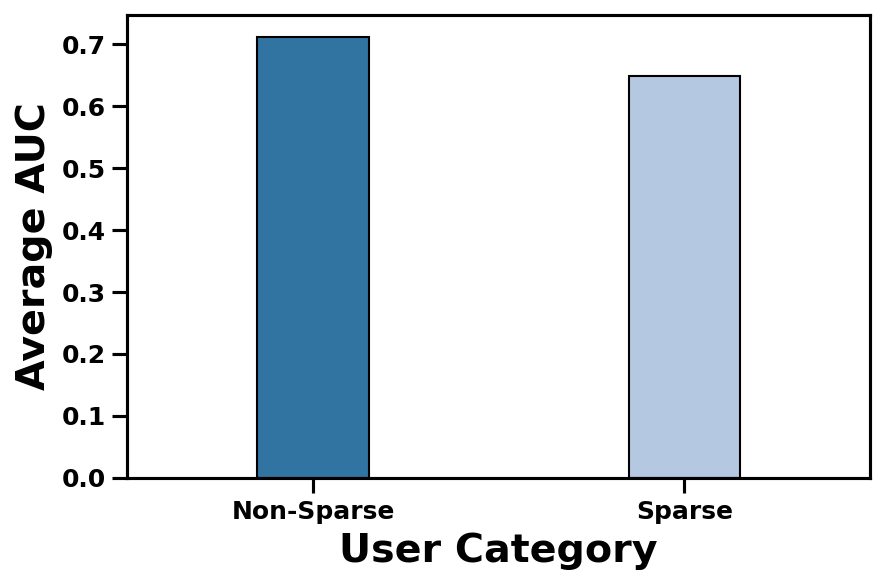}
        \caption{BERT4Rec}
        \label{sfig:Amazon_Video_Games_bar_BERT4Rec}
    \end{subfigure}
    \begin{subfigure}[b]{0.22\textwidth}
        \centering
        \includegraphics[width=\textwidth]{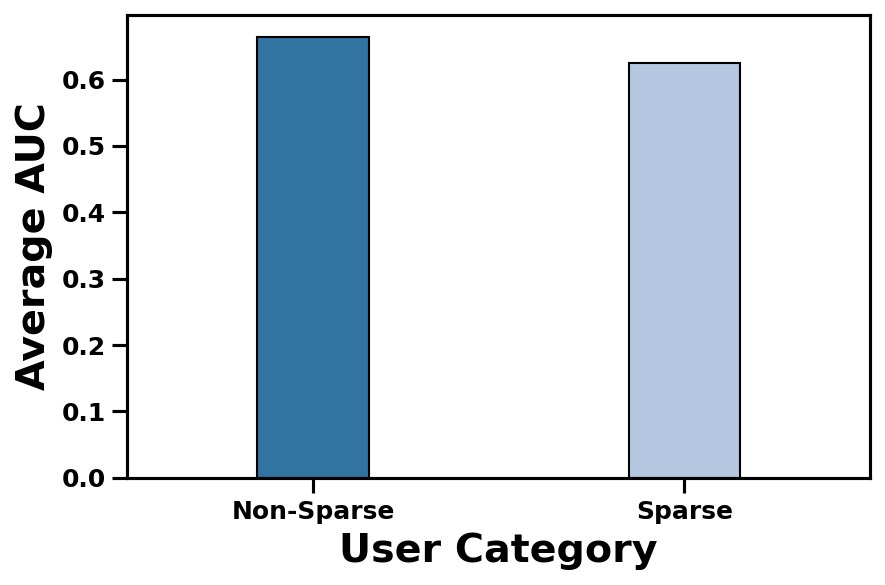}
        \caption{GRU4Rec}
        \label{sfig:Amazon_Video_Games_bar_GRU4Rec}
    \end{subfigure}
    %\hfill
    \caption{Cross-comparison of performance of various recommendation Mmdels (ItemKNN, NeuMF, DMF, NNCF, BPR, SASRec, BERT4Rec, GRU4Rec) on sparse and non-sparse users across three different datasets. The green colored plots from Fig.~\ref{sfig:software_bar_ItemKNN} to Fig.~\ref{sfig:software_bar_GRU4Rec} show results on Amazon Software dataset;  red colored plots from Fig.~\ref{sfig:ml1m_bar_ItemKNN} to Fig.~\ref{sfig:ml1m_bar_GRU4Rec} illustrate results on ML1M dataset; Fig.~\ref{sfig:Amazon_Video_Games_bar_ItemKNN} to Fig.~\ref{sfig:Amazon_Video_Games_bar_GRU4Rec} show results on Amazon Video Games dataset.}
    \label{fig:mean_bar}
\end{figure*}

We then evaluate the AUC score using Eq.~\ref{eq:p(u)} to measure the performance of RS on all these users. In line with the findings of~\citet{li2021user}, on an average, RS performs significantly well on active users as compared to inactive users~\ref{fig:mean_bar}. 
The comparative analysis of the recommendation models across sparse and non-sparse user interactions reveals significant performance variations. In sparse user scenarios, where user-item interactions are limited, models like \textbf{ItemKNN} (Fig.~\ref{sfig:software_bar_ItemKNN},~\ref{sfig:ml1m_bar_ItemKNN},~\ref{sfig:Amazon_Video_Games_bar_ItemKNN}) and \textbf{BPR} (Fig.~\ref{sfig:software_bar_BPR},~\ref{sfig:ml1m_bar_BPR},~\ref{sfig:Amazon_Video_Games_bar_BPR}) exhibit diminished efficacy due to their reliance on sufficient interaction data for accurate predictions. Conversely, advanced models such as \textbf{BERT4Rec} (Fig.~\ref{sfig:software_bar_BERT4Rec},~\ref{sfig:ml1m_bar_BERT4Rec},~\ref{sfig:Amazon_Video_Games_bar_BERT4Rec}) and \textbf{GRU4Rec} (Fig.~\ref{sfig:software_bar_GRU4Rec},~\ref{sfig:ml1m_bar_GRU4Rec},~\ref{sfig:Amazon_Video_Games_bar_GRU4Rec}) demonstrate robust performance, leveraging their sequential learning capabilities to mitigate data sparsity issues effectively. \textbf{NNCF} (Fig.~\ref{sfig:software_bar_NNCF},~\ref{sfig:ml1m_bar_NNCF},~\ref{sfig:Amazon_Video_Games_bar_NNCF}) and \textbf{SASRec} (Fig.~\ref{sfig:software_bar_SASRec},~\ref{sfig:ml1m_bar_SASRec},~\ref{sfig:Amazon_Video_Games_bar_SASRec}) also show commendable adaptability, though their performance slightly lags behind the top sequential models. In non-sparse user contexts, where interaction data is abundant, all models generally perform better, with \textbf{BERT4Rec} (Fig.~\ref{sfig:software_bar_BERT4Rec},~\ref{sfig:ml1m_bar_BERT4Rec},~\ref{sfig:Amazon_Video_Games_bar_BERT4Rec}) and \textbf{GRU4Rec}) (Fig.~\ref{sfig:software_bar_GRU4Rec},~\ref{sfig:ml1m_bar_GRU4Rec},~\ref{sfig:Amazon_Video_Games_bar_GRU4Rec}) consistently leading due to their sophisticated architectures that capture intricate user behavior patterns. \textbf{NNCF} (Fig.~\ref{sfig:software_bar_NNCF},~\ref{sfig:ml1m_bar_NNCF},~\ref{sfig:Amazon_Video_Games_bar_NNCF}) and \textbf{SASRec} (Fig.~\ref{sfig:software_bar_SASRec},~\ref{sfig:ml1m_bar_SASRec},~\ref{sfig:Amazon_Video_Games_bar_SASRec}) maintain strong performance, underscoring their utility in both sparse and non-sparse environments. This analysis highlights a comparatively superior adaptability of sequential models in diverse data scenarios. Additionally, key highlights and trends across various datasets reveal that \textbf{ML1M} (Fig.,~\ref{sfig:ml1m_bar_ItemKNN} to~\ref{sfig:ml1m_bar_GRU4Rec} ), being the least sparse dataset, allows all models to perform comparatively better, with \textbf{BERT4Rec} and \textbf{GRU4Rec} consistently leading. In moderately sparse datasets, the performance gap between these top models and others like \textbf{NNCF} and \textbf{SASRec} narrows, while in the most sparse datasets, the advanced sequential models still outperform others, though their advantage is less pronounced.

\begin{figure*}[t!]
    \centering
    \begin{subfigure}[b]{0.2\textwidth}
        \centering
        \includegraphics[width=\textwidth]{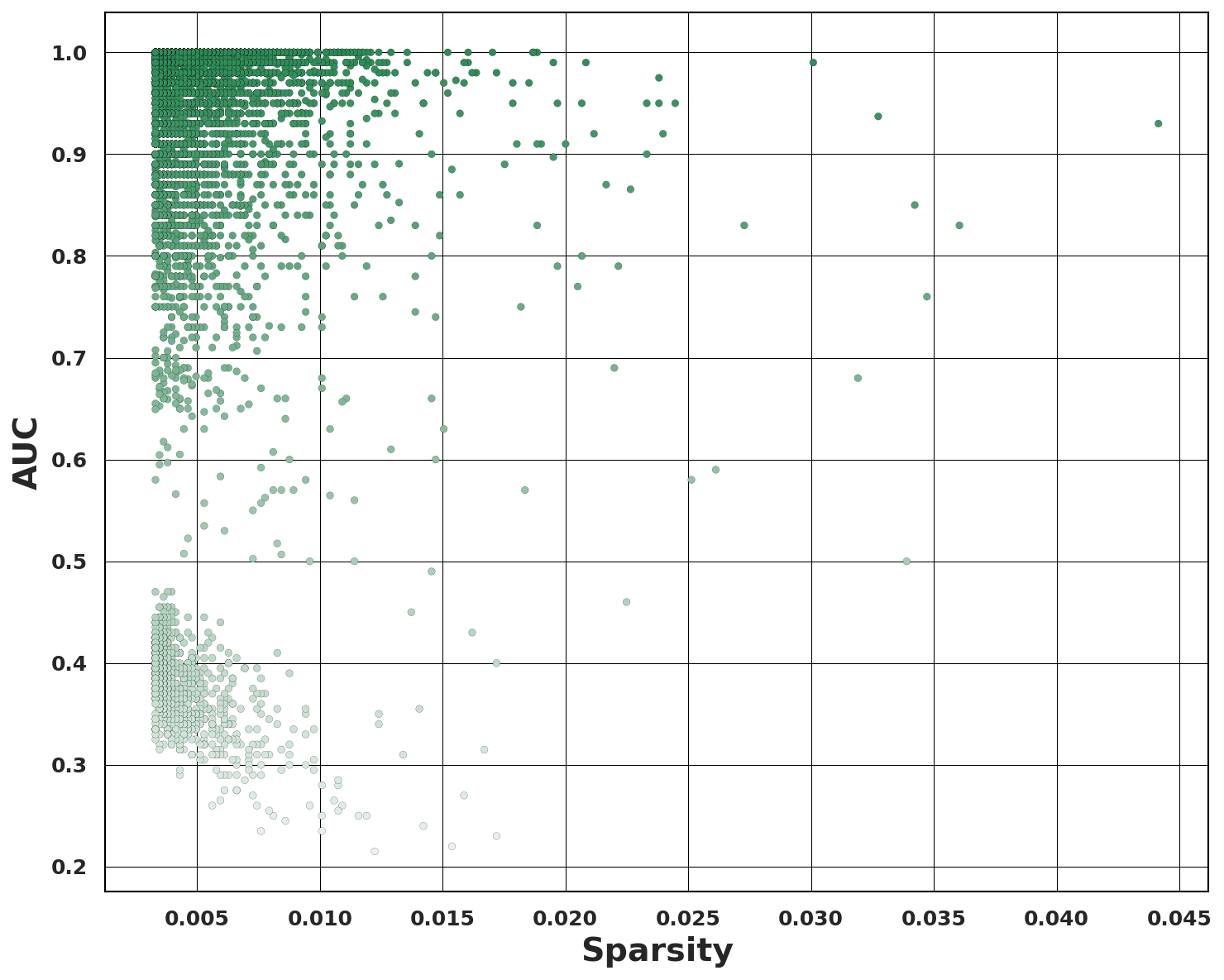}
        \caption{ItemKNN}
        \label{sfig:software_full_ItemKNN}
    \end{subfigure}
    %\hfill
    \begin{subfigure}[b]{0.2\textwidth}
        \centering
        \includegraphics[width=\textwidth]{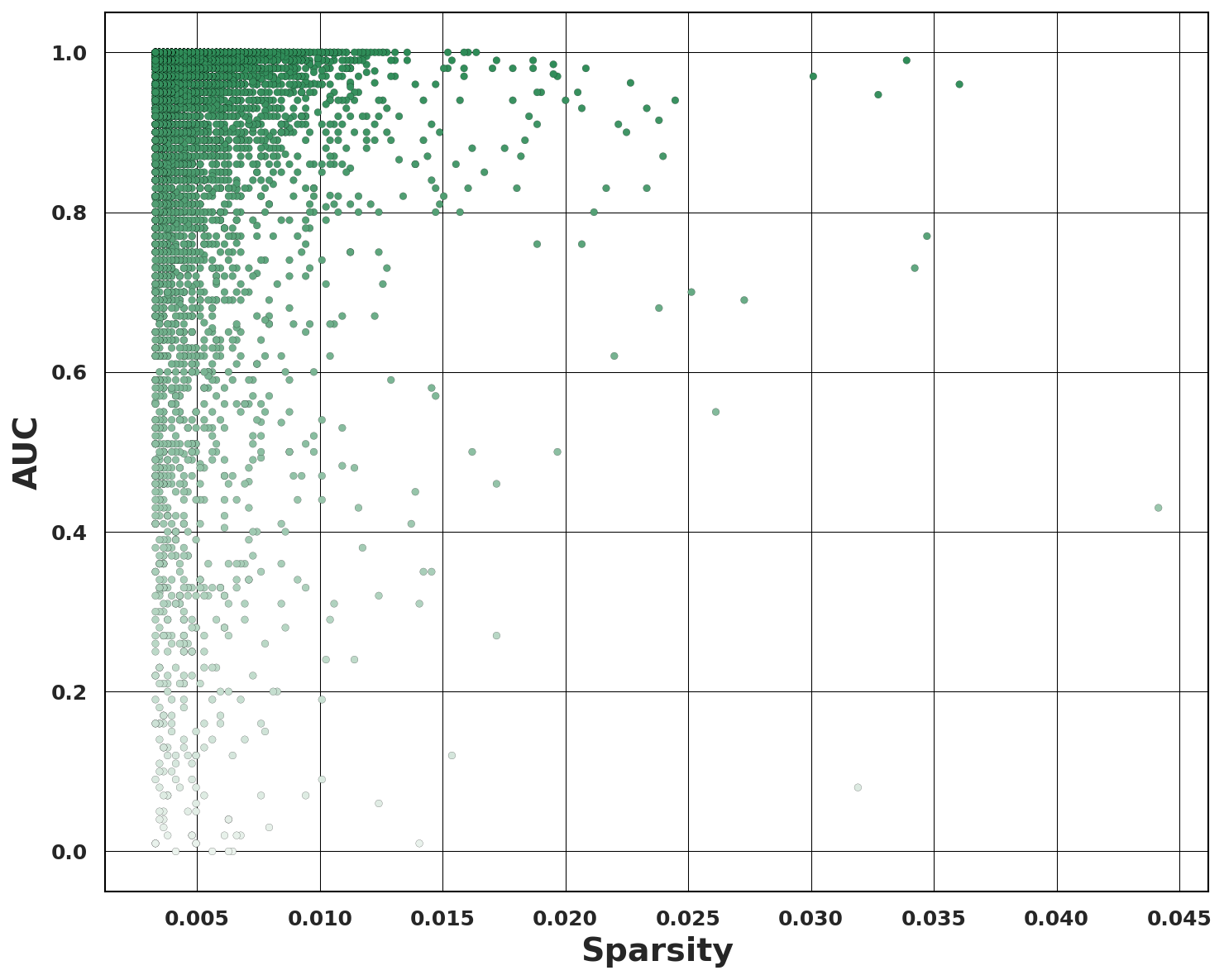}
        \caption{NeuMF}
        \label{sfig:software_full_NeuMF}
    \end{subfigure}
    %\hfill
    \begin{subfigure}[b]{0.2\textwidth}
        \centering
        \includegraphics[width=\textwidth]{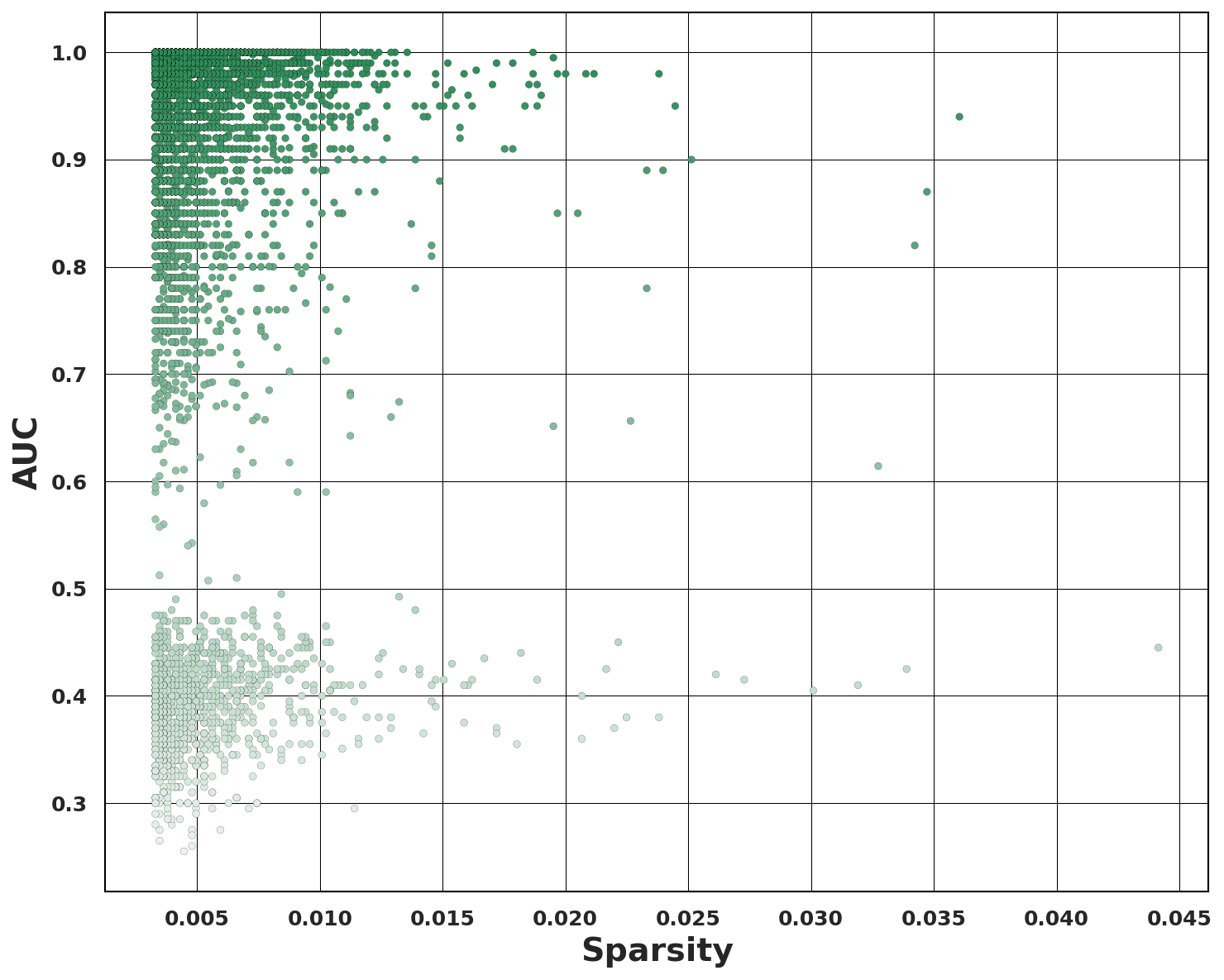}
        \caption{DMF}
        \label{sfig:software_full_DMF}
    \end{subfigure}
    \begin{subfigure}[b]{0.2\textwidth}
        \centering
        \includegraphics[width=\textwidth]{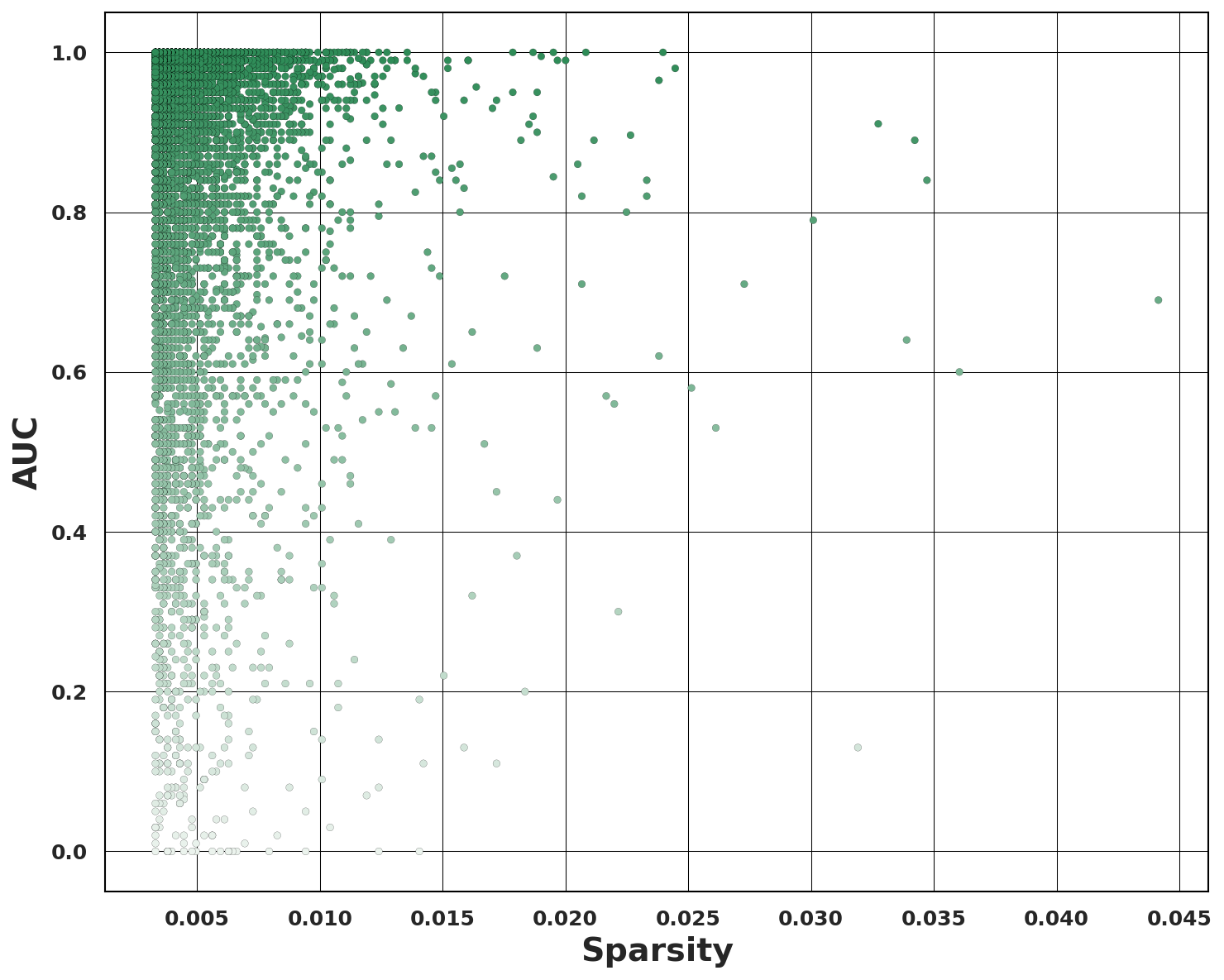}
        \caption{NNCF}
        \label{sfig:software_full_NNCF}
    \end{subfigure}
    %\hfill
    \begin{subfigure}[b]{0.2\textwidth}
        \centering
        \includegraphics[width=\textwidth]{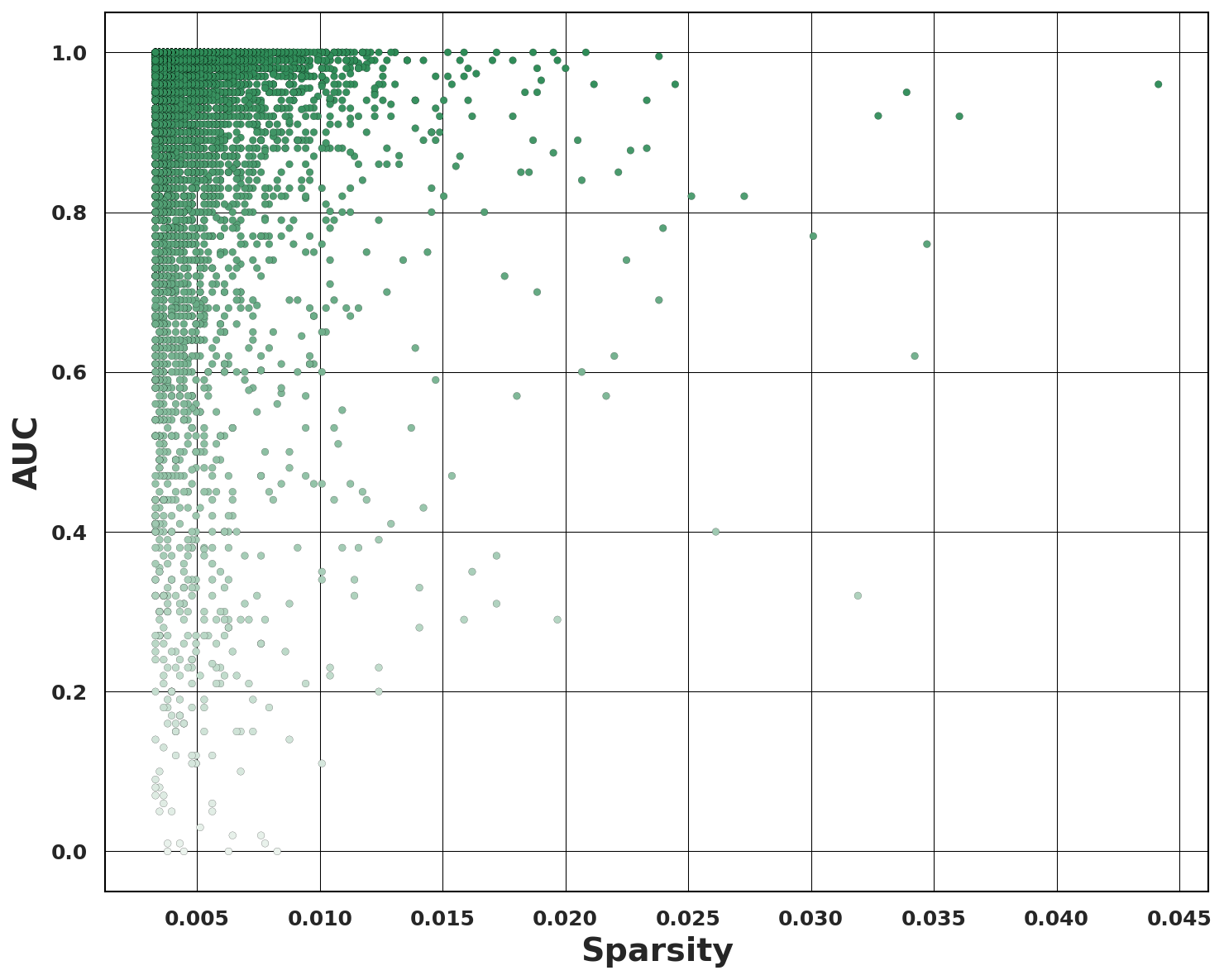}
        \caption{BPR}
        \label{sfig:software_full_BPR}
    \end{subfigure}
    \begin{subfigure}[b]{0.2\textwidth}
        \centering
        \includegraphics[width=\textwidth]{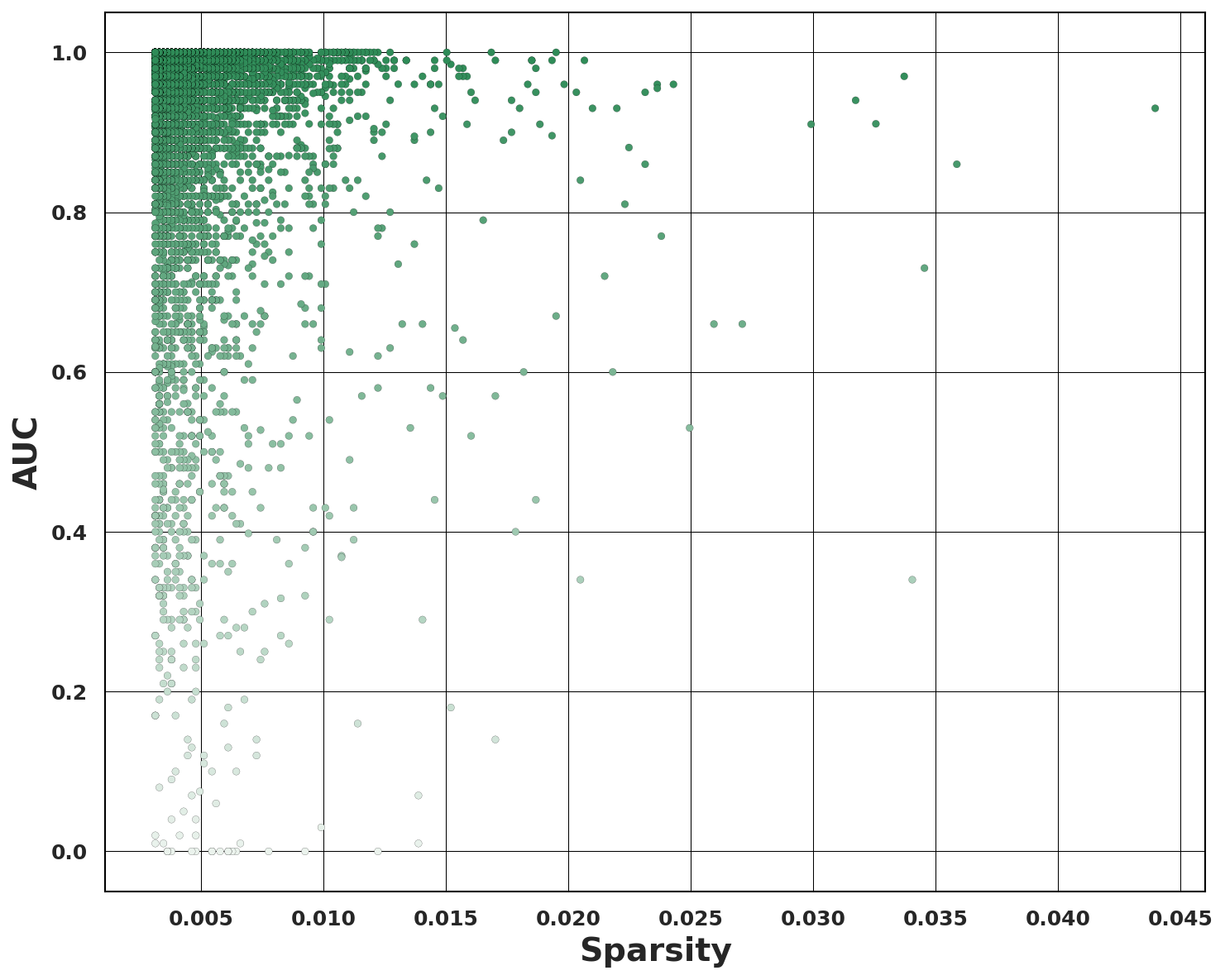}
        \caption{SASRec}
        \label{sfig:software_full_SASRec}
    \end{subfigure}
    %\hfill
    \begin{subfigure}[b]{0.2\textwidth}
        \centering
        \includegraphics[width=\textwidth]{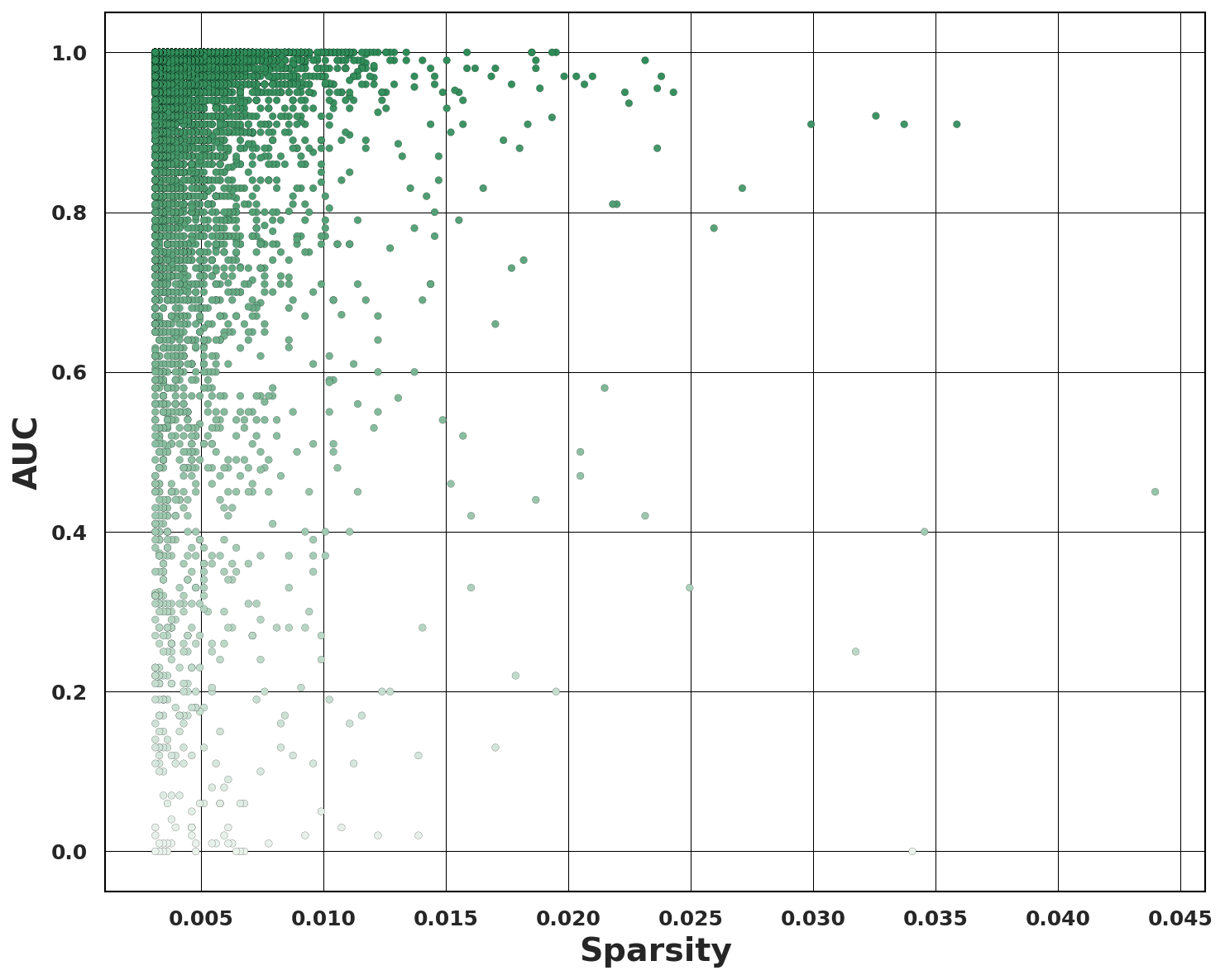}
        \caption{BERT4Rec}
        \label{sfig:software_full_BERT4Rec}
    \end{subfigure}
    \begin{subfigure}[b]{0.2\textwidth}
        \centering
        \includegraphics[width=\textwidth]{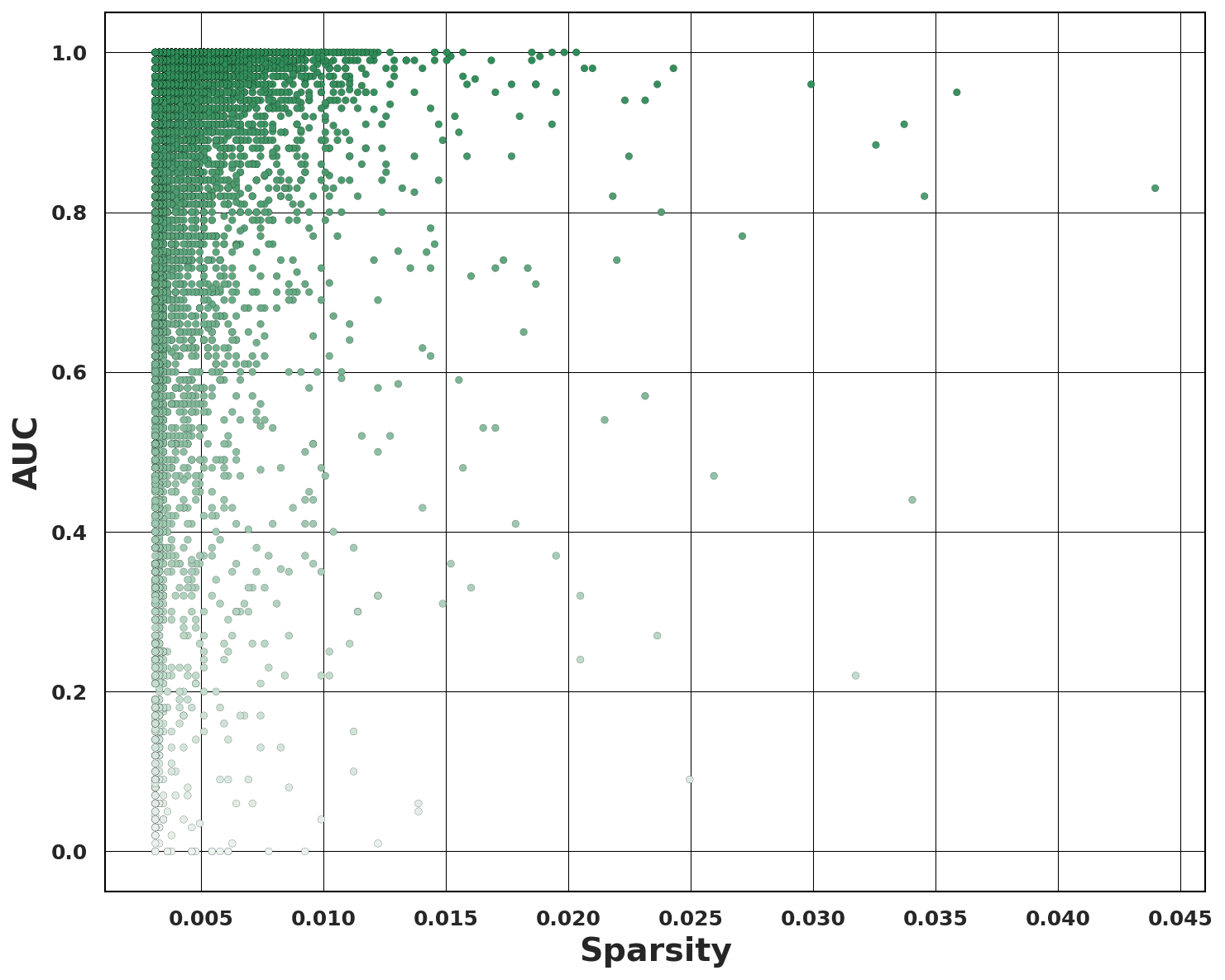}
        \caption{GRU4Rec}
        \label{sfig:software_full_GRU4Rec}
    \end{subfigure}
    %\hfill
    \begin{subfigure}[b]{0.2\textwidth}
        \centering
        \includegraphics[width=\textwidth]{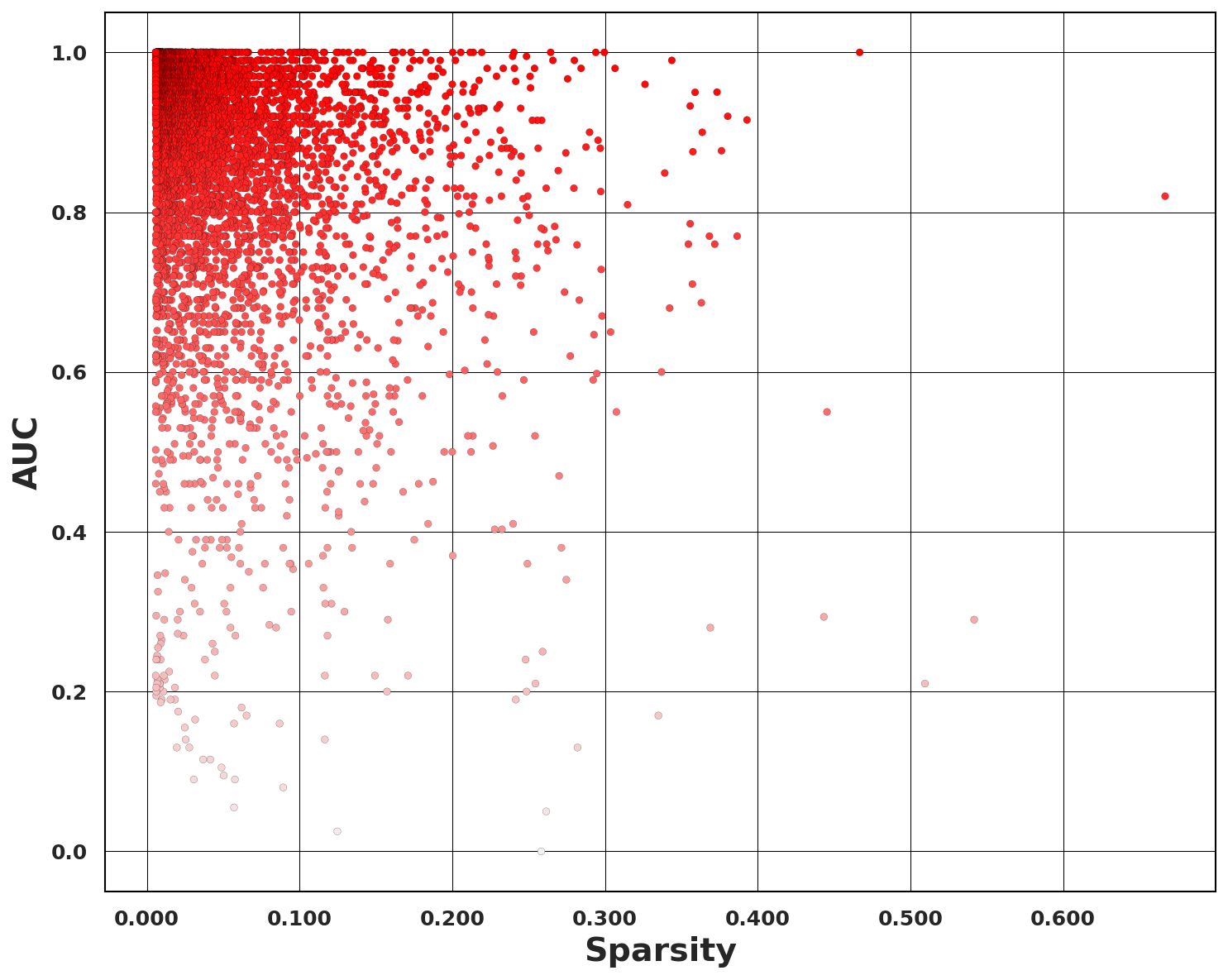}
        \caption{ItemKNN}
        \label{sfig:ml1m_full_ItemKNN}
    \end{subfigure}
    %\hfill
    \begin{subfigure}[b]{0.2\textwidth}
        \centering
        \includegraphics[width=\textwidth]{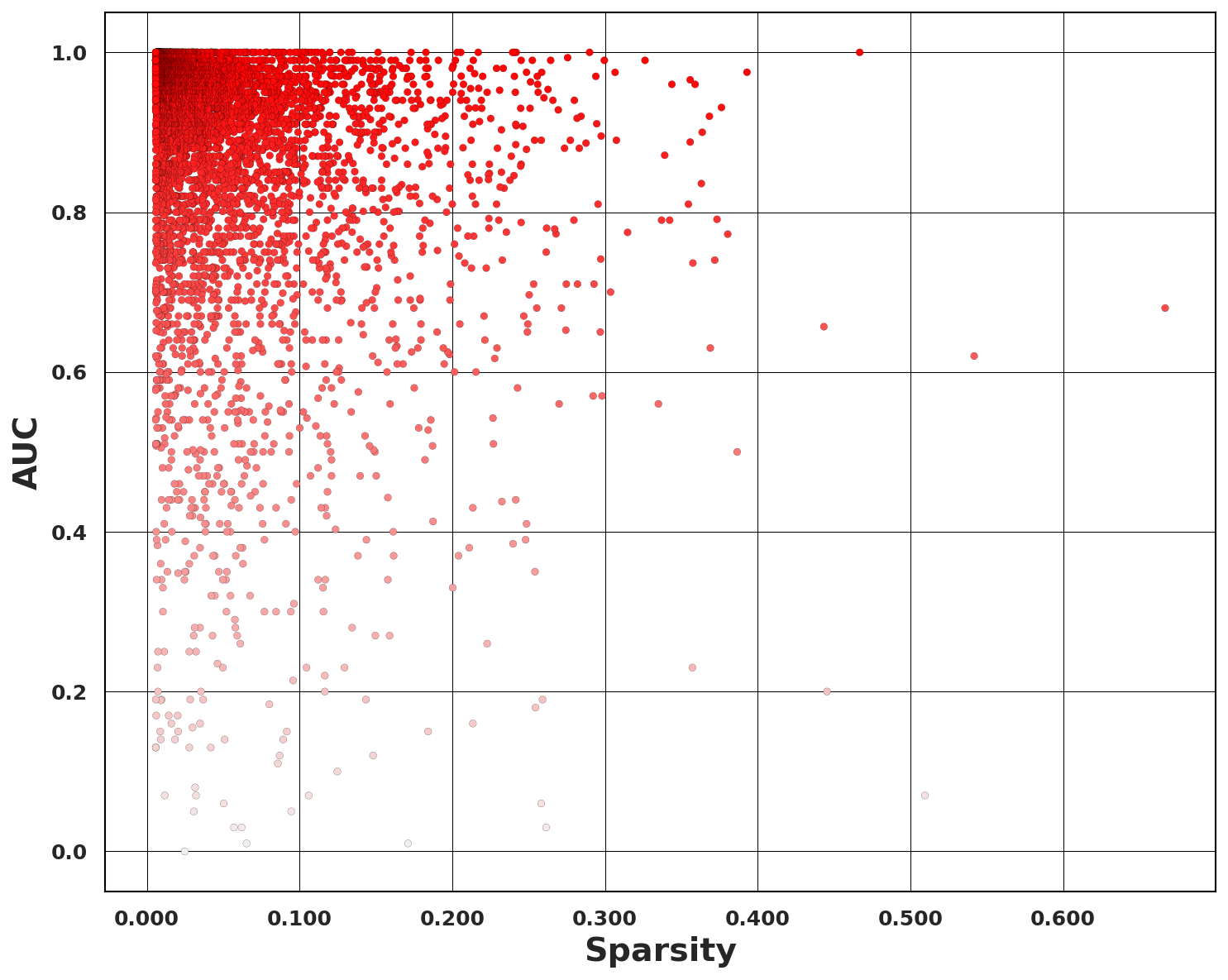}
        \caption{NeuMF}
        \label{sfig:ml1m_full_NeuMF}
    \end{subfigure}
    %\hfill
    \begin{subfigure}[b]{0.2\textwidth}
        \centering
        \includegraphics[width=\textwidth]{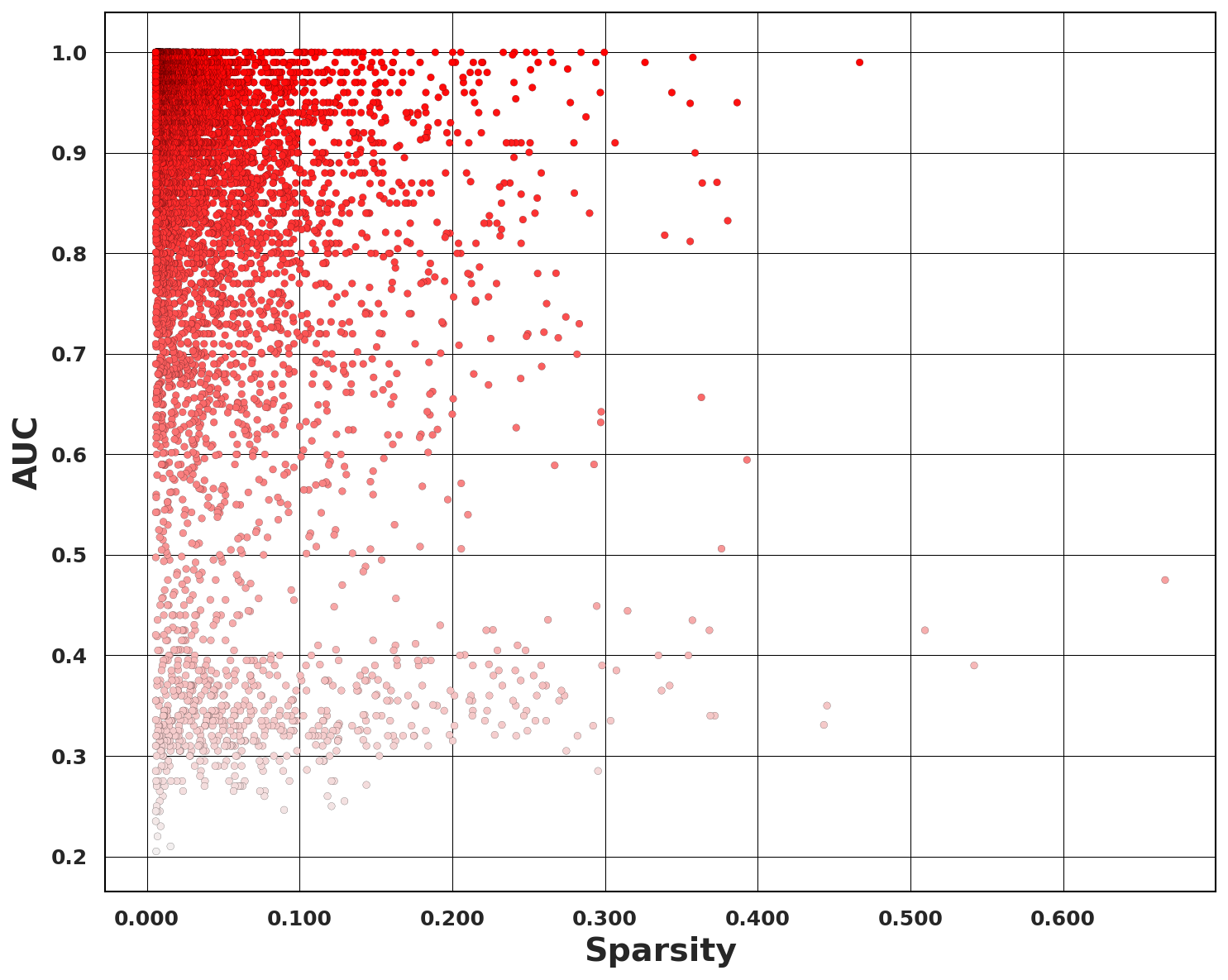}
        \caption{DMF}
        \label{sfig:ml1m_full_DMF}
    \end{subfigure}
    \begin{subfigure}[b]{0.2\textwidth}
        \centering
        \includegraphics[width=\textwidth]{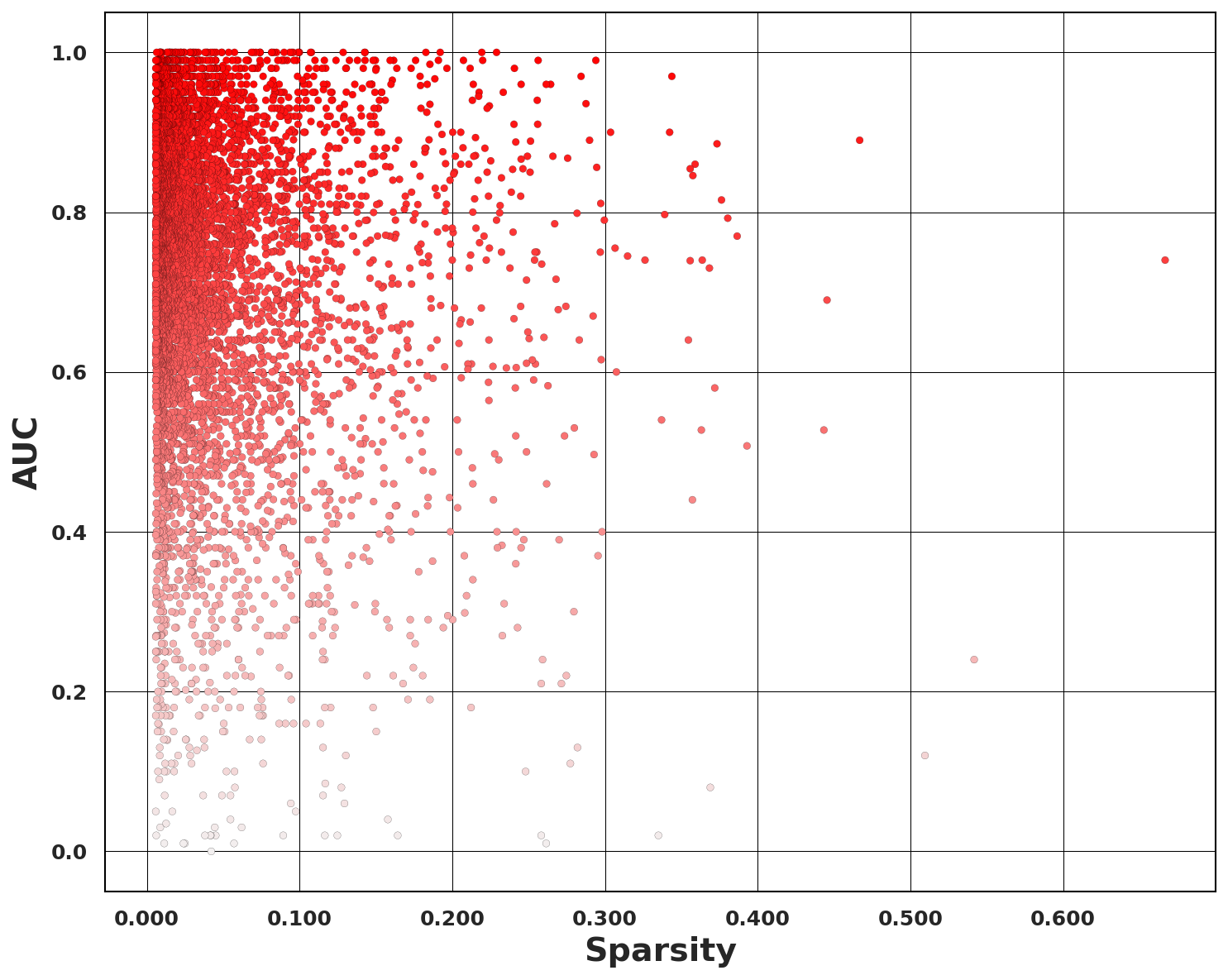}
        \caption{NNCF}
        \label{sfig:ml1m_full_NNCF}
    \end{subfigure}
    %\hfill
    \begin{subfigure}[b]{0.2\textwidth}
        \centering
        \includegraphics[width=\textwidth]{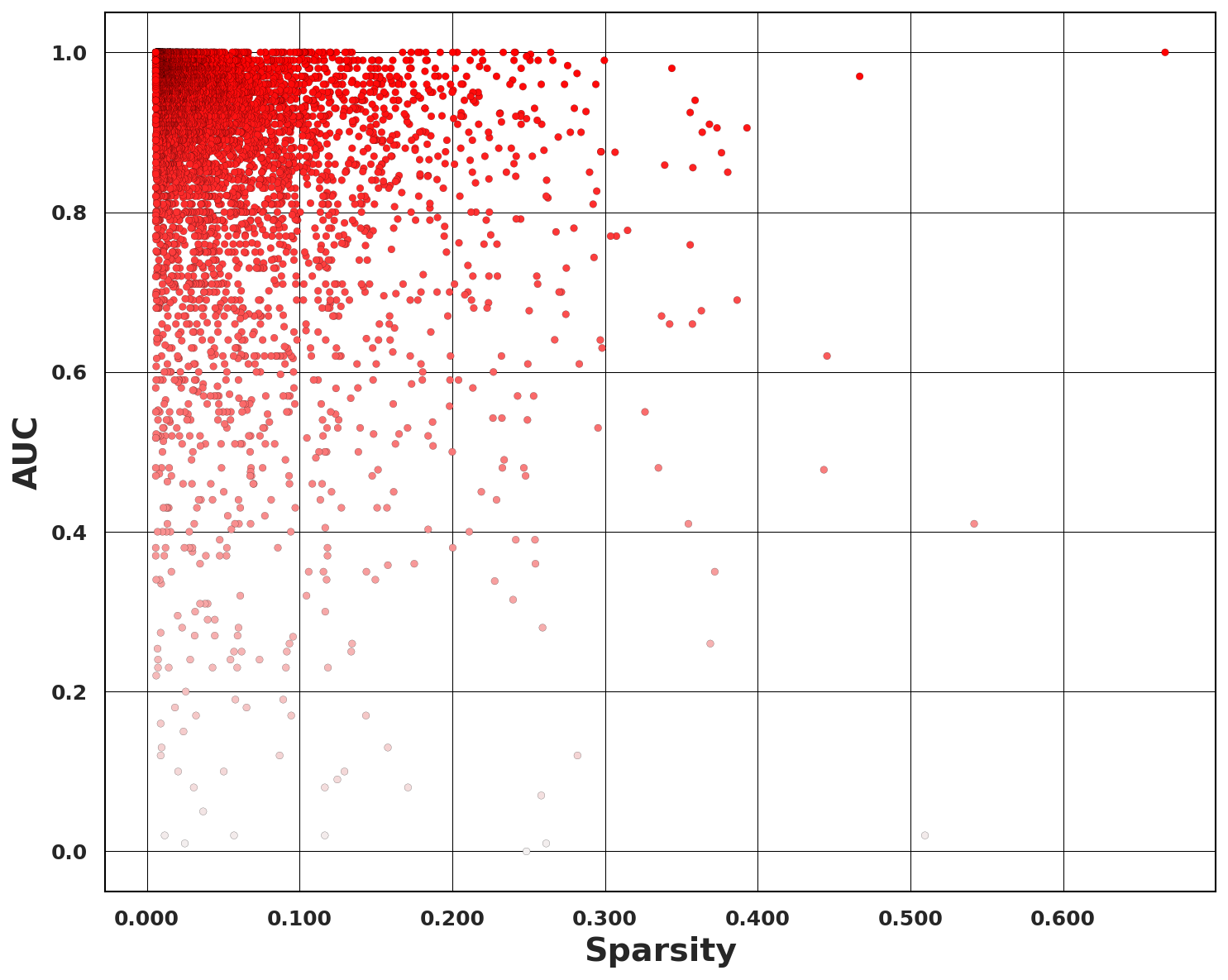}
        \caption{BPR}
        \label{sfig:ml1m_full_BPR}
    \end{subfigure}
    \begin{subfigure}[b]{0.2\textwidth}
        \centering
        \includegraphics[width=\textwidth]{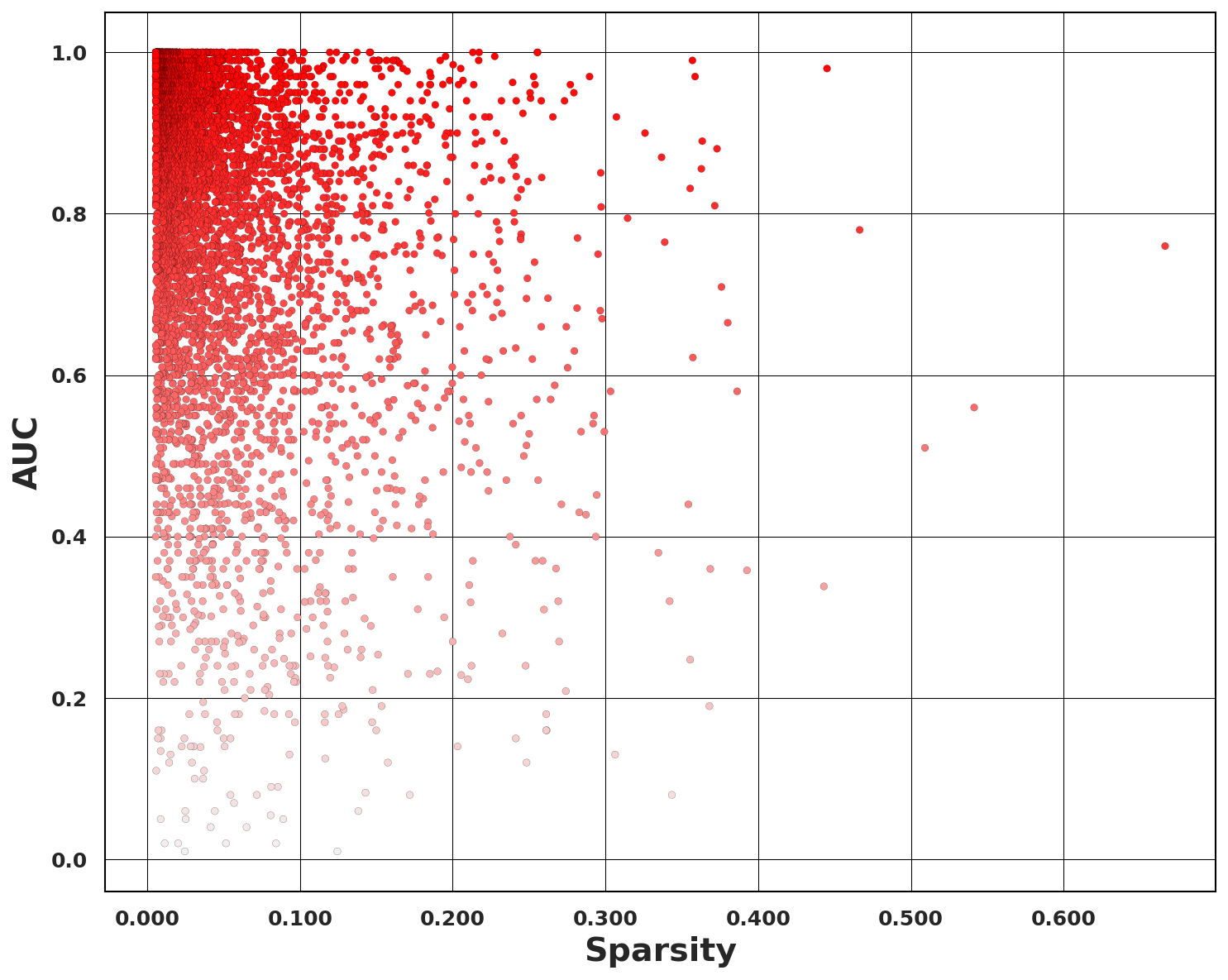}
        \caption{SASRec}
        \label{sfig:ml1m_full_SASRec}
    \end{subfigure}
    %\hfill
    \begin{subfigure}[b]{0.2\textwidth}
        \centering
        \includegraphics[width=\textwidth]{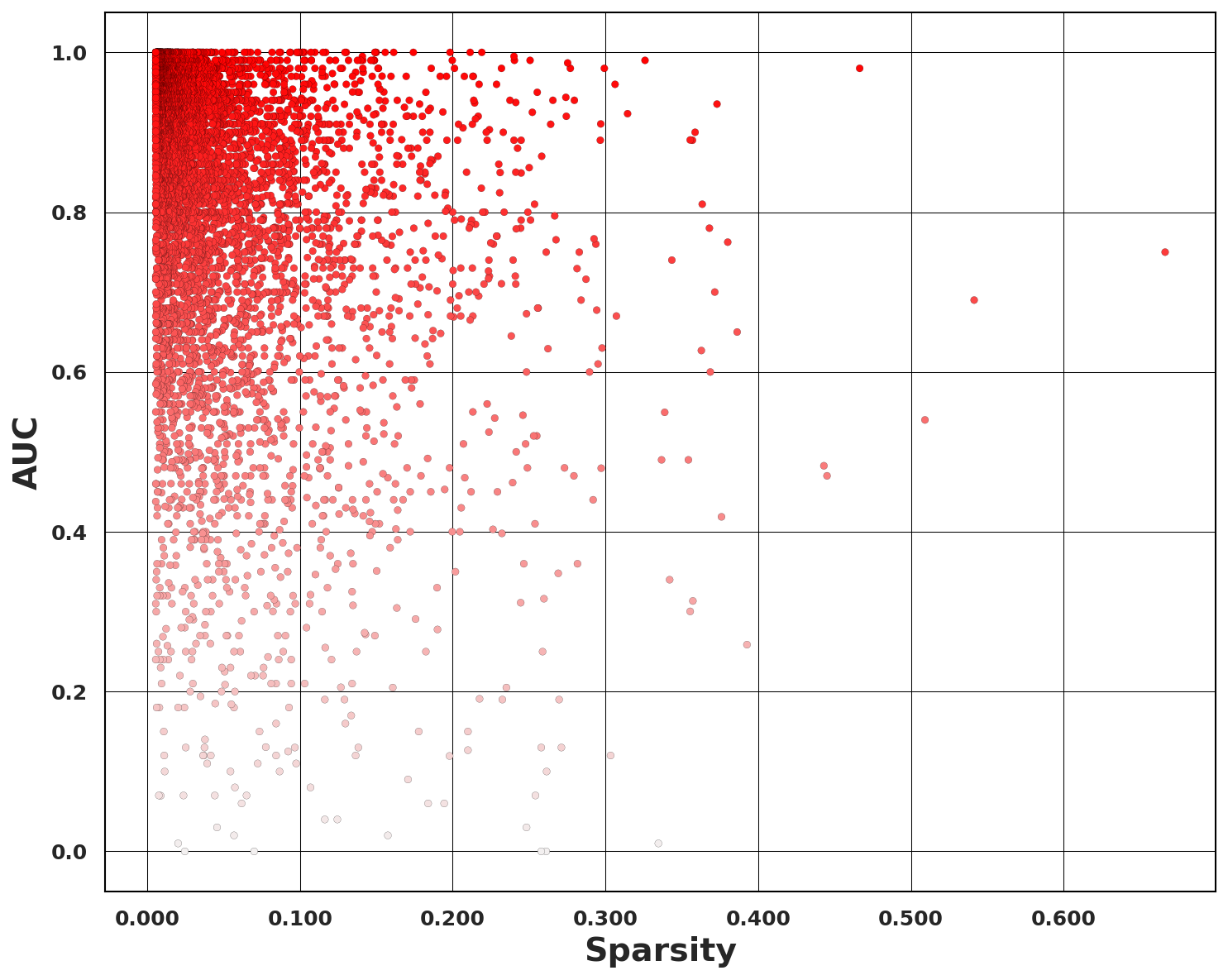}
        \caption{BERT4Rec}
        \label{sfig:ml1m_full_BERT4Rec}
    \end{subfigure}
    \begin{subfigure}[b]{0.2\textwidth}
        \centering
        \includegraphics[width=\textwidth]{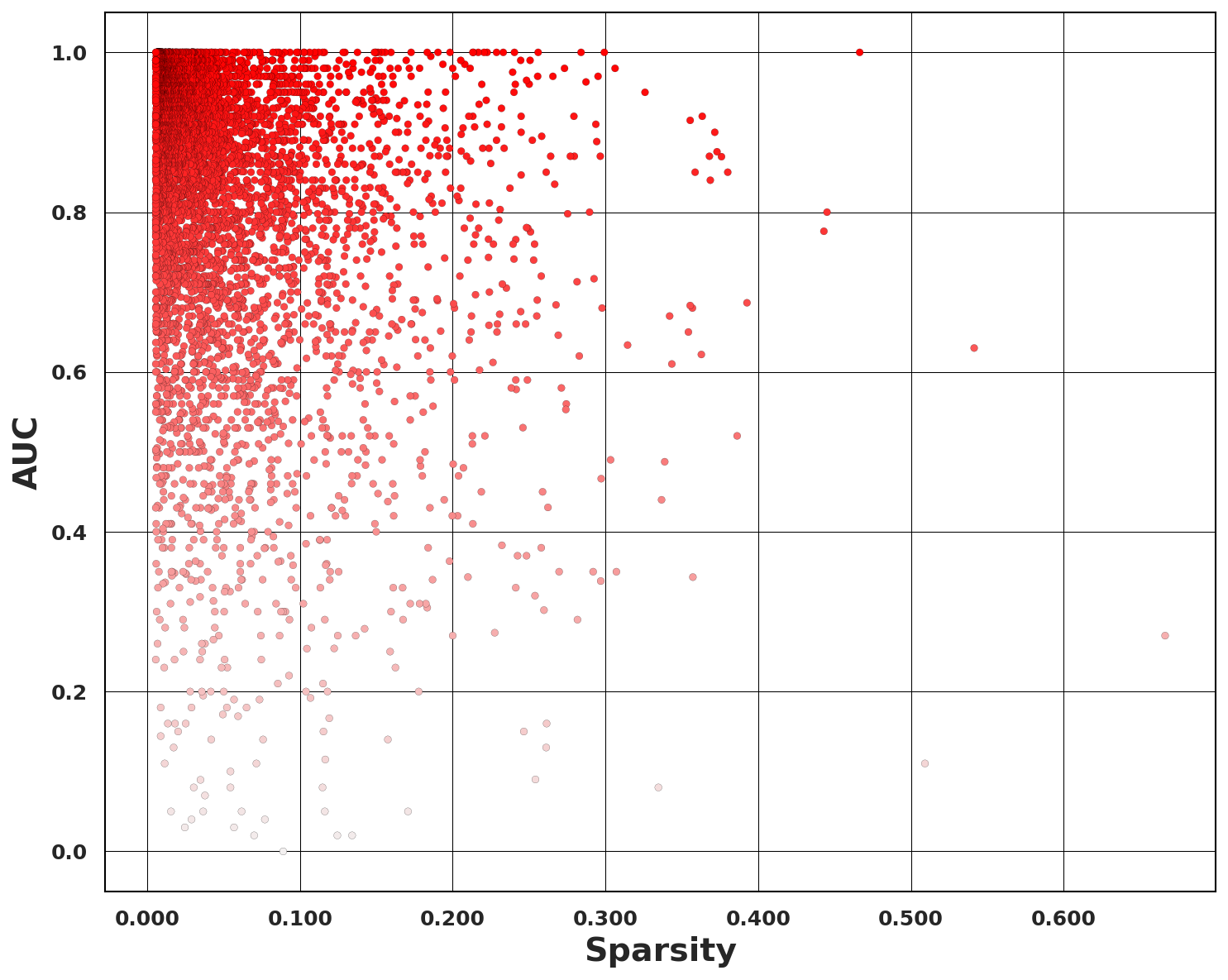}
        \caption{GRU4Rec}
        \label{sfig:ml1m_full_GRU4Rec}
    \end{subfigure}
    %\hfill
    \begin{subfigure}[b]{0.2\textwidth}
        \centering
        \includegraphics[width=\textwidth]{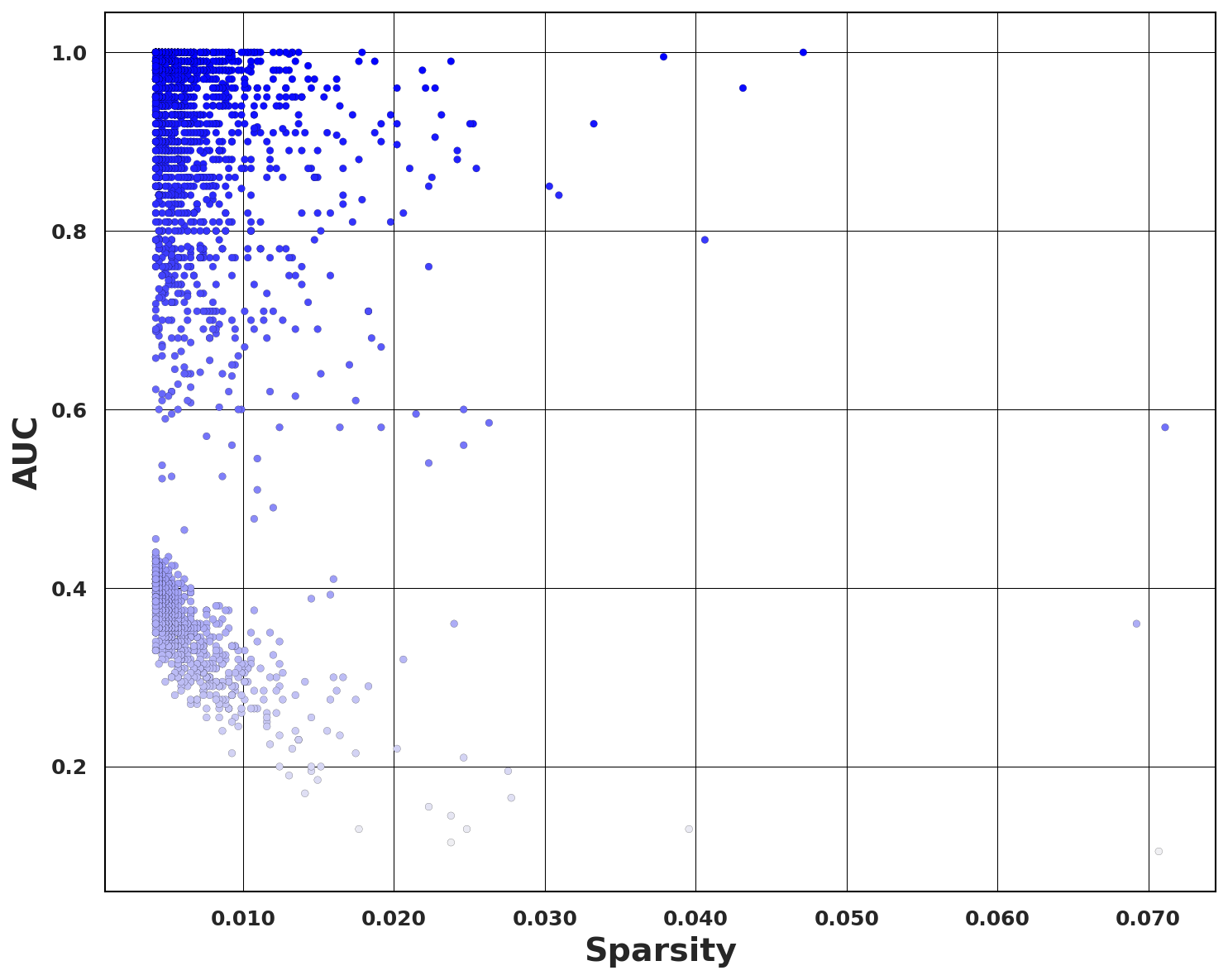}
        \caption{ItemKNN}
        \label{sfig:Amazon_Video_Games_full_ItemKNN}
    \end{subfigure}
    %\hfill
    \begin{subfigure}[b]{0.2\textwidth}
        \centering
        \includegraphics[width=\textwidth]{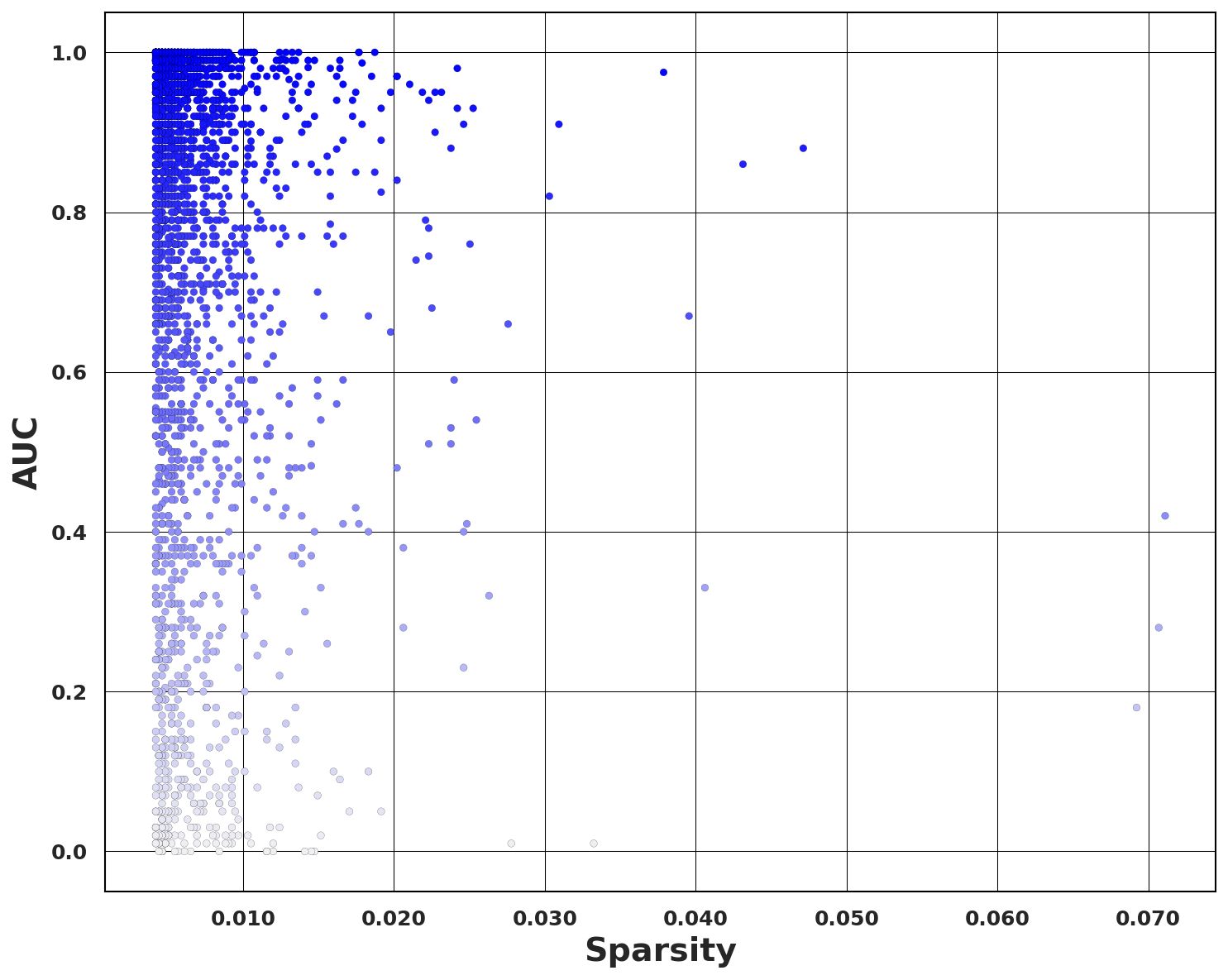}
        \caption{NeuMF}
        \label{sfig:Amazon_Video_Games_full_NeuMF}
    \end{subfigure}
    %\hfill
    \begin{subfigure}[b]{0.2\textwidth}
        \centering
        \includegraphics[width=\textwidth]{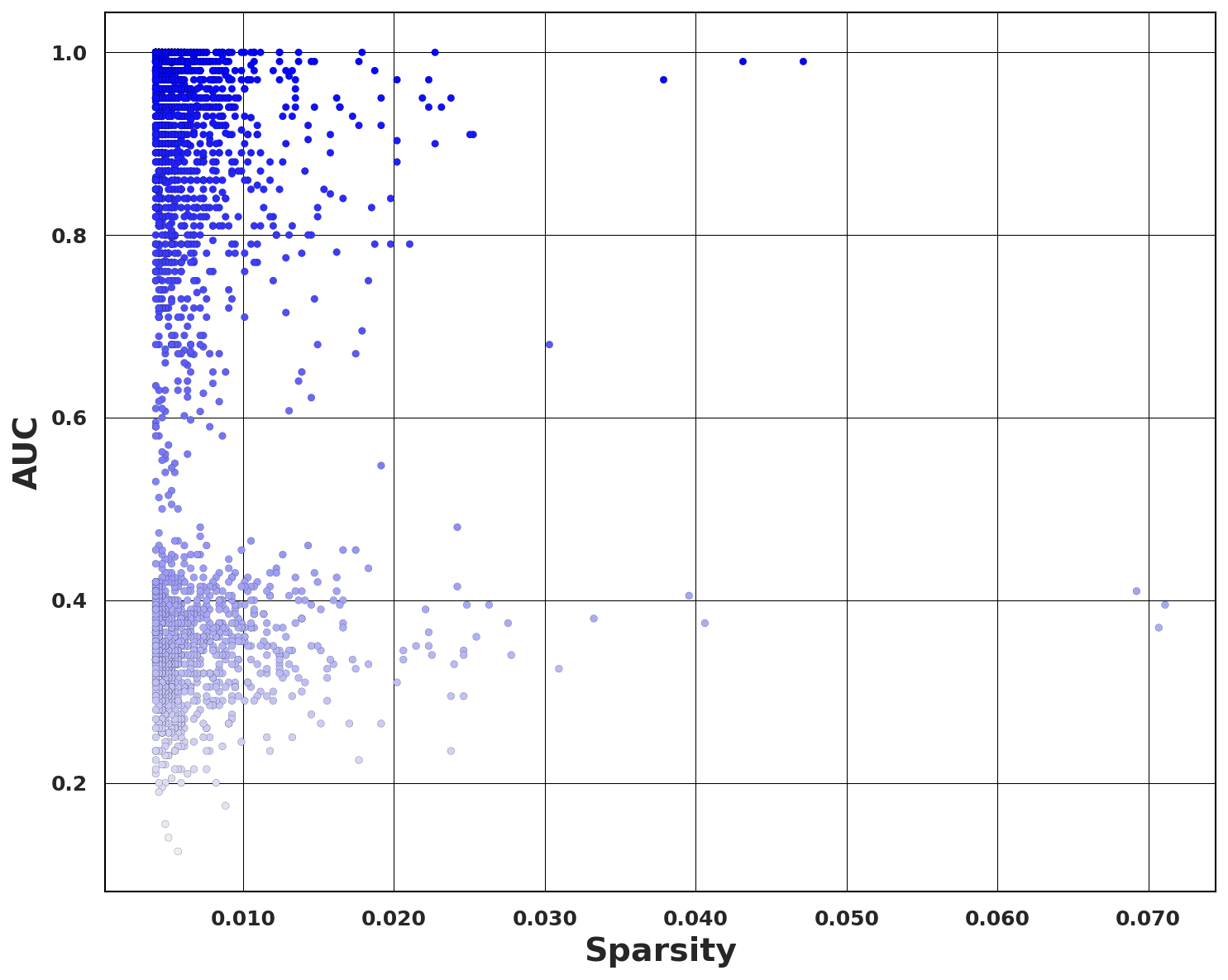}
        \caption{DMF}
        \label{sfig:Amazon_Video_Games_full_DMF}
    \end{subfigure}
    \begin{subfigure}[b]{0.2\textwidth}
        \centering
        \includegraphics[width=\textwidth]{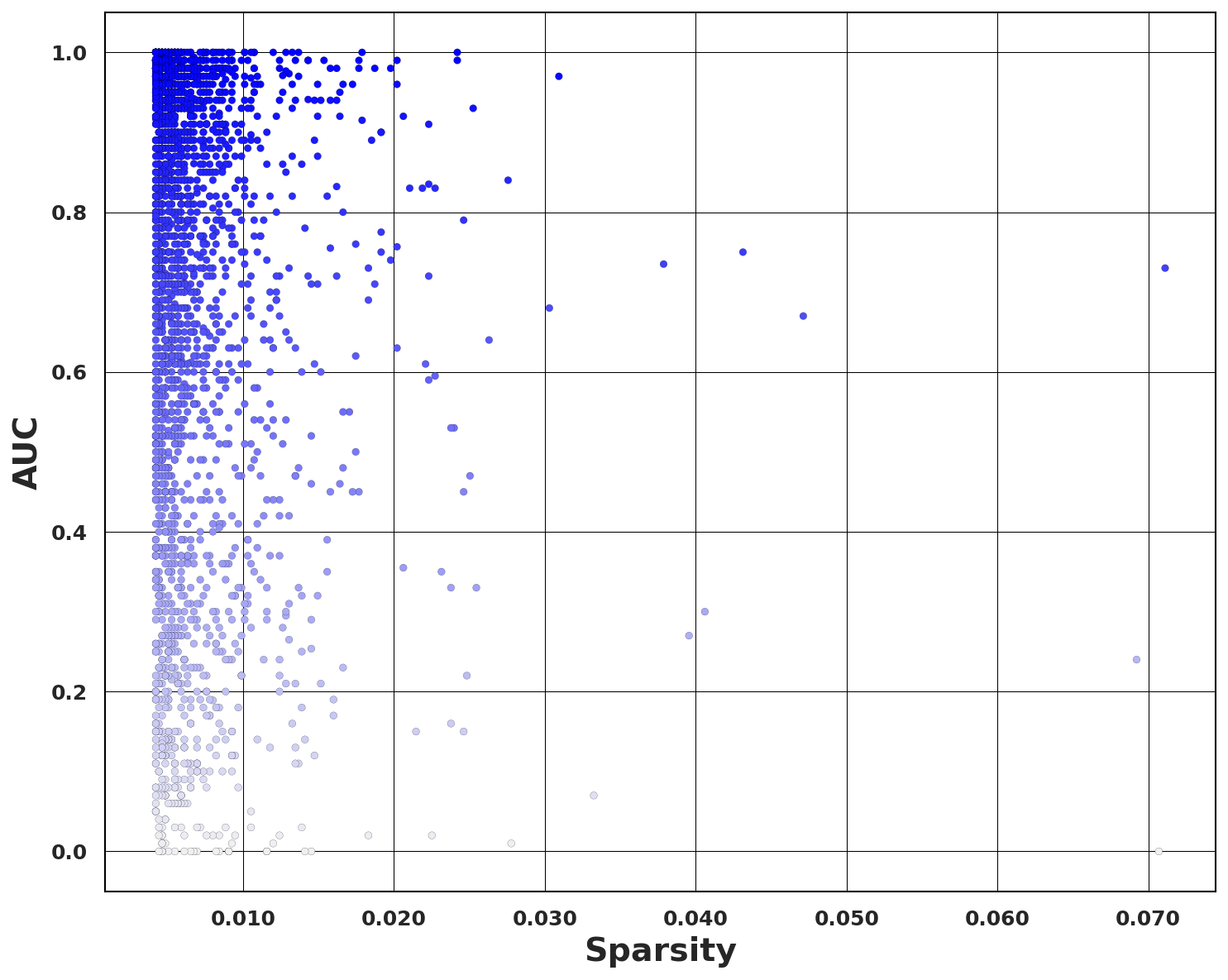}
        \caption{NNCF}
        \label{sfig:Amazon_Video_Games_full_NNCF}
    \end{subfigure}
    %\hfill
    \begin{subfigure}[b]{0.2\textwidth}
        \centering
        \includegraphics[width=\textwidth]{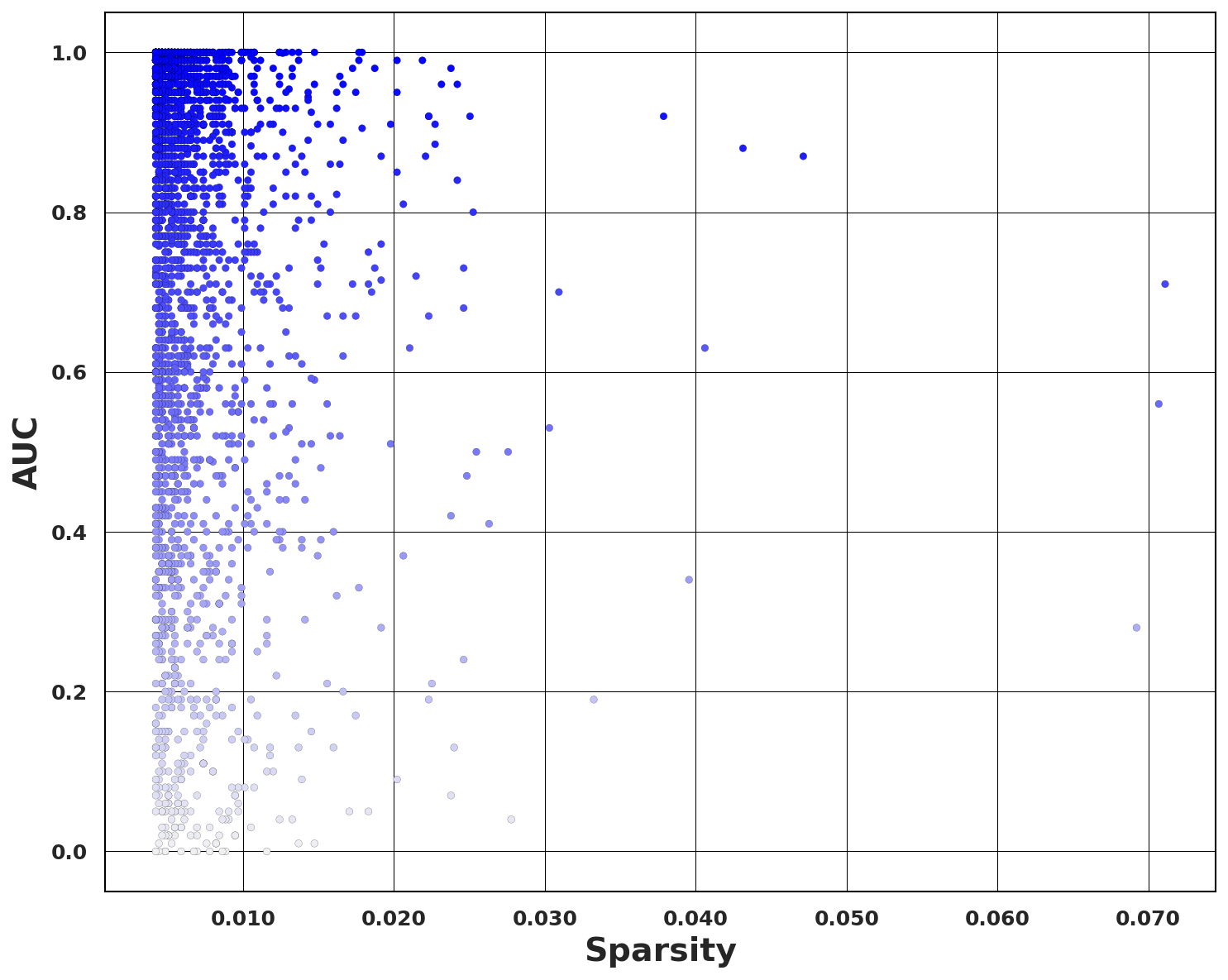}
        \caption{BPR}
        \label{sfig:Amazon_Video_Games_full_BPR}
    \end{subfigure}
    \begin{subfigure}[b]{0.2\textwidth}
        \centering
        \includegraphics[width=\textwidth]{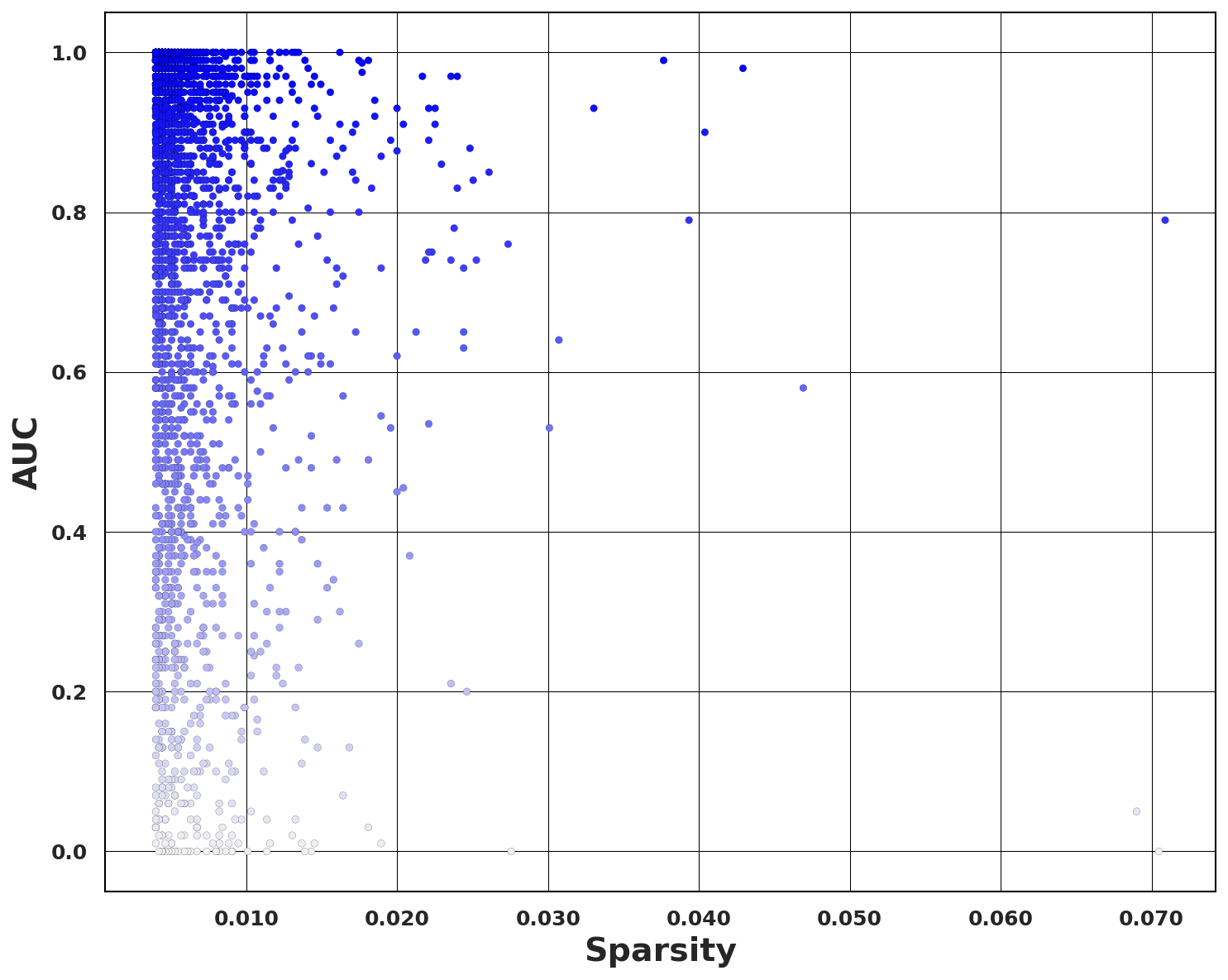}
        \caption{SASRec}
        \label{sfig:Amazon_Video_Games_full_SASRec}
    \end{subfigure}
    %\hfill
    \begin{subfigure}[b]{0.2\textwidth}
        \centering
        \includegraphics[width=\textwidth]{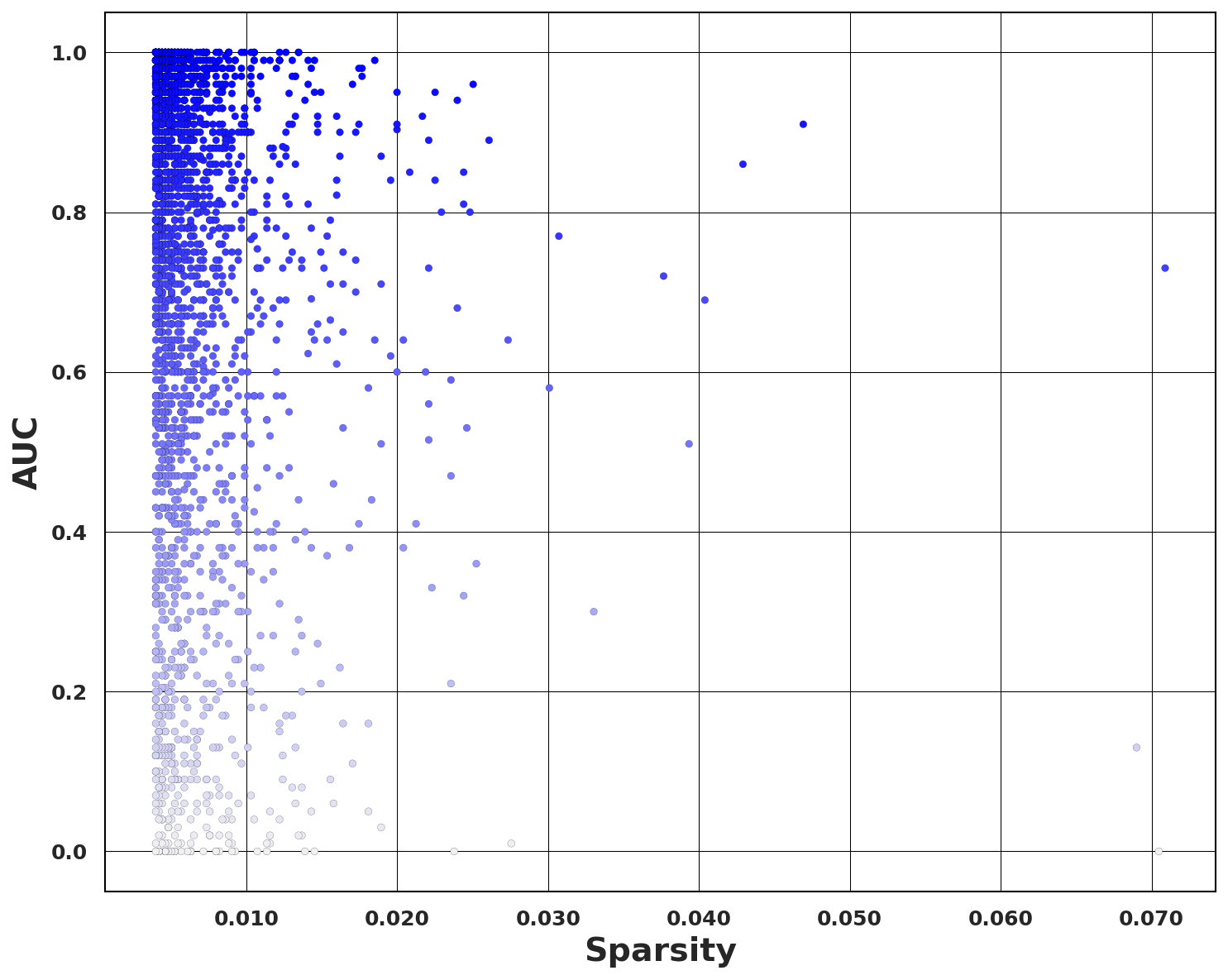}
        \caption{BERT4Rec}
        \label{sfig:Amazon_Video_Games_full_BERT4Rec}
    \end{subfigure}
    \begin{subfigure}[b]{0.2\textwidth}
        \centering
        \includegraphics[width=\textwidth]{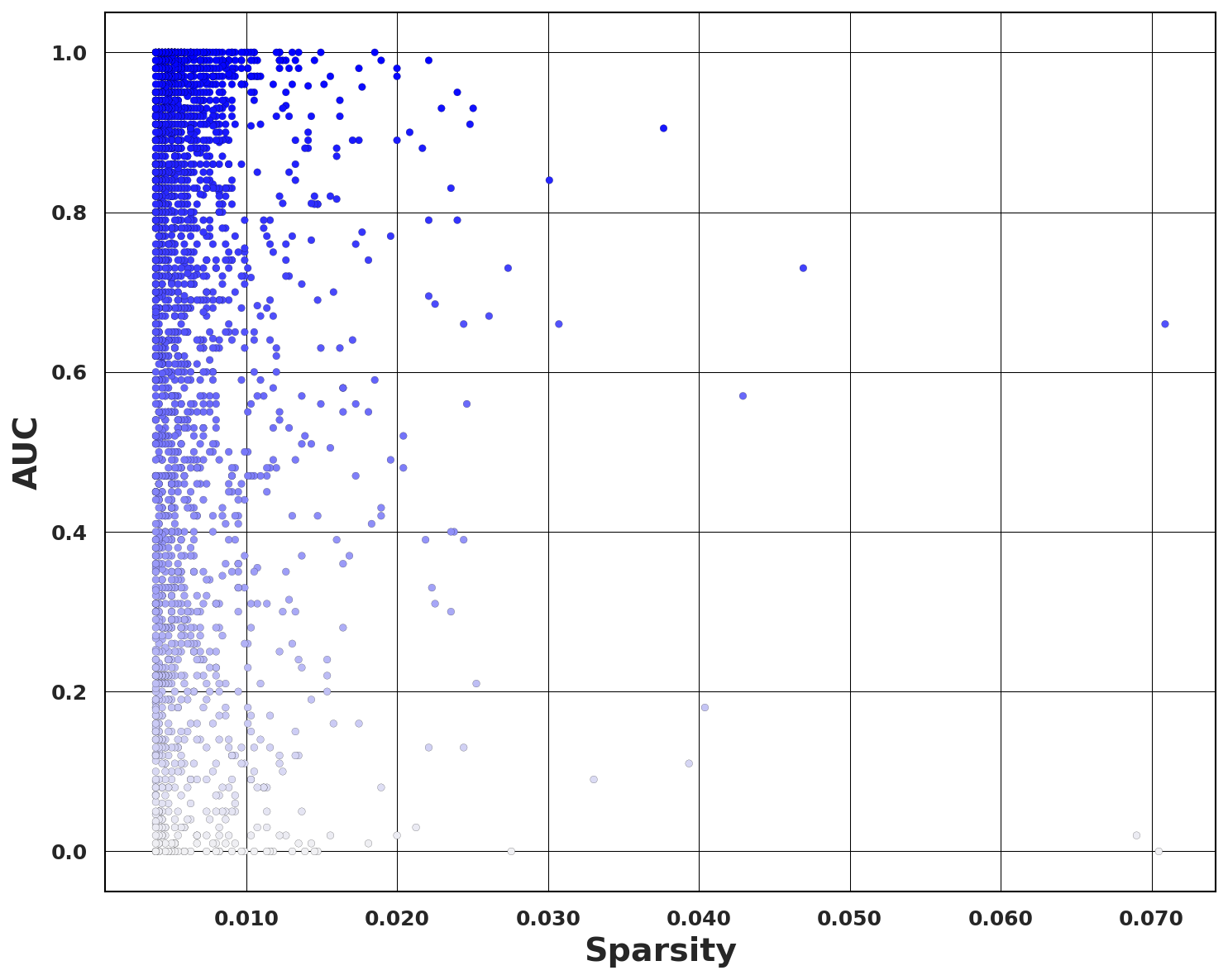}
        \caption{GRU4Rec}
        \label{sfig:Amazon_Video_Games_full_GRU4Rec}
    \end{subfigure}
    %\hfill
    \caption{ AUC vs Sparsity plots of various recommendation Mmdels (ItemKNN, NeuMF, DMF, NNCF, BPR, SASRec, BERT4Rec, GRU4Rec) across three different datasets for \emph{instance-by-instance} evaluation. The green colored plots from Fig.~\ref{sfig:software_full_ItemKNN} to Fig.~\ref{sfig:software_full_GRU4Rec} show results on Amazon Software dataset;  red colored plots from Fig.~\ref{sfig:ml1m_full_ItemKNN} to Fig.~\ref{sfig:ml1m_full_GRU4Rec} illustrate results on ML1M dataset; Fig.~\ref{sfig:Amazon_Video_Games_full_ItemKNN} to Fig.~\ref{sfig:Amazon_Video_Games_full_GRU4Rec} show results on Amazon Video Games dataset.}
    \label{fig:scatterplot}
\end{figure*}

We further strengthen our analysis by leveraging \emph{instance-by-instance} evaluation. For this, we  plot AUC scores against the sparsity index for all users as shown in Fig.~\ref{fig:scatterplot}. The detailed analysis of the plots reveals nuanced insights into the performance of various recommendation models across sparse and non-sparse user interactions. While previous plots indicated that the average AUC for sparse users is comparatively lower, it is crucial to note that not all sparse users receive poor-quality recommendations. Our plots substantiate this by showing that advanced models like \textbf{BERT4Rec} and \textbf{GRU4Rec} can still deliver high-quality recommendations to sparse users by effectively leveraging sequential information. For instance, in the Amazon Software dataset (green plots in Fig.~\ref{sfig:software_full_ItemKNN} to~\ref{sfig:software_full_GRU4Rec}), \textbf{BERT4Rec} (Fig.~\ref{sfig:software_full_BERT4Rec}) and \textbf{GRU4Rec} (Fig.~\ref{sfig:software_full_GRU4Rec})  consistently outperform other models, demonstrating their robustness in handling sparse interactions. This contrasts sharply with simpler models like \textbf{ItemKNN} (Fig.~\ref{sfig:software_full_ItemKNN})  and \textbf{BPR} (Fig.~\ref{sfig:software_full_BPR}), which show a significant drop in performance due to their reliance on ample interaction data. The ML1M dataset (red plots in Fig.~\ref{sfig:ml1m_full_ItemKNN} to~\ref{sfig:ml1m_full_GRU4Rec}), being the least sparse, allows all models to perform comparatively well, with \textbf{BERT4Rec} (Fig. ~\ref{sfig:ml1m_full_BERT4Rec}) and \textbf{GRU4Rec} (Fig. ~\ref{sfig:ml1m_full_GRU4Rec}) leading the results again. This dataset highlights the models' ability to capture intricate user behavior patterns when interaction data is abundant. In the Amazon Video Games dataset (blue plots in Fig.~\ref{sfig:Amazon_Video_Games_full_ItemKNN} to~\ref{sfig:Amazon_Video_Games_full_GRU4Rec}), which is the most sparse, the performance gap between advanced and simpler models widens. \textbf{BERT4Rec} (Fig.~\ref{sfig:Amazon_Video_Games_full_BERT4Rec}) and \textbf{GRU4Rec} (Fig.~\ref{sfig:Amazon_Video_Games_full_GRU4Rec}) maintain their superior performance, underscoring their adaptability and effectiveness in sparse data scenarios. \textbf{NNCF} (Fig.~\ref{sfig:Amazon_Video_Games_full_NNCF}) and \textbf{SASRec} (Fig.~\ref{sfig:Amazon_Video_Games_full_SASRec}) also show commendable performance, though they slightly lag behind the top sequential models.

 This analysis also highlights the presence and impact of outliers in the datasets. Outliers, which are users with exceptionally high or low interaction counts, can significantly influence model performance. In sparse datasets, outliers with high interaction counts can skew the results, making it appear that the model performs better than it does for the majority of users. Conversely, outliers with very low interaction counts can highlight the limitations of simpler models like \textbf{ItemKNN} and \textbf{BPR}, which struggle to provide accurate recommendations with minimal data. Our plots show that advanced models like \textbf{BERT4Rec} and \textbf{GRU4Rec} are less affected by outliers due to their ability to generalize from sequential and contextual patterns. This robustness is particularly evident in the Amazon Software and Amazon Video Games datasets, where these models maintain high performance despite the presence of outliers. This instance-by-instance analysis highlights that though on average, sparse users receive poor recommendation results (AUC), not all sparse users experience uniformly poor results. Some sparse users may still receive relatively accurate recommendations depending on specific factors such as their interaction history or the diversity of their preferences. We thus use our definition to identify such users and mark them as weak.

\begin{figure*}[t!]
    \centering
    \begin{subfigure}[b]{0.23\textwidth}
        \centering
        \includegraphics[width=\textwidth]{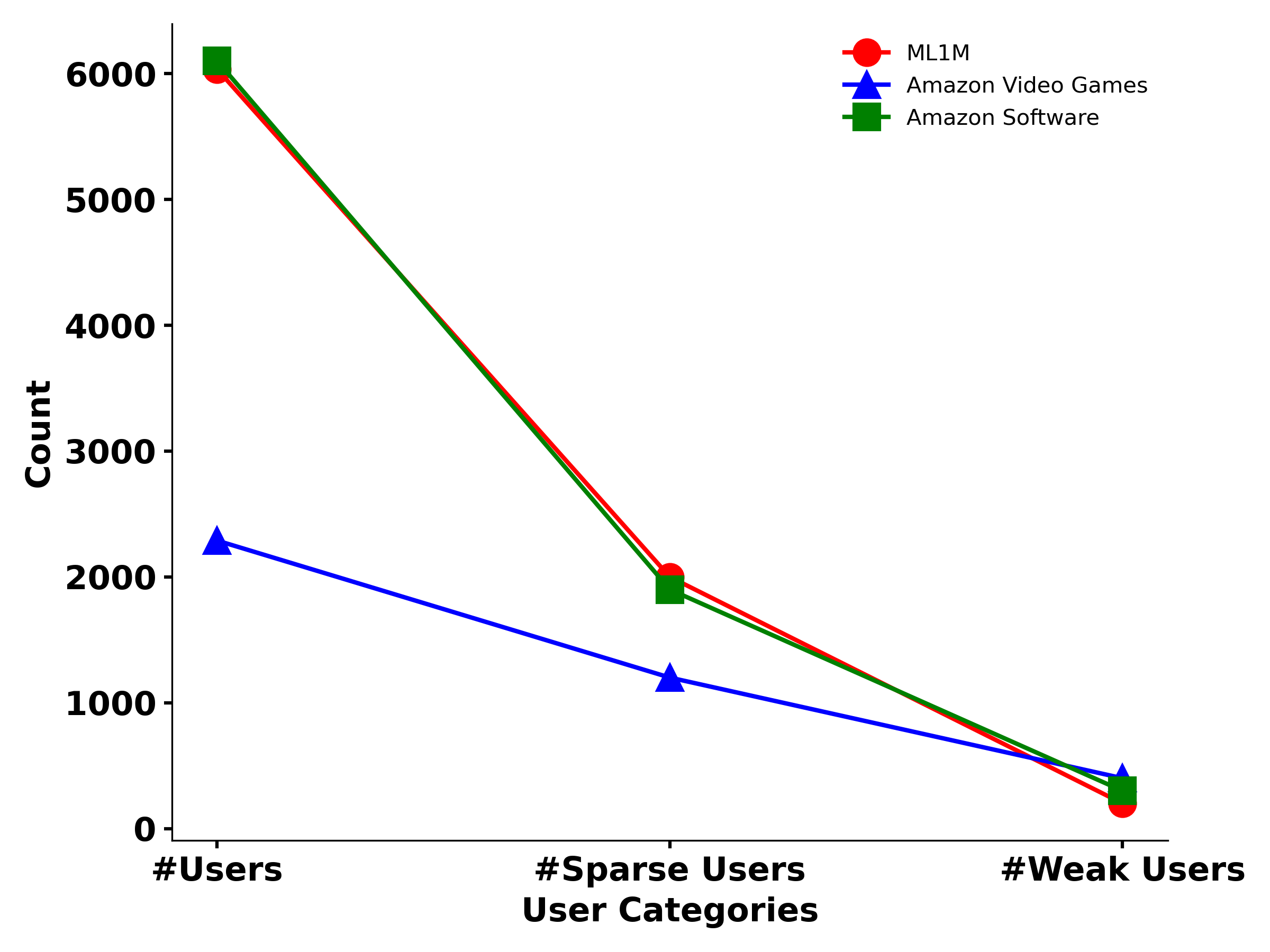}
        \caption{ItemKNN}
        \label{sfig:ItemKNN_userreduction}
    \end{subfigure}
    %\hfill
    \begin{subfigure}[b]{0.23\textwidth}
        \centering
        \includegraphics[width=\textwidth]{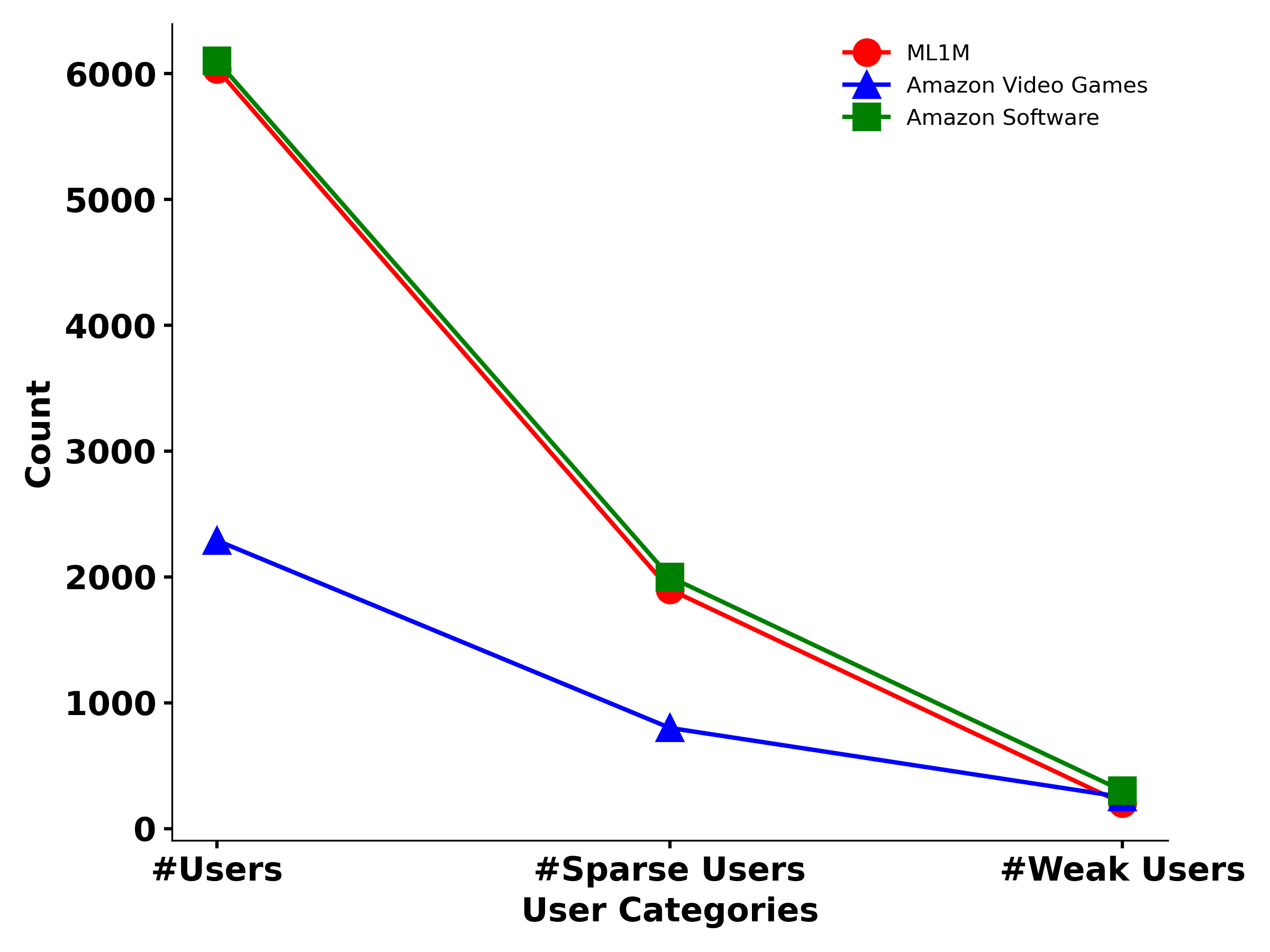}
        \caption{NeuMF}
        \label{sfig:NeuMF_userreduction.png}
    \end{subfigure}
    %\hfill
    \begin{subfigure}[b]{0.23\textwidth}
        \centering
        \includegraphics[width=\textwidth]{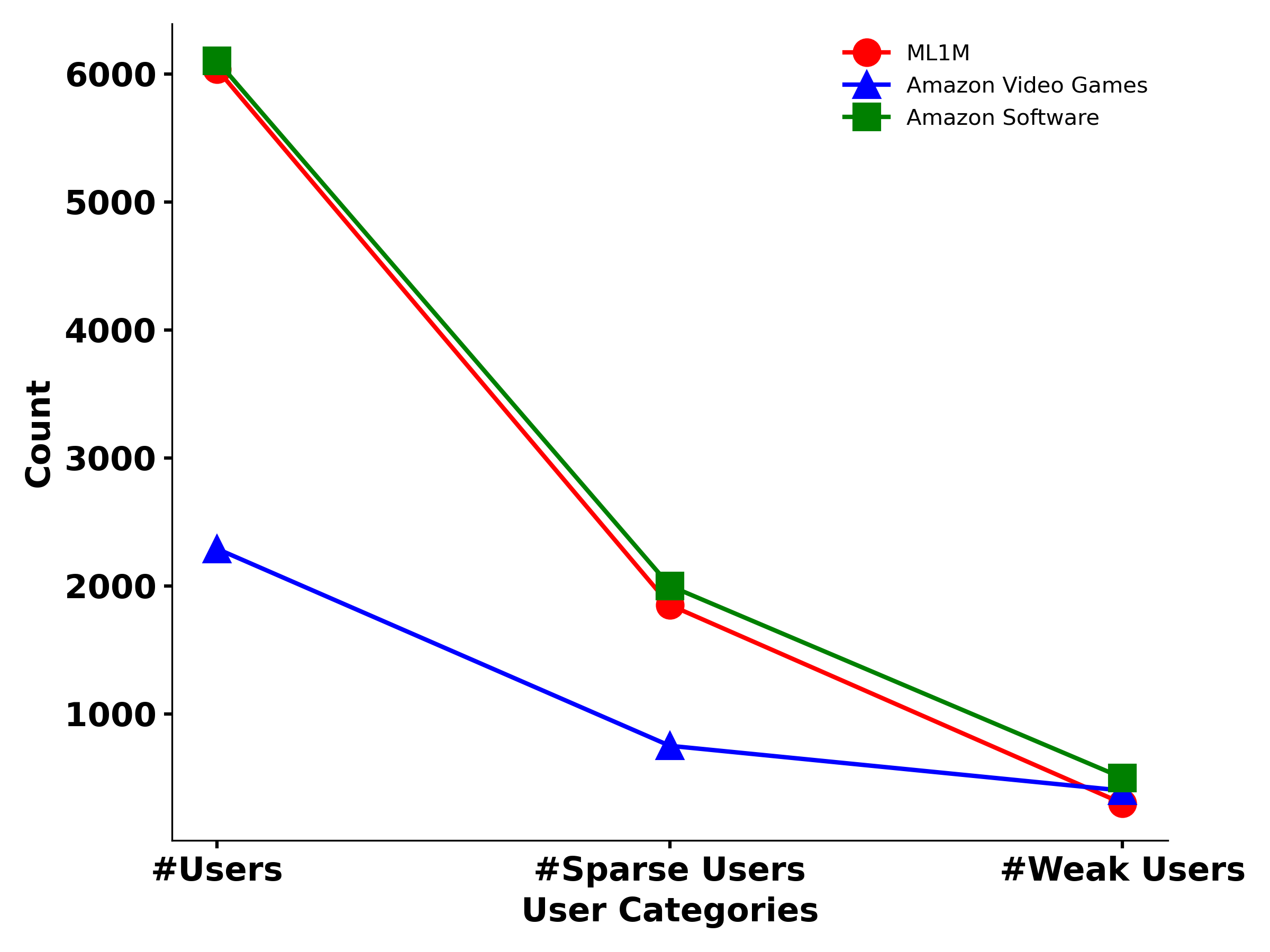}
        \caption{DMF}
        \label{sfig:DMF_userreduction.png}
    \end{subfigure}
    \begin{subfigure}[b]{0.23\textwidth}
        \centering
        \includegraphics[width=\textwidth]{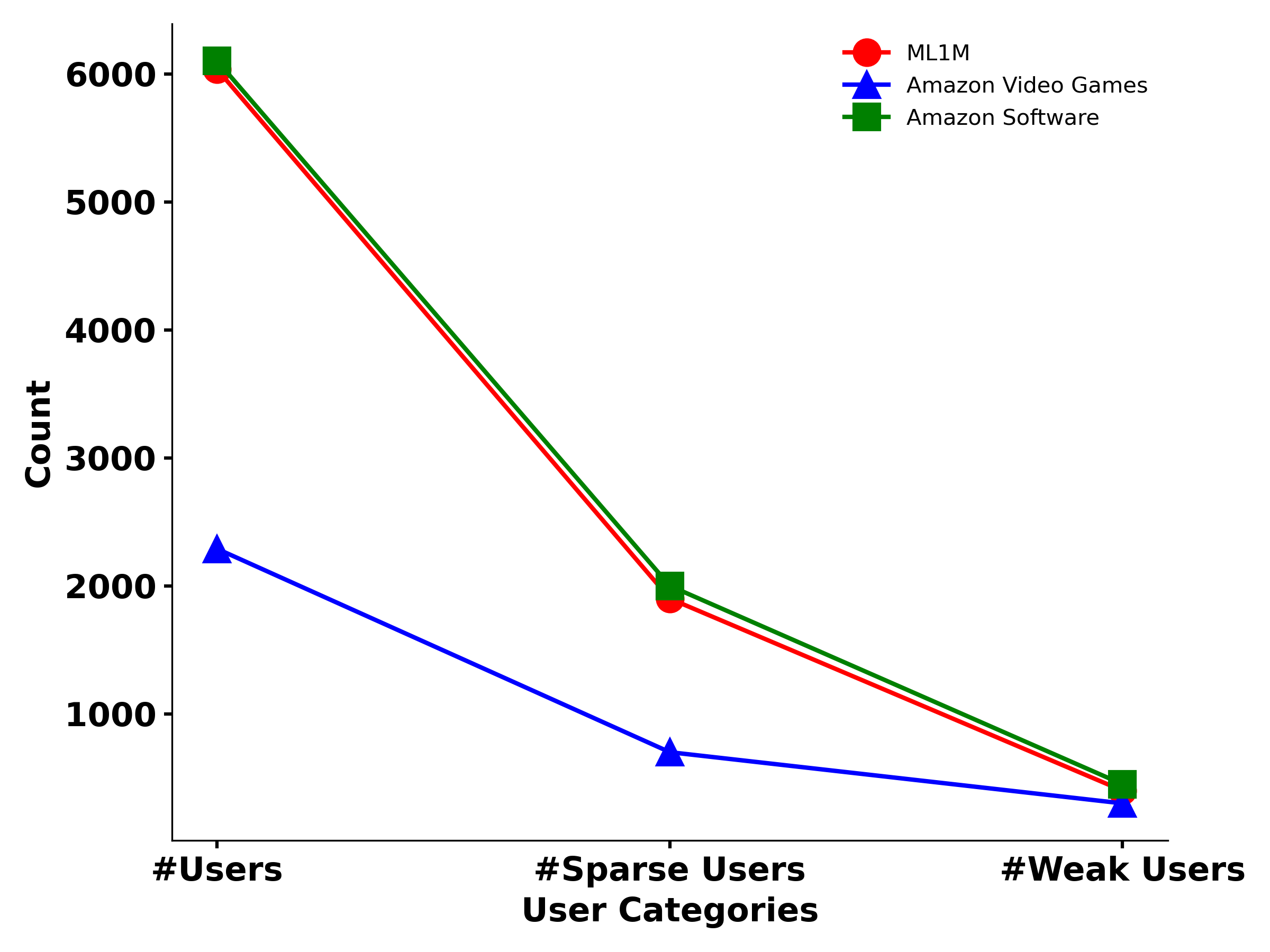}
        \caption{NNCF}
        \label{sfig:NNCF_userreduction.png}
    \end{subfigure}
    %\hfill
    \begin{subfigure}[b]{0.23\textwidth}
        \centering
        \includegraphics[width=\textwidth]{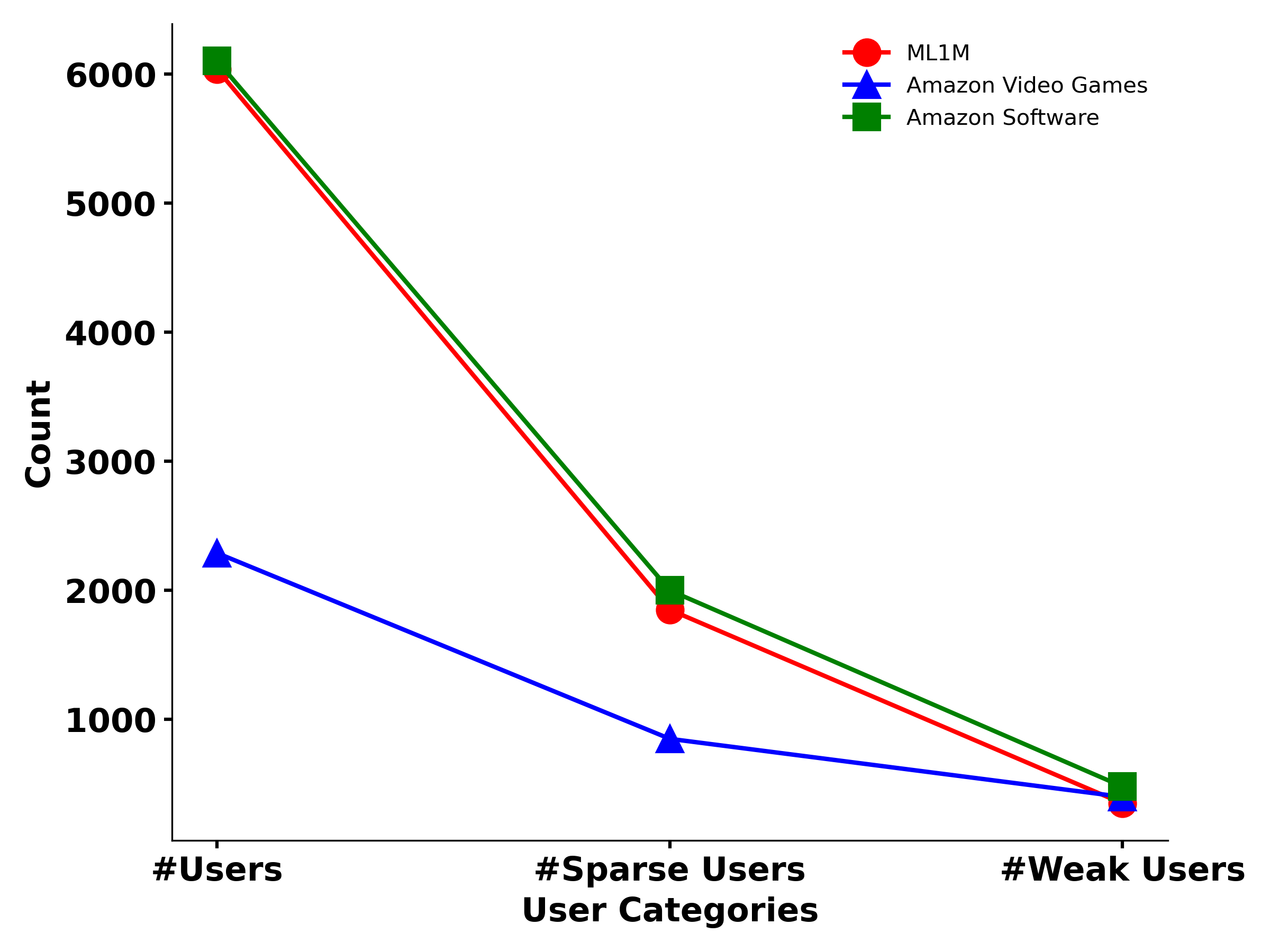}
        \caption{BPR}
        \label{sfig:BPR_userreduction.png}
    \end{subfigure}
    \begin{subfigure}[b]{0.23\textwidth}
        \centering
        \includegraphics[width=\textwidth]{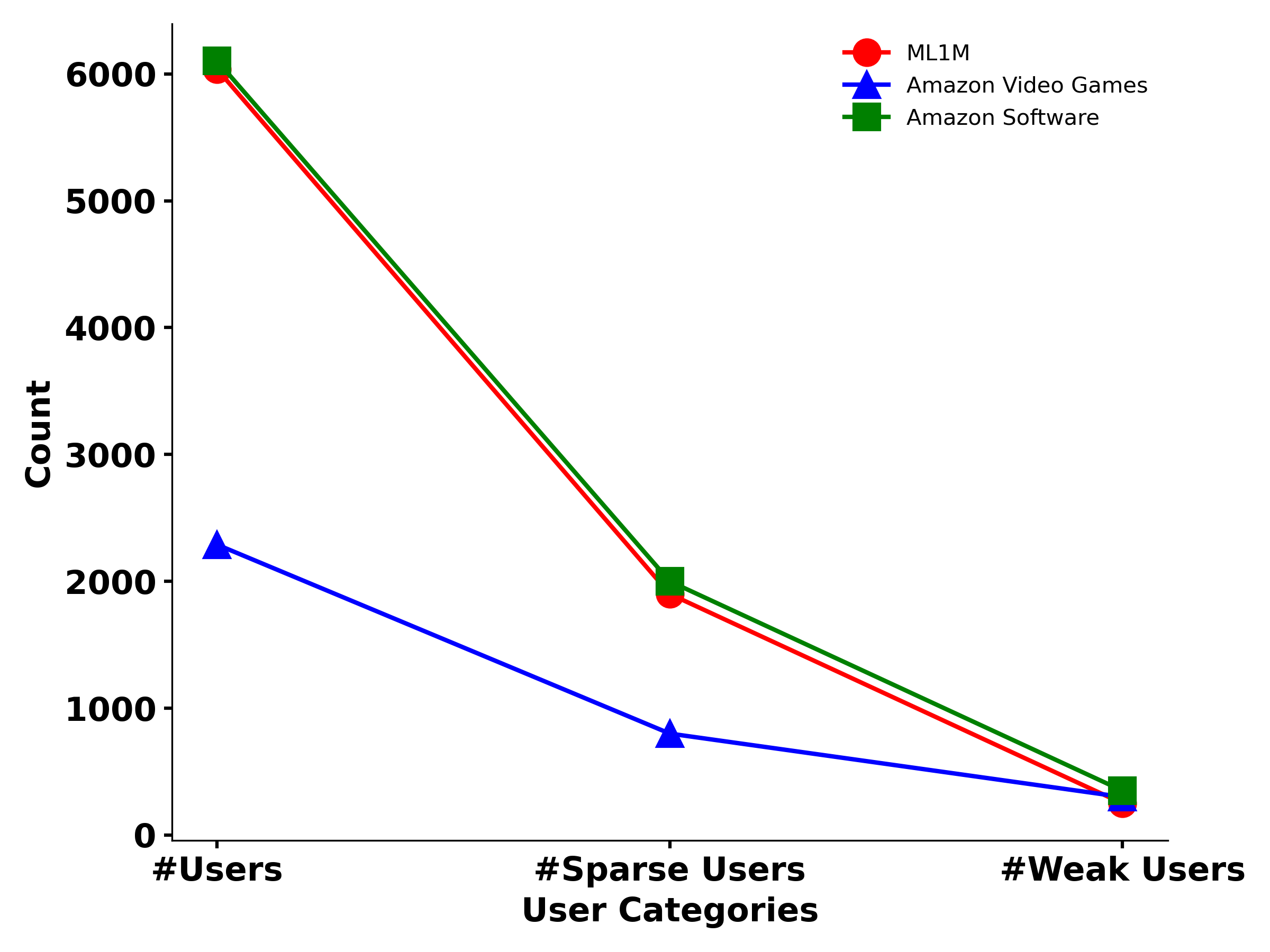}
        \caption{SASRec}
        \label{sfig:SASRec_userreduction.png}
    \end{subfigure}
    %\hfill
    \begin{subfigure}[b]{0.23\textwidth}
        \centering
        \includegraphics[width=\textwidth]{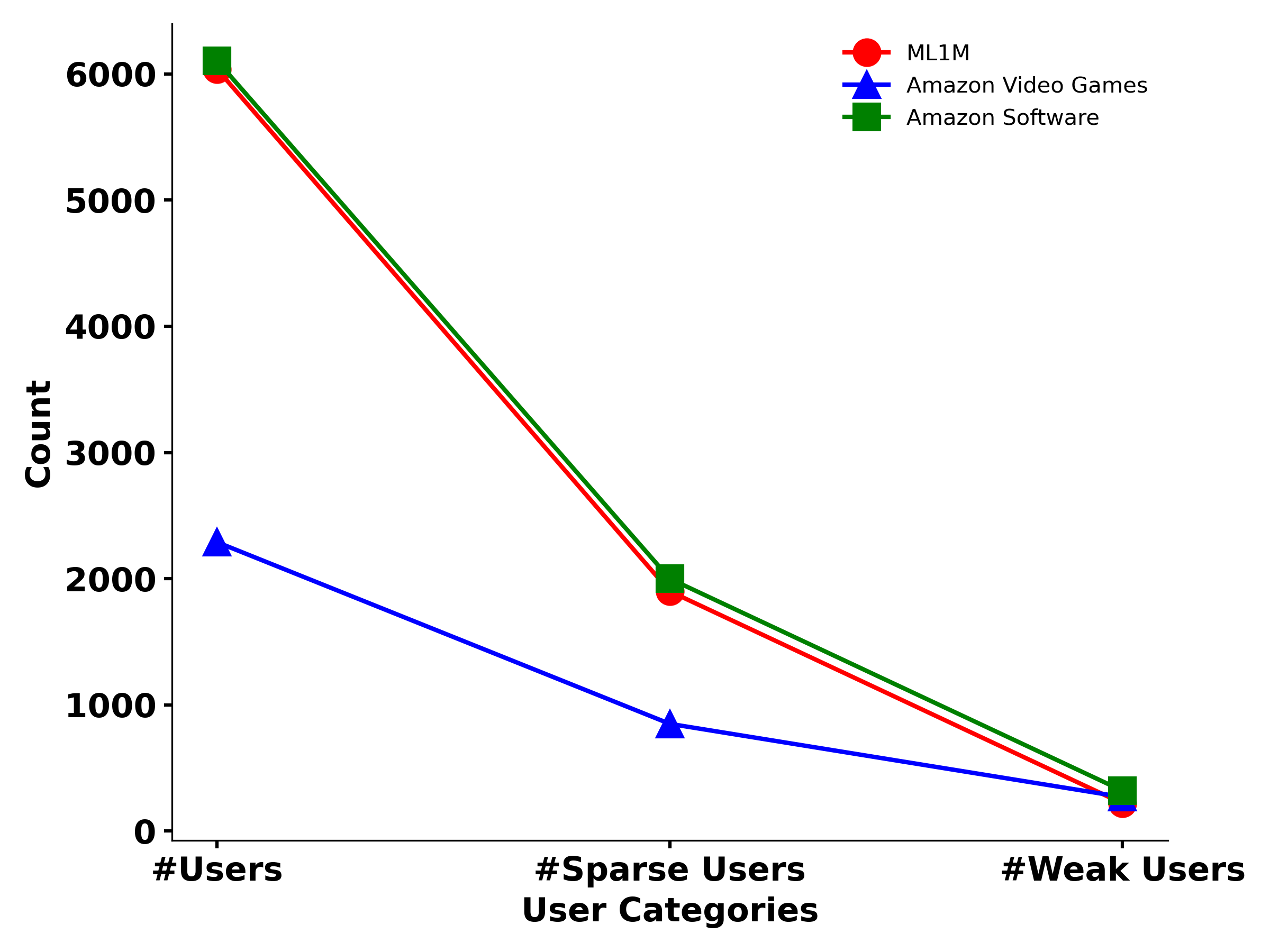}
        \caption{BERT4Rec}
        \label{sfig:BERT4Rec_userreduction.png}
    \end{subfigure}
    \begin{subfigure}[b]{0.23\textwidth}
        \centering
        \includegraphics[width=\textwidth]{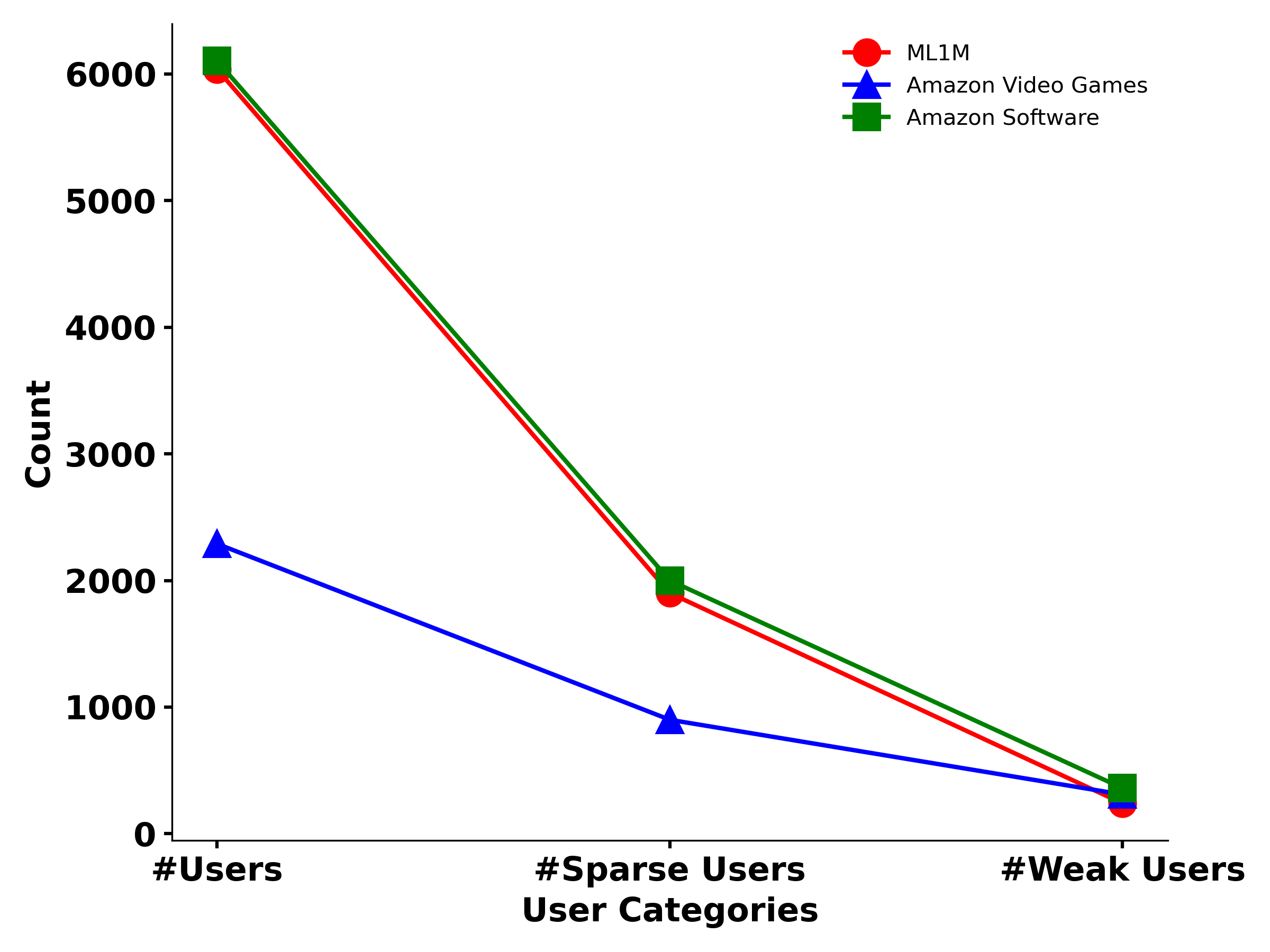}
        \caption{GRU4Rec}
        \label{sfig:GRU4Rec_userreduction.png}
    \end{subfigure}
    \caption{ Detailed analysis of User Sparsity and Weak User Identification across three Datasets and RSs.}
    \label{fig:lineplot_comparison}
\end{figure*}

In Fig.~\ref{fig:lineplot_comparison} we compare how our instance-by-instance criteria elicits extreme weak users across all three datasets. The ML1M dataset (red line in all plots) analysis reveals a consistent pattern in user sparsity across all models, with a $68.68\%$ reduction from total users to sparse users. This uniformity suggests that the dataset has a stable distribution of user interactions, making it a reliable benchmark for evaluating recommendation systems. However, the reduction from sparse users to weak users varies significantly among models. \textbf{BPR} (Fig.~\ref{sfig:BPR_userreduction.png}) and \textbf{ItemKNN} (Fig.~\ref{sfig:ItemKNN_userreduction}) exhibit the highest reduction rates, indicating their effectiveness in identifying weak users within the sparse user group. In contrast, models like \textbf{DMF} (Fig.~\ref{sfig:DMF_userreduction.png})and \textbf{NNCF} (Fig.~\ref{sfig:NNCF_userreduction.png})show lower reduction rates, suggesting they may struggle more with distinguishing weak users from sparse ones. In the Amazon Video Games dataset (blue line in all plots), the reduction from total users to sparse users is consistently $70.40\%$ across all models and  the reduction from sparse users to weak users shows more variation. \textbf{SASRec} (Fig.~\ref{sfig:SASRec_userreduction.png}) stands out with the highest reduction rate, suggesting it is particularly effective in identifying weak users within this dataset. On the other hand, \textbf{DMF} (Fig.~\ref{sfig:DMF_userreduction.png}) shows the lowest reduction rate, indicating it may not be as adept at distinguishing weak users from sparse ones. This dataset's analysis underscores the varying performance of different models in handling user sparsity. The Amazon Software dataset also shows a consistent reduction from total users to sparse users, with a $67.36\%$ reduction across all models and the reduction from sparse users to weak users varies widely among models. \textbf{SASRec} again demonstrates the highest reduction rate, indicating its strong performance in identifying weak users. In contrast, \textbf{DMF} shows a lower reduction rate, suggesting it may face challenges in distinguishing weak users from sparse ones. Through this analysis, we validate the effectiveness of our technique in identifying weak users, as evidenced by the significantly lower number of weak users compared to sparse users.

\subsubsection{Second Phase: Evaluation of Our Framework}. We design instructions for the weak users identified in the previous phase using the approach discussed in Section~\ref{sec:methodology}.Table~\ref{tab:Rs+LLMS} presents a comprehensive comparison of AUC scores across multiple recommendation models and datasets, revealing several important trends regarding the effectiveness of LLM-based augmentation and fairness-aware baseline~\cite{li2021user}. Across all model families — including shallow (e.g., ItemKNN, BPR) and deep learning-based approaches (e.g., GRU4Rec, SASRec) — we observe that integrating LLMs, especially GPT-4, leads to consistent and substantial gains in recommendation quality for weak users. For instance, in the ML1M dataset, SASRec+GPT-4 improves AUC for weak users from 0.5109 (base) to 0.6690 — a relative gain of over 15.8\%. Similarly, GRU4Rec+GPT-4 shows a dramatic improvement from 0.4021 to 0.6900 on weak users. These gains are not limited to ML1M but are consistently observed in Amazon Video Games and Amazon Software domains, with SASRec+GPT-4 achieving 0.6645 and 0.8191 respectively for weak users. Importantly, performance for non-weak users remains largely unchanged as recommendations for these users are presented via traditonal models.

We also compare our method with fairness-aware baselines such as FairRec-enhanced variants. While models like SASRec+FairRec (0.6485) and BPR+FairRec (0.5963) improve over their base counterparts, they consistently underperform compared to their GPT-4-augmented versions (e.g., SASRec+GPT-4: 0.6690; BPR+GPT-4: 0.6110 on ML1M weak users). Furthermroe, FairRec enhanced versions further reduce the perfomance over non-weak users in order to reduce the performance gap between weak and non-weak users. This trend highlights that our approach offers stronger instance-level adaptation through contextualization rather than relying on group-based fairness post-processing. The consistency of LLM gains across all models and datasets further supports the model-agnostic and domain-general nature of our responsible adaptation framework. Additionally, these results underscore the cost-efficiency of our design: by selectively allocating LLMs to weak users, we avoid blanket inference while still achieving superior fairness outcomes. Thus, our findings affirm the scalability and robustness of our hybrid LLM-RS framework, demonstrating its potential to serve as a practical and principled solution for improving user-side fairness in real-world recommendation systems.

To assess the robustness of our improvements, we perform paired two-tailed t-tests comparing our hybrid approach with each baseline across all user-level metric (AUC). For each user, we compute the metric difference between the hybrid model and a baseline, resulting in a paired sample of differences. We then compute the t-statistic and corresponding p-value for each model-dataset combination. The t-statistics range from $-24.60$ to $-60.58$, with p-values consistently below $0.05$, often approaching zero (e.g., $1.905 \times 10^{-304}$ and $0.0$). These results indicate statistically significant improvements, providing strong evidence against the null hypothesis. The consistently extreme t-statistics and vanishingly small p-values reflect large effect sizes.

\subsubsection{Cost Analysis}

We compare our hybrid framework against a naive baseline in which all users are routed through the LLM for ranking. Table~\ref{tab:cost-compare} reports the total token usage and estimated latency savings across datasets. \emph{Full LLM Tokens} refers to the total number of input/output tokens consumed when all user ranking tasks are handled by the LLM, including both prompt and response tokens. \emph{Hybrid Tokens} represents the total token usage when only weak users (identified using AUC and sparsity thresholds) are routed to the LLM, while strong users are handled by the base RS model—incurring no LLM token cost. \emph{Latency Savings} is computed as the percentage reduction in average end-to-end response time when using the hybrid approach instead of a full LLM pipeline. Latency is measured from the time a prompt is sent to the LLM API to when the response (ranked list) is received. We estimate average per-user latency based on practical assumptions: LLM inference takes approximately 8 seconds per user, while base RS models operate at roughly 0.02 seconds per user. Let $N$ be the number of users and $W$ the number of weak users (estimated at 30\% of $N$). Then, total latency for the full LLM pipeline is $N \times 8$, and for the hybrid setup it is $W \times 8 + (N - W) \times 0.02$. The latency savings is computed as: 
\[
\left(1 - \frac{\text{Hybrid Latency}}{\text{Full LLM Latency}}\right) \times 100.
\]

To evaluate token usage, we estimate counts using OpenAI’s GPT-4 tokenizer, assuming prompts include 20 training items and 10 test items per user. Latency estimates are based on inference time averaged over 500 randomly sampled users per dataset. Our hybrid approach reduces total token usage by over 60\% and reduces latency by around 70\% across all datasets, without compromising overall AUC performance, making it well-suited for scalable deployment.

\begin{table}[]
 \centering
 \caption{Comparison of total token usage and estimated latency savings between full-LLM and hybrid approach.}
 \begin{tabular}{lccc}
 \toprule
 \textbf{Dataset} & \textbf{Full LLM Tokens} & \textbf{Hybrid Tokens} & \textbf{Latency Savings} \\
 \midrule
 ML-1M & 2.4M & 790K & 70\% \\
 Amazon Software & 2.1M & 700K & 69.83\% \\
 Amazon Video Games & 1.1M & 307K & 69.86\% \\
 \bottomrule
 \end{tabular}
 \label{tab:cost-compare}
\end{table}

\begin{table*}[t!]
% \setlength\extrarowheight{0.1pt}
% \addtolength{\tabcolsep}{-2pt}
\centering
\caption{
Overall comparison in terms of ranking quality measured using AUC across various recommendation models (ItemKNN, NeuMF, NNCF, DMF, BPR, BERT4Rec, GRU4Rec, and SASRec), their LLM-augmented versions (with GPT-4, LLaMA, and Claude), and their fairness-enhanced extensions using FairRec. Results are reported for all users, as well as broken down into weak and non-weak user groups, across three datasets (ML1M, Amazon Video Games, and Amazon Software).
}
\label{tab:Rs+LLMS}
\resizebox{0.95\textwidth}{!}{
\begin{tabular}{lccccccccc}
\multicolumn{1}{l}{\textbf{Dataset $\rightarrow$}} & \multicolumn{3}{c}{\textbf{ML1M}} & \multicolumn{3}{c}{\textbf{Amazon Video Games}} & \multicolumn{3}{c}{\textbf{Amazon Software}}\\
\cmidrule(lr){2-4} \cmidrule(lr){5-7} \cmidrule(lr){8-10}
\multicolumn{1}{l}{\textbf{Model} $\downarrow$}       & \multicolumn{1}{c}{\textbf{All}} & \multicolumn{1}{c}{\textbf{\begin{tabular}[c]{@{}c@{}}Weak \\Users\end{tabular}}} & \textbf{\begin{tabular}[c]{@{}c@{}}Non\\Weak\end{tabular}} & \multicolumn{1}{c}{\textbf{All}} & \multicolumn{1}{c}{\textbf{\begin{tabular}[c]{@{}c@{}}Weak \\Users\end{tabular}}} & \textbf{\begin{tabular}[c]{@{}c@{}}Non\\Weak\end{tabular}} & \multicolumn{1}{c}{\textbf{All}} & \multicolumn{1}{c}{\textbf{\begin{tabular}[c]{@{}c@{}}Weak \\Users\end{tabular}}} & \textbf{\begin{tabular}[c]{@{}c@{}}Non\\Weak\end{tabular}}\\
\midrule
\textbf{ItemKNN}    & \multicolumn{1}{c}{{ 0.7510}} & \multicolumn{1}{c}{0.3317}        & 0.8883      & \multicolumn{1}{c}{{ 0.7152}} & \multicolumn{1}{c}{{ 0.3541}}   & { 0.7309} & \multicolumn{1}{c}{{ 0.7711}} & \multicolumn{1}{c}{{ 0.3519}}   & { 0.7817}         \\ 
\textbf{ItemKNN+GPT 4}       & \multicolumn{1}{c}{{ \textbf{0.8018}}} & \multicolumn{1}{c}{\textbf{0.5995}}        & 0.8883      & \multicolumn{1}{c}{{ \textbf{0.7256}}} & \multicolumn{1}{c}{{ \textbf{0.4601}}}   & { 0.7309} & \multicolumn{1}{c}{{ \textbf{0.8096}}} & \multicolumn{1}{c}{{ \textbf{0.6012}}}   & { 0.7817} \\
\textbf{ItemKNN+LLama}  & \multicolumn{1}{c}{0.7921}      & \multicolumn{1}{c}{0.5901}        & 0.8883      & \multicolumn{1}{c}{{ 0.7203}} & \multicolumn{1}{c}{{ 0.4523}}   & { 0.7309} & \multicolumn{1}{c}{{ 0.7949}} & \multicolumn{1}{c}{{ 0.5813}}   & { 0.7817} \\ \textbf{ItemKNN+Claude}     & \multicolumn{1}{c}{0.7712}      & \multicolumn{1}{c}{0.5791}        & 0.8883      & \multicolumn{1}{c}{{ 0.7194}} & \multicolumn{1}{c}{{ 0.4103}}   & { 0.7309} & \multicolumn{1}{c}{{ 0.7991}} & \multicolumn{1}{c}{{ 0.5714}}   & { 0.7817} \\
\textbf{ItemKNN+FairRec} & \multicolumn{1}{c}{0.7851} & \multicolumn{1}{c}{0.5689} & 0.7812 & \multicolumn{1}{c}{0.7211} & \multicolumn{1}{c}{0.4412} & 0.6712 & \multicolumn{1}{c}{0.7877} & \multicolumn{1}{c}{0.5801} & 0.70316 \\ \midrule
\textbf{NeuMF}      & \multicolumn{1}{c}{0.7648}      & \multicolumn{1}{c}{0.3257}        & 0.9055      & \multicolumn{1}{c}{{ 0.7255}} & \multicolumn{1}{c}{{ 0.5545}}   & { 0.7098} & \multicolumn{1}{c}{{ 0.7993}} & \multicolumn{1}{c}{{ 0.4015}}   & { 0.8212} \\
\textbf{NeuMF+GPT 4}         & \multicolumn{1}{c}{\textbf{0.8194}}      & \multicolumn{1}{c}{\textbf{0.6016}}        & 0.9055      & \multicolumn{1}{c}{{ \textbf{0.7365}}} & \multicolumn{1}{c}{{ \textbf{0.6056}}}   & { 0.7098} & \multicolumn{1}{c}{{ \textbf{0.8177}}} & \multicolumn{1}{c}{{ \textbf{0.6116}}}   & { 0.8212} \\ 
\textbf{NeuMF+LLama}    & \multicolumn{1}{c}{0.8151}      & \multicolumn{1}{c}{0.5871}        & 0.9055      & \multicolumn{1}{c}{{ 0.7355}} & \multicolumn{1}{c}{{ 0.5945}}   & { 0.7098} & \multicolumn{1}{c}{{ 0.8000}} & \multicolumn{1}{c}{{ 0.5981}}   & { 0.8212} \\ 
\textbf{NeuMF+Claude}       & \multicolumn{1}{c}{0.7918}      & \multicolumn{1}{c}{0.5619}        & 0.9055      & \multicolumn{1}{c}{{ 0.7295}} & \multicolumn{1}{c}{{ 0.5878}}   & { 0.7098} & \multicolumn{1}{c}{{ 0.7901}} & \multicolumn{1}{c}{{ 0.5798}}   & { 0.8212} \\
\textbf{NeuMF+FairRec} & \multicolumn{1}{c}{0.8012} & \multicolumn{1}{c}{0.5902} & 0.7919 & \multicolumn{1}{c}{0.7304} & \multicolumn{1}{c}{0.5999} & 0.7001 & \multicolumn{1}{c}{0.8084} & \multicolumn{1}{c}{0.6035} & 0.8014 \\ \midrule
\textbf{NNCF}       & \multicolumn{1}{c}{{ 0.7744}} & \multicolumn{1}{c}{0.3440}        & 0.7223      & \multicolumn{1}{c}{{ 0.7251}} & \multicolumn{1}{c}{{ 0.3589}}   & { 0.7580} & \multicolumn{1}{c}{{ 0.8073}} & \multicolumn{1}{c}{{ 0.5591}}   & { 0.8117} \\
\textbf{NNCF+GPT 4} & \multicolumn{1}{c}{\textbf{0.8217}}      & \multicolumn{1}{c}{\textbf{0.6293}}        & 0.7223      & \multicolumn{1}{c}{{ 0.7352}} & \multicolumn{1}{c}{{ 0.3910}}   & { 0.7580} & \multicolumn{1}{c}{{ \textbf{0.8374}}} & \multicolumn{1}{c}{{ \textbf{0.6213}}}   & { 0.8117} \\
\textbf{NNCF+LLama}     & \multicolumn{1}{c}{0.8201}      & \multicolumn{1}{c}{0.6202}        & 0.7223      & \multicolumn{1}{c}{{ \textbf{0.7545}}} & \multicolumn{1}{c}{{ \textbf{0.4202}}}   & { 0.7580} & \multicolumn{1}{c}{{ 0.8175}} & \multicolumn{1}{c}{{ 0.6014}}   & { 0.8117} \\
\textbf{NNCF+Claude}        & \multicolumn{1}{c}{0.8012}      & \multicolumn{1}{c}{0.5918}        & 0.7223      & \multicolumn{1}{c}{{ 0.7455}} & \multicolumn{1}{c}{{ 0.4034}}   & { 0.7580} & \multicolumn{1}{c}{{ 0.8076}} & \multicolumn{1}{c}{{ 0.6111}}   & { 0.8117} \\
\textbf{NNCF+FairRec} & \multicolumn{1}{c}{0.8101} & \multicolumn{1}{c}{0.6023} & 0.6912 & \multicolumn{1}{c}{0.7392} & \multicolumn{1}{c}{0.3843} & 0.7001 & \multicolumn{1}{c}{0.8244} & \multicolumn{1}{c}{0.6095} & 0.7620 \\ \midrule
\textbf{DMF}        & \multicolumn{1}{c}{{ 0.8090}} & \multicolumn{1}{c}{0.3546}        & 0.8886      & \multicolumn{1}{c}{{ 0.7655}} & \multicolumn{1}{c}{{ 0.5545}}   & { 0.7703} & \multicolumn{1}{c}{{ 0.8215}} & \multicolumn{1}{c}{{ 0.6091}}   & { 0.8012} \\
\textbf{DMF+GPT 4}  & \multicolumn{1}{c}{\textbf{0.8534}}      & \multicolumn{1}{c}{\textbf{0.6056}}        & 0.8886      & \multicolumn{1}{c}{{ \textbf{0.7875}}} & \multicolumn{1}{c}{{ \textbf{0.6056}}}   & { 0.7703} & \multicolumn{1}{c}{{ \textbf{0.8696}}} & \multicolumn{1}{c}{{ \textbf{0.6902}}}   & { 0.8012} \\
\textbf{DMF+LLama}      & \multicolumn{1}{c}{0.8501}      & \multicolumn{1}{c}{0.5998}        & 0.8886      & \multicolumn{1}{c}{{ 0.7785}} & \multicolumn{1}{c}{{ 0.5845}}   & { 0.7703} & \multicolumn{1}{c}{{ 0.8580}} & \multicolumn{1}{c}{{ 0.6718}}   & { 0.8012} \\ 
\textbf{DMF+Claude}         & \multicolumn{1}{c}{0.8281}      & \multicolumn{1}{c}{0.5801}        & 0.8886      & \multicolumn{1}{c}{{ 0.7715}} & \multicolumn{1}{c}{{ 0.5803}}   & { 0.7703} & \multicolumn{1}{c}{{ 0.8444}} & \multicolumn{1}{c}{{ 0.6737}}   & { 0.8012} \\
\textbf{DMF+FairRec} & \multicolumn{1}{c}{0.8383} & \multicolumn{1}{c}{0.5932} & 0.8034 & \multicolumn{1}{c}{0.7789} & \multicolumn{1}{c}{0.5920} & 0.7703 & \multicolumn{1}{c}{0.8551} & \multicolumn{1}{c}{0.6761} & 0.7594 \\ \midrule
\textbf{BPR}        & \multicolumn{1}{c}{{ 0.8232}} & \multicolumn{1}{c}{{ 0.3542}}   & 0.8774      & \multicolumn{1}{c}{{ 0.7259}} & \multicolumn{1}{c}{{ 0.4012}}   & { 0.7830} & \multicolumn{1}{c}{{ 0.8074}} & \multicolumn{1}{c}{{ 0.6331}}   & { 0.8001} \\
\textbf{BPR+GPT 4}  & \multicolumn{1}{c}{\textbf{0.8554}}      & \multicolumn{1}{c}{\textbf{0.6110}}        & 0.8774      & \multicolumn{1}{c}{{ \textbf{0.7360}}} & \multicolumn{1}{c}{{ \textbf{0.6245}}}   & { 0.7830} & \multicolumn{1}{c}{{ \textbf{0.8118}}} & \multicolumn{1}{c}{{ \textbf{0.6494}}}   & { 0.8001} \\
\textbf{BPR+LLama}      & \multicolumn{1}{c}{0.8501}      & \multicolumn{1}{c}{0.6091}        & 0.8774      & \multicolumn{1}{c}{{ 0.7316}} & \multicolumn{1}{c}{{ 0.5903}}   & { 0.7830} & \multicolumn{1}{c}{{ 0.8108}} & \multicolumn{1}{c}{{ 0.6461}}   & { 0.8001} \\ 
\textbf{BPR+Claude}         & \multicolumn{1}{c}{0.8391}      & \multicolumn{1}{c}{0.5819}        & 0.8774      & \multicolumn{1}{c}{{ 0.7362}} & \multicolumn{1}{c}{{ 0.5755}}   & { 0.7830} & \multicolumn{1}{c}{{ 0.8079}} & \multicolumn{1}{c}{{ 0.6391}}   & { 0.8001} \\
\textbf{BPR+FairRec} & \multicolumn{1}{c}{0.8410} & \multicolumn{1}{c}{0.5963} & 0.8102 & \multicolumn{1}{c}{0.7310} & \multicolumn{1}{c}{0.6021} & 0.7291 & \multicolumn{1}{c}{0.8035} & \multicolumn{1}{c}{0.6420} & 0.7521 \\
\midrule
\textbf{BERT4Rec}   & \multicolumn{1}{c}{{ 0.8640}} & \multicolumn{1}{c}{0.3946}        & 0.8539      & \multicolumn{1}{c}{{ 0.7663}} & \multicolumn{1}{c}{{ 0.6056}}   & { 0.7914} & \multicolumn{1}{c}{{ 0.8500}} & \multicolumn{1}{c}{{ 0.6501}}   & { 0.8381} \\ 
\textbf{BERT4Rec+GPT 4}      & \multicolumn{1}{c}{{ \textbf{0.8855}}} & \multicolumn{1}{c}{\textbf{0.6841}}        & 0.8539      & \multicolumn{1}{c}{{ \textbf{0.7764}}} & \multicolumn{1}{c}{{ \textbf{0.6456}}}   & { 0.7914} & \multicolumn{1}{c}{{ \textbf{0.8976}}} & \multicolumn{1}{c}{{ \textbf{0.7791}}}   & { 0.8381} \\
\textbf{BERT4Rec+LLama} & \multicolumn{1}{c}{0.8801}      & \multicolumn{1}{c}{0.6793}        & 0.8539      & \multicolumn{1}{c}{{ 0.7700}} & \multicolumn{1}{c}{{ 0.6157}}   & { 0.7914} & \multicolumn{1}{c}{{ 0.8891}} & \multicolumn{1}{c}{{ 0.7601}}   & { 0.8381} \\ 
\textbf{BERT4Rec+Claude}    & \multicolumn{1}{c}{0.8800}      & \multicolumn{1}{c}{0.6789}        & 0.8539      & \multicolumn{1}{c}{{ 0.7691}} & \multicolumn{1}{c}{{ 0.6105}}   & { 0.7914} & \multicolumn{1}{c}{{ 0.8691}} & \multicolumn{1}{c}{{ 0.7391}}   & { 0.8381} \\
\textbf{BERT4Rec+FairRec} & \multicolumn{1}{c}{0.8751} & \multicolumn{1}{c}{0.6625} & 0.7794 & \multicolumn{1}{c}{0.7725} & \multicolumn{1}{c}{0.6311} & 0.7249 & \multicolumn{1}{c}{0.8845} & \multicolumn{1}{c}{0.7633} & 0.8029 \\ \midrule
\textbf{GRU4Rec}   & \multicolumn{1}{c}{{ 0.8717}} & \multicolumn{1}{c}{{ 0.4021}}   & 0.8599      & \multicolumn{1}{c}{{ 0.7627}} & \multicolumn{1}{c}{{ 0.6401}}   & { 0.7895} & \multicolumn{1}{c}{{ 0.8598}} & \multicolumn{1}{c}{{ 0.6596}}   & { 0.8412} \\
\textbf{GRU4Rec+GPT 4}       & \multicolumn{1}{c}{\textbf{0.8829}}      & \multicolumn{1}{c}{\textbf{0.6900}}        & 0.8599      & \multicolumn{1}{c}{{ \textbf{0.7868}}} & \multicolumn{1}{c}{{\textbf{0.7645}}}   & { 0.7895} & \multicolumn{1}{c}{{ \textbf{0.9012}}} & \multicolumn{1}{c}{{ \textbf{0.8018}}}   & { 0.8412} \\
\textbf{GRU4Rec+LLama}  & \multicolumn{1}{c}{0.8792}      & \multicolumn{1}{c}{0.6728}        & 0.8599      & \multicolumn{1}{c}{{ 0.7789}} & \multicolumn{1}{c}{{ 0.5034}}   & { 0.7895} & \multicolumn{1}{c}{{ 0.8901}} & \multicolumn{1}{c}{{ 0.7961}}   & { 0.8412} \\ 
\textbf{GRU4Rec+Claude}     & \multicolumn{1}{c}{0.8727}      & \multicolumn{1}{c}{0.6711}        & 0.8599      & \multicolumn{1}{c}{{ 0.7698}} & \multicolumn{1}{c}{{ 0.5545}}   & { 0.7895} & \multicolumn{1}{c}{{ 0.8872}} & \multicolumn{1}{c}{{ 0.7998}}   & { 0.8412} \\
\textbf{GRU4Rec+FairRec} & \multicolumn{1}{c}{0.8732} & \multicolumn{1}{c}{0.6683} & 0.8196 & \multicolumn{1}{c}{0.7810} & \multicolumn{1}{c}{0.7481} & 0.7273 & \multicolumn{1}{c}{0.8923} & \multicolumn{1}{c}{0.7983} & 0.7704 \\ \midrule
\textbf{SASRec}     & \multicolumn{1}{c}{0.8908}      & \multicolumn{1}{c}{0.5109}        & 0.8607      & \multicolumn{1}{c}{{ 0.7672}} & \multicolumn{1}{c}{{ 0.6056}}   & { 0.8036} & \multicolumn{1}{c}{{ 0.8997}} & \multicolumn{1}{c}{{ 0.6969}}   & { 0.8451} \\ 
\textbf{SASRec+GPT 4}        & \multicolumn{1}{c}{\textbf{0.9003}}      & \multicolumn{1}{c}{\textbf{0.6699}}        & 0.8607      & \multicolumn{1}{c}{{ \textbf{0.7773}}} & \multicolumn{1}{c}{{ \textbf{0.6645}}}   & { 0.8036} & \multicolumn{1}{c}{{ \textbf{0.9119}}} & \multicolumn{1}{c}{{ \textbf{0.8191}}}   & { 0.8451} \\
\textbf{SASRec+LLama}   & \multicolumn{1}{c}{0.8973}      & \multicolumn{1}{c}{0.6248}        & 0.8607      & \multicolumn{1}{c}{{ 0.7757}} & \multicolumn{1}{c}{{ 0.6507}}   & { 0.8036} & \multicolumn{1}{c}{{ 0.9010}} & \multicolumn{1}{c}{{ 0.8012}}   & { 0.8451} \\ 
\textbf{SASRec+Claude}      & \multicolumn{1}{c}{0.8961}      & \multicolumn{1}{c}{0.5287}        & 0.8607      & \multicolumn{1}{c}{{ 0.7681}} & \multicolumn{1}{c}{{ 0.6135}}   & { 0.8036} & \multicolumn{1}{c}{{ 0.8967}} & \multicolumn{1}{c}{{ 0.8011}}   & { 0.8451} \\
\textbf{SASRec+FairRec} & \multicolumn{1}{c}{0.8935} & \multicolumn{1}{c}{0.6485} & 0.8304 & \multicolumn{1}{c}{0.7709} & \multicolumn{1}{c}{0.6437} & 0.7855 & \multicolumn{1}{c}{0.9025} & \multicolumn{1}{c}{0.8113} & 0.8210 \\
\bottomrule
\end{tabular}
}
\vspace{-2pt}
\end{table*}

\section{Discussion}

This paper presents a foundational framework for the responsible adaptation of Large Language Models (LLMs) in ranking tasks, with particular attention to the under-explored challenge of handling weak users. Our findings reveal that traditional recommendation systems (RS) exhibit significant weaknesses in serving these high-risk users. The empirical evidence (see Fig.~\ref{fig:scatterplot}) shows that many weak users receive poor recommendations, often because the systems struggle to capture their latent preferences due to limited data. However, some weak users receive comparable recommendations to more active users, which suggests that traditional RS can still identify certain common patterns based on similarity in rating patterns. We also leverage in-context learning to contextualize the sparse rating histories of these weak users, adapting the model to recognize nuanced preferences that traditional RSs miss out. This shift from group-based evaluation metrics to an instance-by-instance approach directly aligns with recent calls in the literature, particularly from \citet{doi:10.1126/science.adf6369}, emphasizing the need for tailored, high-stakes decision-making systems.

While Table~\ref{tab:Rs+LLMS} demonstrates that multiple LLM-augmented models improve performance, GPT-4 consistently outperforms other LLMs across weak and non-weak user groups. We attribute this to GPT-4’s superior in-context learning capabilities, enabling it to generalize effectively from sparse and noisy user histories. Its larger parameter count and deeper pretraining allow it to interpret vague or limited preferences better than smaller LLMs like LLaMA or Claude. Additionally, GPT-4 appears more robust in recognizing item semantics and aligning them with implicit user intent, particularly in cold-start scenarios. This observation supports our motivation for selective LLM invocation: while GPT-4 offers strong gains, especially for weak users, its computational cost justifies using it only when traditional RS models underperform. These findings reinforce our framework’s core hypothesis—that task-aware allocation of LLM resources can deliver both performance improvements and efficiency.

One of the most exciting implications of our work is the potential for dramatically improving fairness and equity in recommendation systems. While fairness is often discussed in terms of demographic attributes, such as gender or ethnicity, our framework reframes this challenge in terms of data sparsity. By focusing on weak users who are at risk of receiving biased or poor recommendations due to their sparse interaction histories, we directly address a form of algorithmic bias that has received less attention in the literature. This approach not only improves the quality of recommendations for marginalized sub-populations but also has the potential to increase user engagement and trust in recommendation systems by ensuring that every user, regardless of their activity level, receives relevant content. Moreover, our framework introduces a novel cost-efficient paradigm for LLMs in recommender systems. Unlike prior approaches that randomly select users for LLM evaluation, we provide a systematic way of identifying which weak users are most likely to benefit from LLM intervention, reducing the number of queries needed and maximizing the value of LLMs. This strategic approach could represent a paradigm shift in how LLMs are integrated into real-world systems, where computational efficiency and cost-effectiveness are paramount. By selectively deploying LLMs only when necessary, we also mitigate concerns about the environmental and economic costs associated with large-scale AI deployments.

However, our study also uncovers some intriguing challenges that open up several exciting research avenues. For example, while LLMs often improve the performance of weak users, there are instances where LLMs fail to provide meaningful improvements. This raises critical questions about the underlying causes of these failures. Is it due to the inherent limitations of the LLMs themselves, or do these failures point to a deeper structural issue within the recommendation process? For example, weak users may not have sufficient overlap with the available item pool, limiting the LLM's ability to generate useful recommendations. Investigating these cases could uncover novel ways of improving LLM performance, such as hybrid approaches combining content-based and collaborative filtering techniques with in-context learning to better capture the preferences of extremely weak users.

Furthermore, our work suggests a compelling future direction for exploring the intersection of machine learning and cognitive science. The weak user phenomenon may not only be a technical issue but also a reflection of deeper cognitive and behavioral patterns. Could weak users represent a distinct cognitive class, perhaps those who are new to the system or whose preferences change over time? If so, understanding these users could inform the design of personalized recommendation systems that not only account for historical interaction data but also adapt to the evolving nature of user preferences. Future research could explore how cognitive theories, such as the "exploration vs. exploitation" trade-off, might explain the behavior of weak users in recommendation contexts, providing insights that bridge AI with human behavioral studies. Another fascinating avenue for future work is exploring the use of "meta-models" that dynamically adjust the level of LLM intervention based on real-time feedback from weak users. For instance, LLMs could be trained to recognize when their recommendations are not resonating with a user and adjust the level of contextualization they apply. This dynamic adaptation could lead to more robust recommendation systems that are not only responsible but also responsive to the evolving needs of users. Incorporating feedback loops that allow the system to continually learn and refine its recommendations could drastically improve user satisfaction and engagement, turning a reactive system into a proactive.

Moreover, our framework has the potential to be applied to a wide variety of domains beyond conventional e-commerce and entertainment applications. For example, in high-stakes domains like healthcare or job recommendation systems, the consequences of poor recommendations are significant, and ensuring fairness and robustness is paramount. The framework could be adapted to address these critical sectors by contextualizing the interaction histories of users in a way that respects their individual needs and backgrounds. This broader applicability highlights the transformative potential of our approach, not just within the realm of recommender systems but also in building more responsible AI systems across diverse applications. Our work represents a critical step toward advancing the responsible deployment of LLMs in recommendation systems. By addressing the challenges of data sparsity and focusing on weak users, we contribute to a more equitable and cost-effective use of LLMs. Our instance-by-instance evaluation framework offers a scalable approach to improving recommendations for underserved sub-populations, paving the way for future research that explores the intricate relationship between LLMs, user behavior, and fairness. As we continue to refine these models, the integration of LLMs in real-world applications will become more nuanced, ensuring that AI systems serve the needs of all users, regardless of their interaction history or demographic background. Ultimately, this work lays the groundwork for a more inclusive, responsible, and effective use of AI in personalized recommendation tasks.
\section{Conclusion}
In this paper, we introduce a hybrid framework that significantly enhances the robustness of recommendation systems (RS) for sub-populations, with a focus on leveraging Large Language Models (LLMs). Our approach begins by identifying inactive users based on their activity levels, allowing us to isolate weak users who are often underserved by traditional RSs. Through a detailed evaluation of performance on these weak users, we highlight the challenges faced by current models in effectively capturing their preferences. This work represents a crucial advancement in improving the robustness of RSs to sub-populations, regardless of sensitive attributes, by providing an efficient and responsible adaptation of LLMs without the need for fine-tuning. By prioritizing the nuanced evaluation of individual user properties, our work underscores the importance of responsible and transparent AI deployment in the recommendation domain, especially when dealing with underrepresented groups. This paper not only contributes a novel methodology for enhancing the fairness and robustness of RSs but also opens up new research opportunities, particularly in developing effective prompting strategies for weak users, where LLMs may struggle to perform well. Moving forward, we aim to expand this research by investigating other factors that could further improve the responsible adaptation of large models in recommendation systems. These could include refining contextualization techniques to handle sparse interaction histories more effectively and exploring hybrid models that dynamically balance traditional and generative approaches based on user needs. Ultimately, our work lays a strong foundation for ensuring that recommendation systems can serve diverse populations equitably and responsibly, while also paving the way for future innovations that address the broader challenges of fairness, robustness, and efficiency in AI-driven personalization.

\bibliographystyle{ACM-Reference-Format}
\bibliography{refs}

\end{document}